\documentclass{aa}  
\usepackage[colorlinks=true,citecolor=blue]{hyperref}

\usepackage{breqn}

\usepackage{graphicx}
\usepackage{tabularx}
\usepackage{adjustbox}
\usepackage{caption}
\usepackage{txfonts}
\usepackage{subfig}
\begin{document}

   \title{Modelling chemical clocks}
   \subtitle{Theoretical evidences of the space and time evolution  of [s/$\alpha$] in the Galactic disc with {\em Gaia}-ESO survey}


   \author{M. Molero
          \inst{1,2}\fnmsep
          \and
          L. Magrini\inst{3}
          \and
          M. Palla\inst{4,5}
          \and
          G. Cescutti\inst{6,2,10}
          \and
          C. Viscasillas V{\'a}zquez\inst{7, 3}
          \and
          G. Casali\inst{8,9,4}
          \and
          E. Spitoni\inst{2}
          \and
          F. Matteucci\inst{2,6,10}
          \and
          S.Randich\inst{3}
          }

   \institute{Institut f\"ur Kernphysik, Technische Universit\"at Darmstadt, Schlossgartenstr. 2, Darmstadt 64289, Germany\\
              \email{mmolero@theorie.ikp.physik.tu-darmstadt.de; marta.molero@inaf.it}
        \and
            INAF, Osservatorio Astronomico di Trieste, Via Tiepolo 11, I-34131 Trieste, Italy
        \and
            INAF, Osservatorio Astrofisico di Arcetri, Largo E. Fermi 5, I-50125 Firenze, Italy
        \and
            Dipartimento di Fisica e Astronomia “Augusto Righi”, Alma Mater Studiorum, Università di Bologna, Via Gobetti 93/2, I-40129 Bologna, Italy
        \and
            INAF, Osservatorio di Astrofisica e Scienza dello Spazio, Via Gobetti 93/3, I-40129 Bologna, Italy
        \and
            Dipartimento di Fisica, Sezione di Astronomia, Universitá degli studi di Trieste, Via G.B. Tiepolo 11, I-34143 Trieste, Italy
        \and
            Institute of Theoretical Physics and Astronomy, Vilnius University, Sauletekio av. 3, 10257 Vilnius, Lithuania
        \and
            Research School of Astronomy \& Astrophysics, Australian National University, Cotter Rd., Weston, ACT2611, Australia
        \and
            ARC Centre of Excellence for All Sky Astrophysics in 3 Dimensions (ASTRO3D), Stromlo, Australia
        \and
            INFN - Sezione di Trieste, via A. Valerio 2, I-34100, Trieste, Italy
             }


 
  \abstract
   {Chemical clocks based on [s-process elements/$\alpha$ elements] ratios are widely used to estimate ages of Galactic stellar populations. However, the [s/$\alpha$] vs. age relations are not universal, varying with metallicity, location in the Galactic disc, and specific s-process elements. Moreover, current Galactic chemical evolution models struggle to reproduce the observed [s/$\alpha$] increase at young ages, particularly for Ba.}
   {Our aim is to provide chemical evolution models for different regions of the Milky Way disc in order to identify the conditions required to reproduce the observed [s/H], [s/Fe], and [s/$\alpha$] vs. age relations.}
   {We adopt a detailed, multi-zone chemical evolution model for the Milky Way including state-of-the-art nucleosynthesis prescriptions for neutron-capture elements: s-process elements are synthesized in asymptotic giant branch (AGB) stars and rotating massive stars, while r-process elements originate from neutron star mergers and magneto-rotational supernovae. Starting from a baseline model that successfully reproduces a wide range of neutron-capture element abundance patterns, we explore variations in gas infall/star formation history scenarios, AGB yield dependencies on progenitor stars, and rotational velocity distributions for massive stars. Results of our model are compared with the open clusters dataset from the sixth data release of the \textit{Gaia}-ESO survey.}
   {A three-infall scenario for disc formation aligns better with observed trends. The models capture the rise of [s/$\alpha$] with age in the outer regions but fail towards the inner regions, with larger discrepancies for second s-process peak elements. Specifically, Ba production in the last 3 Gyr of chemical evolution would need to increase by slightly more than half to match observations. S-process contribution from low-mass ($\mathrm{\sim 1.1\ M_\odot}$) AGB stars help reconcile predictions with data but require a too strong increase which is not predicted by current nucleosynthesis calculations, even with potential i-process contribution. Variations in the metallicity dependence of AGB yields either worsen the agreement or show inconsistent effects across elements, while distributions of massive star rotational velocities with lower velocity at high metallicities fail to improve results due to balanced effects on different elements.}
   {The predictions of our model confirm, as expected, that there is no single relationship [s/$\alpha$] vs. age, but that it varies along the MW disc. However, the current prescriptions for neutron-capture element yields are not able to fully capture the complexity of evolution, particularly in the inner disc. }


   \maketitle
%

\section{Introduction}
\label{sec: introduction}
The time dimension broadens our horizon in understanding the processes of formation and evolution of our Galaxy. 
The main tools for going back in time are stellar ages. 
The usual method adopted to derive stellar ages consists in comparing observed properties, such as magnitudes and colours, or inferred quantities, such as effective temperature (T$_{\rm eff}$) and surface gravity (log~g), with the output of stellar evolution models, the so-called isochrone fitting \citep[see, e.g.][]{Mints2018A&A...618A..54M}.  
However, ages of field stars are difficult to obtain with precision, particularly in some phases of stellar evolution, such as the low main sequence or the red giant branch, where we have strong degeneration among isochrones used to derive ages.

To overcome this problem, there are at least two possible alternative approaches: the use of asteroseismology \citep[see, e.g.][]{ChaplinMiglio2013} and the use of empirical methods that link some properties of stars with their age. The former is still model dependent, but it allows us to derive more precise ages (with typical errors of 20\%, \citealp{Miglio2021}; or as low as 10\%, \citealp{Montalbal2021}), while the latter needs to be calibrated on stars with known ages. The quantities used to infer stellar ages can be, e.g.,  stellar rotation, magnetic activity, photospheric abundances \citep[see, e.g.][]{Soderblom2010ARA&A..48..581S} and rocky exoplanet composition (\citealp{Weeks2024}). 
Several abundances and abundances ratios have been used in the literature to estimate stellar ages, such as, e.g. lithium abundance --A(Li)-- \citep[e.g.][]{Jeffries2023MNRAS.523..802J}, [C/N] \citep[e.g.][]{Masseron2015MNRAS.453.1855M, Casali2019A&A...629A..62C, Spoo2022AJ....163..229S}, and the abundance ratio between a slow neutron-capture element and an $\alpha$ element, [s/$\alpha$] \citep[e.g.][]{Feltzing2017MNRAS.465L.109F, Delgado2019A&A...624A..78D,  Casali2020A&A...639A.127C, Casali2023A&A...677A..60C, Tautvaisiene2021, Moya2022A&A...660A..15M, Viscasillas2022A&A...660A.135V, Berger2022,  Ratcliffe2024MNRAS.528.3464R, Sheejeelammal2024}. Some of these chemical age indicators depend on stellar evolution, such as A(Li) or [C/N], others are the result of Galactic chemical evolution, such as [$\alpha$/Fe] or [s/$\alpha$].
The calibration process is usually based on the measurement of stellar properties in member stars of open star clusters (OCs), whose ages are known from the isochrone fitting of the cluster sequence. In this way, the chemical clocks can be calibrated and the calibrated relations can be applied to field stars. 
In particular, chemical clocks based on [s/$\alpha$] have been widely used to derive `statistical' ages of the Galactic stellar populations \citep[e.g.][]{Casali2020A&A...639A.127C, Manea2023arXiv231015257M, Ratcliffe2024MNRAS.528.3464R, Boulet2024A&A...685A..66B}. 

The production of s-process material comes from rotating massive stars (e.g., \citealp{pignatari2010, Frischknecht2012, 2016MNRAS.456.1803F, Limongi2018}) and asymptotic giant branch (AGB) stars (e.g., \citealp{Gallino1998, Lugaro2003, Cristallo2015, Karakas2016, Busso2021}). In particular, rotating massive stars are strongly responsible for the production of the first peak s-process elements (Y, Sr, Zr) via weak s-process; while the main and the strong s-processes are due to low- and intermediate-mass stars (LIMSs) during their AGB phase. The main and the strong s-processes are mainly responsible for the production of elements belonging to the second (Ba, La, Ce) and to the third (Pb, Au, Bi) s-process peak, respectively. Because of the longer lifetimes of low-mass stars one should expect an increasing trend of the [s/H] with decreasing stellar ages, in particular for the elements belonging to the second and third s-process peak. Such a trend is observed for the [s/Fe] ratios as well. The increasing trend towards younger ages has been first observed by \citet{d'orazi2009} for the abundance of Ba in a large sample of OCs, and then confirmed by a large number of works (e.g., \citealp{Maiorca2011, Maiorca2012, Jacobson2013, Mishenina2015, Casali2020A&A...639A.127C, Baratella2021, Viscasillas2022A&A...660A.135V, Sheminova2024}). Not only Ba, but also other neutron-capture elements belonging to both the first and second s-process peaks show a similar increase (\citealp{Magrini2018, Frasca2019}, but see also \citealp{Yong2012, d'orazi2017}). The same reasoning is applied to the chemical clocks [s/$\alpha$]. Since $\alpha$-elements are mainly produced by massive stars on short timescales while s-process elements are released to the interstellar medium (ISM) at later times, their ratio should be higher at young ages. If such a trend with age is tight, then these abundance ratios can be used to infer stellar ages by means of empirical relations.

The discussion on the universality of the relationship between age and [s/$\alpha$] derived from the stars in the solar neighborhood and its dependence on the Galactic stellar populations --thin or thick disc--, on the position across the Galactic disc, and  on metallicity remained open for a long time \citep[see, e.g.][]{Feltzing2017MNRAS.465L.109F, Delgado2019A&A...624A..78D}. The works based on the large sample of open clusters observed in the {\em Gaia}-ESO survey --spanning a wide range in age and Galactocentric distances-- confirmed a strong radial variation along the disc and a parameterisable dependence on metallicity \citep[see][]{Casali2020A&A...639A.127C, Viscasillas2022A&A...660A.135V}. A first attempt to understand the variations in the age-[s/$\alpha$] relationships along the Galactic disc was made in \citet{Magrini2021A&A...646L...2M} by taking into account the inside-out formation of the disc and the metallicity dependence of the yields, particularly for the s-process \citep{Cristallo2011ApJS..197...17C, Vescovi2021Univ....8...16V, Vescovi2023EPJWC.27906001V}. To investigate the different hypotheses, it is important to rely on a self-consistent chemical evolution model. 

From a modeling perspective, chemical evolution has encountered difficulties in reproducing the observed rise in the [s/Fe] and [s/$\alpha$] ratios at very young stellar ages. Models frequently show a flat trend in younger stars, which could result from either a balanced delay in enrichment timing between second-peak s-process elements from AGB stars and Fe from Type Ia supernovae (affecting [s/Fe] ratios), and/or a reduced production of s-process elements during the most recent billion years of chemical evolution (impacting both [s/Fe] and [s/$\alpha$] ratios). Different solutions have been proposed in literature to reconcile the model predictions with the observed trend, such as increasing Ba production at higher metallicities (\citealp{Ratcliffe2024MNRAS.528.3464R}) or in lower-mass AGB stars (\citealp{d'orazi2009, Maiorca2012}). These solutions represent promising progress in understanding the production and distribution of s-process elements. However, they often depend on comparisons with a single element, Ba, and focus solely on the solar neighborhood. In this study, we test some of these existing solutions along with new approaches, examining both the production in AGB and in rotating massive stars. We compare our models with both Ba abundances (as representatives of the second peak s-process elements) and Y abundances (representing the first peak) across three different Galactocentric regions. We adopt a well-tested chemical evolution model that includes state-of-the-art nucleosynthesis calculations for neutron-capture elements, including contributions from AGB and rotating massive stars for the s-process, as well as neutron star mergers and magneto-rotational supernovae for the r-process. In fact, it is important to note that most elements typically classified as s-/r-process elements actually have dual (or even more) production pathways. Specifically, the fractions of Y and Ba produced by the s-process at solar metallicities are 78$\%$ and 89$\%$, respectively (\citealp{Prantzos2020}), with the remaining fractions primarily accountable to the r-process (and only negligible or no contribution from the p-process). Here, our model results are compared to the open clusters set of the last data release of the \textit{Gaia}-ESO survey (\citealp{Randich2022}). Our baseline model has been shown to provide a satisfactory fit to the standard abundance patterns of [El/Fe] versus [Fe/H] for a wide range of neutron capture elements in different Galactocentric regions observed by the \textit{Gaia}-ESO survey (see \citealp{Molero2023}).

The paper is structured as follows: in Section \ref{sec:data}, we describe the \textit{Gaia}-ESO OC sample. In Section \ref{sec: GCE model}, we present the chemical evolution framework including the infall and the nucleosynthesis prescriptions. In Section \ref{sec: results}, we present results of models with different prescriptions for infall, AGB and rotating massive stars, and their impact on the observed chemical clocks. Finally, in Section \ref{sec: discussion and conclusions}, we address the limitations of the model, suggest potential improvements, and provide our conclusions.

\section{Observational data}
\label{sec:data}
The {\it Gaia}-ESO survey \citep{Randich2022,  Gilmore2022A&A...666A.120G} is a large public spectroscopic survey. It observed for 340 nights at the Very Large Telescope (VLT) from the end of 2011 to 2018, gathering  $\sim$190000 spectra, for nearly 115000 targets belonging to all the main Galactic populations. 
{\it Gaia}-ESO observations were carried on  at two different resolving powers, R: the medium-resolution sample was observed with GIRAFFE at R$\sim$20000 and the high-resolution sample with UVES at 
R$\sim$47000. In particular, high resolution spectra,  covering a wide spectral range from 480.0~nm to 680.0~nm (U580) or from 420.0~nm to 620.0~nm (U520),  provide abundances for about 30 different ions and including numerous neutron-capture elements, such as Y, Zr, La, Ce, Ba, Eu, Nd, Pr and Sm, discussed, e.g., in \citet{Molero2023}.
{\em Gaia}-ESO dedicated about 40\% of its observation time to the the population of open star clusters, used both to determine their specific properties and as calibrators \citep{Pancino2017A&A...598A...5P, Bragaglia2022A&A...659A.200B}.  
The sample clusters covered a wide range in age, Galactocentric distance, mass and metallicity \citep[see][]{Randich2022}.
Stellar parameters and chemical abundances from {\em Gaia}-ESO have been combined with homogeneous ages and distances from {\em Gaia} \citep[e.g.][]{Cantat2020A&A...640A...1C}. 

In the present work, we adopt the same sample of open clusters used in \citet{Magrini2023A&A...669A.119M} and in \citet{Palla2024}.  In this sample of 62 OCs, only  clusters older than 100 Myr were considered, to avoid observational biases in the chemical study of young stars. 
For the distribution in age and distances of this sample clusters, we refer to \citet[see their Fig.~1]{Viscasillas2022A&A...660A.135V}, in which the membership selection is also described. In addition, following the discussion in \citet{Magrini2023A&A...669A.119M} and in \citet{Palla2024}, in each clusters we made a further selection of member stars based on their stellar parameters.
From an evaluation of internal trends between abundances and stellar parameters, present in different spectroscopic surveys \citep[see][for a discussion]{Magrini2023A&A...669A.119M}, we consider only stars with $\log g$  > 2.5 and microturbulent velocity $\xi$ < 1.8 km~$^{-1}$  to compute the mean cluster abundances. 

We complemented our sample of OCs, with a sample of field stars from the same database, selected as in \citet{Viscasillas2022A&A...660A.135V}. 
Due to the wide metallicity range covered by the field stars, we use them to study the evolution in the [El/Fe] vs [Fe/H] planes. 
However, owing to the very large uncertainties in determining their ages, we do not  use them in the  age vs. abundance ratios diagrams. 

The most studied chemical clocks in the literature are [Y/Mg] and [Y/Al] (\citealp{TucciMaia2016, Nissen2017, Spina2018, Tautvaisiene2021, Magrini2021A&A...646L...2M, Berger2022, Sheejeelammal2024},, but see also \citealp{DelgadoMena2019, Jofre2020, Ratcliffe2024MNRAS.528.3464R}). For this  work, we selected two s-process elements: Y for the first peak elements and Ba for the second peak. To compute the abundance ratios [s/$\alpha$] we used Si as representative of the class of the $\alpha$ elements, due to the larger sampling of OCs relative to Al and the more precise nucleosynthesis prescriptions from massive stars relative to Mg (see discussion in e.g., \citealp{Romano2010, palla2022, Jost2024}). 


\section{The chemical evolution models}
\label{sec: GCE model}

In this Section, we present the chemical evolution models adopted to study the variations in the [s/$\alpha$] vs. Age relations along the Galactic disc. The models are as follows:
\begin{itemize}
    \item The revised two-infall model proposed by \citet{Palla2020} (see also \citealp{Spitoni2019} and \citealp{Spitoni2021}) and previously adopted in \citet{Molero2023} to study the distribution of the neutron-capture elements along the MW disc.
    \item The three-infall model from \citet{Spitoni2023}. In particular, here we will use the version proposed by \citet{Palla2024}, extended to the whole disc.
\end{itemize}

The revised two-infall model is a variation of the classical two-infall model of \citet{Chiappini1997} (see also \citealp{Chiappini2001}) designed to explain the dichotomy in $\alpha$-element abundances observed both in the solar vicinity (e.g., \citealp{Hayden2014, recio-blanco2014, Mikolaitis2017}) and at various radii (e.g., \citealp{Hayden2015}). The model suggests that the initial primordial gas infall event formed the chemically thick disc (corresponding to the high-$\alpha$ sequence), while the second infall event, occurring approximately 3-4 Gyr years later, formed the chemically thin disc (the low-$\alpha$ sequence). It is important to note that the two-infall model used here does not aim to distinguish the thick and thin disc populations based on geometric or kinematic criteria (see \citealp{Kawata2016} for further discussion). The three-infall model, on the other hand, is an extension of the two-infall one, designed to replicate the low-$\alpha$ sequence through two distinct gas infall episodes (the most recent of which began approximately $2.7$ Gyr ago). This approach aims to account for the recent chemical impoverishment in metallicity with low [$\alpha$/Fe] values identified by \citet{PVP_Ale} (reported for the first time by \citealp{Spina2017}), as well as the recent increase in star formation activity described by \citet{Ruiz-Lara2020} (see also \citealp{Isern2019, Mor2019}). 

\subsection{Basic equations}

The basic equations that describe the evolution of the fraction of gas mass in the form of a generic chemical element $i$ are:
\begin{equation}
        \dot{G}_i(R,t)=-\psi(R,t)X_i(R,t)+\dot{G}_{i,inf}(R,t)+R_i(R,t),
\label{eq: GCE equation}
\end{equation}
where $X_i(R,t)$ represents the abundance by mass of a given element $i$. The term $\psi(R,t)X_i(R,t)$ is the rate at which chemical elements are subtracted by the interstellar medium (ISM) to be included in stars. The star formation rate (SFR) is parameterised following the Schmidt-Kennicutt law (\citealp{Kennicutt1998}), as:
\begin{equation}
    \psi(R,t)\propto\nu \sigma_{gas}(R,t)^k,
\end{equation}
with a law-index $k=1.5$ and $\sigma_{\mathrm{gas}}$ and $\nu$ being the surface gas density and the star formation efficiency.

$R_{\mathrm{i}}(R,t)$ is the rate of restitution of matter from the stars with different masses into the ISM in the form of the element $i$. It takes into account the nucleosynthesis from a variety of stars and phenomena, including stellar winds, SN explosions of all kinds, novae and neutron-star mergers. For the complete expression we refer to \citet{Matteucci2012}. Here, we discuss in more detail the nucleosynthesis from the different sources in section \ref{sec: Nucleosynthesis prescriptions}.

For the two-infall model, the accretion term on the right-hand side of Equation \ref{eq: GCE equation} is computed as:
\begin{equation}
   \dot{G}_{i,inf}(R,t)=A(R)X_{i,inf}e^{-\frac{t}{\tau_1}}+\theta(t-t_{max})B(R)X_{i,inf}e^{\frac{t-t_{max}}{\tau_2}},
\end{equation}
where $X_{\mathrm{i,inf}}$ is the composition of the infalling gas, here assumed to be primordial for both the infall episodes. $\tau_1\simeq1\ \mathrm{Gyr}$ is the infall timescale for the first infall event, assumed to be constant for all radii. During the first infall event, the star formation efficiency is set to $\nu=2\ \mathrm{Gyr^{-1}}$, fixed at all Galactocentric distances as well. On the contrary, for the second gas infall event, the timescale $\tau_2$ is assumed to vary with the radius according to the inside-out scenario (see e.g., \citealp{Matteucci1989, Romano2000, Chiappini2001}), with longer timescales towards larger distances. The star formation efficiency during the second infall episode is assumed to be a function of the Galactocentric distances as well, with $\nu \simeq1\ \mathrm{Gyr^{-1}}$ at $R_{\mathrm{GC}}=8\ \mathrm{kpc}$. In order to correctly reproduce the slope of the abundance gradients of Fe and $\alpha$-elements, the star formation efficiency assumes larger values toward the inner part of the disc (see e.g., \citealp{Grisoni2018}). $t_{max}\simeq3.25\ \mathrm{Gyr}$ is the time for the maximum infall on the disc and it corresponds to the start of the second infall episode. $\theta$ is the Heavyside step function. The parameters $A(R)$ and $B(R)$ are fixed to reproduce the surface mass density of the MW disc at the present time in the solar neighborhood as provided by \citet{McKee2015}, equal to $\mathrm{47.1 \pm\ 3.4\ M_\odot \mathrm{pc^{-2}}}$. At different Galactocentric radii, as discussed in \citet{Palla2020}, the surface mass densities of the chemically thick and thin discs are assumed to follow an exponential profile, as:
\begin{equation}
    \Sigma_{thick;thin}(R)=\Sigma_{0,thick;thin}e^{-R/{R_{d,thick;thin}}},
\end{equation}
where $R_{\mathrm{d;thick}}=2.3\ \mathrm{kpc}$ and $R_{\mathrm{d;thin}}=3.5\ \mathrm{kpc}$ are the disc scale lengths for the high-$\alpha$ and for the low-$\alpha$ disc, respectively.

For the three-infall case, the accretion term in Equation \ref{eq: GCE equation} has the following form:
\begin{dmath}
    \dot{G}_{i,inf}(R,t)=A(R)X_{i,inf}e^{-\frac{t}{\tau_1}}+
    \theta(t-t_{max,1})B(R)X_{i,inf}e^{\frac{t-t_{max,1}}{\tau_2}}
    +\theta(t-t_{max,2})C(R)X_{i,inf}e^{\frac{t-t_{max,2}}{\tau_3}},
\end{dmath}
where $\tau_3=1\ \mathrm{Gyr}$ is the timescale of the third infall event and $t_{\mathrm{max,2}}=11\ \mathrm{Gyr}$ is the Galactic time of the start of the third accretion event. The coefficient $C(R)$ is set in order to fit the present-day total surface density of the third accretion phase, $\Sigma_3$. Since the three-infall model splits the low-$\alpha$ sequence in two gas accretion sequences, the sum between $\Sigma_2$ and $\Sigma_3$ is equal to the density profile of the low-$\alpha$ sequence of the two-infall model. In particular, the ratio $\Sigma_2/\Sigma_3$ as a function of the Galactocentric distance as well as all the variables of the third-infall event are fixed following the prescriptions of the best model of \citet{Palla2024}. All the other parameters for the first and the second infall phases are as in the two-infall model case. In \citet{Spitoni2023}, the three-infall model has been first introduced to predict the recent chemical impoverishment characteristic of the young, massive stellar populations observed in Gaia DR3. Similarly, in \citet{Palla2024}, the model was employed to account for the properties of young OCs of the \textit{Gaia}-ESO survey.

\begin{figure}
    \centering
    \includegraphics[width=1\columnwidth]{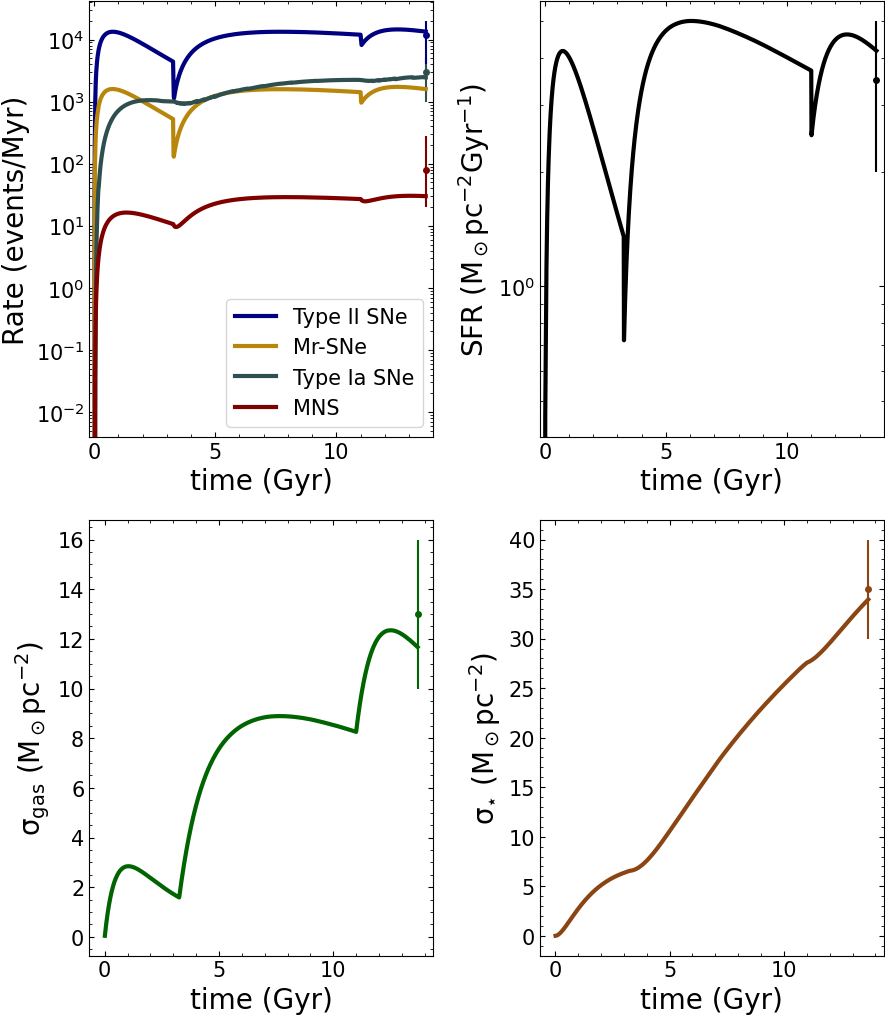}
    \caption{Time evolution of the rate of Type Ia SNe, Type II SNe, MR-SNe and MNS (upper left panel), SFR (upper right panel), surface densities of gas (lower left panel) and of stars (lower right panel) as predicted by the three-infall model. Predictions of the model are compared to present day values from \citet{Cappellaro1999} (for SNe rates), from \citet{Abbott2021} (for MNS rate) and from \citet{Prantzos2018} (for SFR and surface gas and stars densities).}
    \label{fig: rates}
\end{figure}

Comparisons of the evolution of some important quantities predicted by the three-infall model to present-day observations are reported in Figure \ref{fig: rates}. The predicted SFR, surface densities of stars and gas are computed in the solar neighbourhood and compared with present-day estimates as suggested by \citet{Prantzos2018}. Rates of Type Ia, Type II SNe, MR-SNe and MNS are averaged over the whole disc and compared with the observational values estimated of \citet{Cappellaro1999} and \citet{Abbott2021}. 

\subsection{Nucleosynthesis prescriptions}
\label{sec: Nucleosynthesis prescriptions}

In this work, we adopt yield prescriptions similar to those used by \citet{Molero2023}, where a detailed description can be found. Here, we provide a brief recap.

Yields for low- and intermediate-mass stars (LIMSs; with initial masses $\mathrm{1.0\leq M/M_\odot \le 8.0}$) are taken from the FRUITY database (\citealp{Cristallo2009, Cristallo2011ApJS..197...17C, Cristallo2015}). We adopt the non-rotational set which provides yield grids for progenitors with 8 initial stellar masses, from $\mathrm{1.3\ M_\odot}$ to $\mathrm{6.0\ M_\odot}$, and 12 values of metallicities from $\mathrm{Z = 4.8\times10^{-5}}$ to $\mathrm{Z = 2.0\times10^{-2}}$. Yields for the $\mathrm{8\ M_\odot}$ have been obtained extrapolating the FRUITY yields for all metallicity values to reduce the gap between LIMSs and massive stars yields. Both the main and the strong s-process component are produced in AGB stars. However, as underlined in \citet{Magrini2021A&A...646L...2M} (see \citealp{Vescovi2021}), isotopic ratio measurements in presolar SiC grains showed that the neutron density in FRUITY models is likely overestimated (\citealp{Liu2018}). Consequently, s-process yields from FRUITY AGB models are often reduced by a factor of two or more in chemical evolution studies (e.g., \citealp{rizzuti2019, Rizzuti2021, Molero2024}, see \citealp{Cescutti2022} for a recent review). In this work, we relax this reduction across all metallicities unless stated otherwise.

Massive stars (with initial masses $\mathrm{8\leq M/M_\odot \leq 120}$) are assumed to explode either as normal core-collapse supernovae (CC-SNe) or as  magneto-rotational supernovae (MR-SNe). For normal CC-SNe, yields are taken from \citet{Limongi2018} set R. We run models with either a constant initial rotational velocity of $\mathrm{150\ km~s^{-1}}$ or a velocity distribution. About $\mathrm{20\%}$ of massive stars in the $\mathrm{10-25\ M_\odot}$ range are assumed to explode as MR-SNe, representing a source of r-process material with yields from \citet{Nishimura2017} (model $\mathrm{L0.75}$). 

The second r-process site is represented by merging neutron stars (MNS). They are computed as systems of two neutron stars with masses of $\mathrm{1.4\ M_\odot}$, originating from progenitors in the $\mathrm{9-50\ M_\odot}$ range. Mergers follow the DTD of \citet{Simonetti2019} (see also \citealp{Greggio2021} for a detailed discussion) and their rate is constrained to reproduce the latest estimation from \cite{Abbott2021}. Yields of r-process material from MNS are obtained by scaling to solar the yield of Sr measured in the re-analysis of the spectra of the kilonova $\mathrm{AT2017gfo}$ by \cite{Watson2019} (after having considered uncertainties in its derivation; see \citealp{Molero2021} for details). The Eu yield so obtained is consistent with the theoretical calculation of \citet{Korobkin} and with estimates from \citet{Matteucci2014}.

It is however important to stress that some of the main uncertainties in the production of r-process elements in chemical evolution models are the adopted prescriptions for MR-SNe. While for MNS we can rely on observations for the yields (\citealp{Watson2019}), for the delay-time distribution (DTDs, from observations of sGRBs, e.g., \citealp{G16}) and for the rate (\citealp{Kalogera2004, Abbott2021}), for MR-SNe we have to arbitrarily choose (i) the fraction of normal CC-SNe which can die as a MR-SNe, (ii) the mass range and (iii) the yields (since different authors often obtain different results). In \citet{Molero2023}, we show some possibilities in which these free parameters can be fixed in a self-consistent manner. Nevertheless, ambiguities persist, and the possibility of a substantial level of degeneracy remains a significant consideration. In particular, while for normal CC-SNe yield grids for a wide number of different progenitors exist, this is not the case for MR-SNe for which the prescription for one single stellar mass (e.g., $\mathrm{35\ M_\odot}$) is interpolated to all the range of masses considered. Therefore, in our chemical evolution model, MR-SNe are included solely as sources of r-process material, excluding their contribution to Fe and other elements.

Finally, for Type Ia SNe, we assume the single-degenerate scenario for the progenitors, with stellar yields from \citet{Iwamoto1999} (model W7).

\begin{figure*}
    \includegraphics[width=1\textwidth]{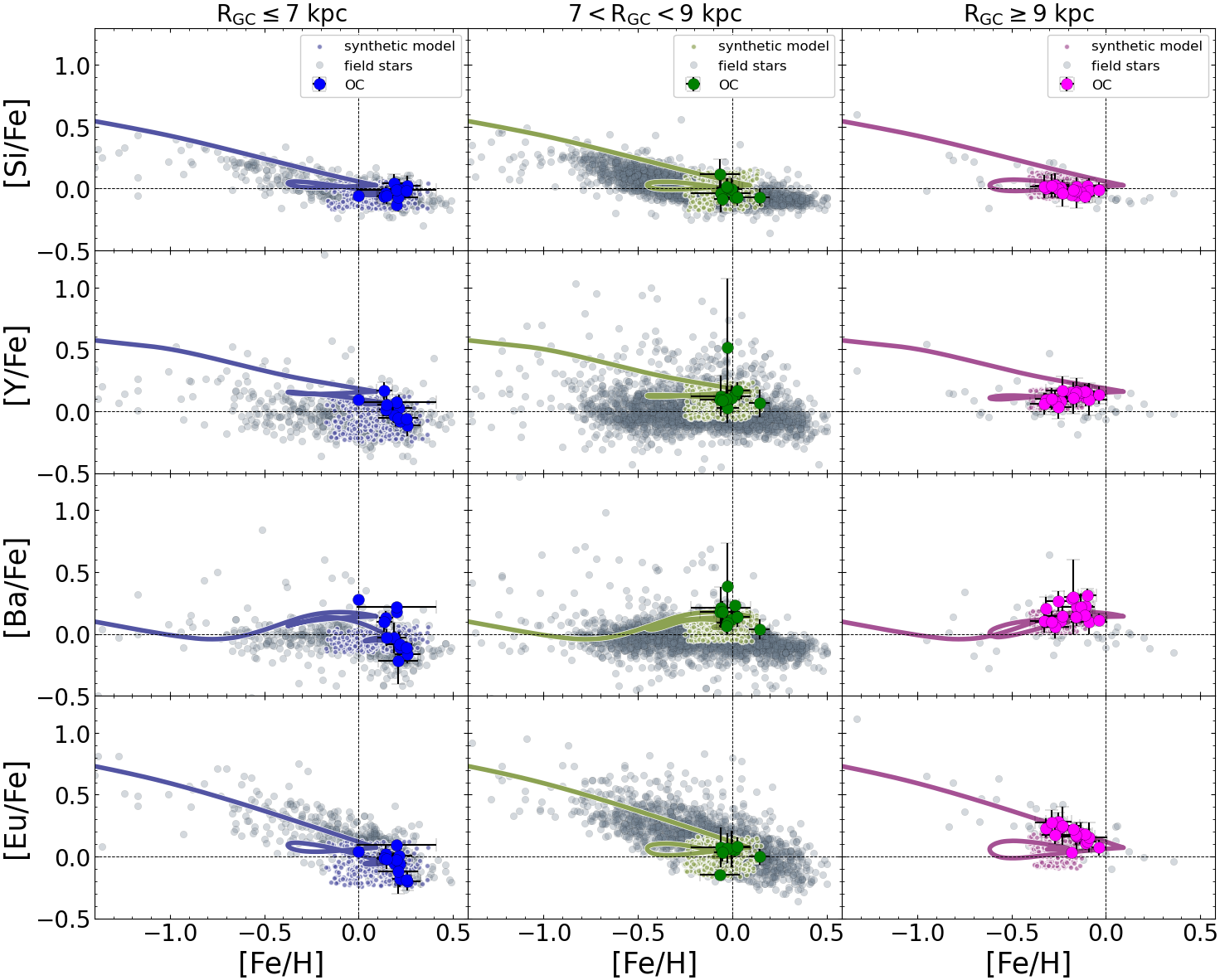}
     \caption{[X/Fe] vs. [Fe/H] abundance patterns for Si, Y, Ba and Eu as predicted by the three-infall model at $\mathrm{R_{GC}=6,\ 8,\ 12\ kpc}$ compared to the sample of OCs (colored dots) and field (grey dots) stars in the inner, solar and outer regions, respectively. Results of the synthetic model as well as OCs are plotted only for $\mathrm{Age \leq 3 Gyr}$.}%
 \label{fig: EuFe SiFe vs FeH}%
\end{figure*}

As a sanity check for the nucleosynthesis prescriptions of r-process, Fe and $\alpha$-elements, we show the evolution of the Si, Y, Ba and Eu abundances in the typical [X/Fe] vs. [Fe/H] diagnostic diagrams (see Figure \ref{fig: EuFe SiFe vs FeH}) predicted by the three-infall model for the three different Galactocentric distances considered. From a nucleosynthesis point of view, here prescriptions are identical to the one adopted in \citet{Molero2023}, with reduced s-process AGB production. Note that, following \citet{Palla2024} (see also \citealp{Spitoni2024}), we account for observational uncertainties in the model predicted abundances by adding a random error at each time step, $t$: 
\begin{equation} [X/H]_{new}(t) = [X/H]_{old}(t) \pm N([0,\ \sigma_{[X/H]}]), 
\end{equation} 
where $N$ is a random generator with a normal distribution, and $\sigma_{\mathrm{[X/H]}}$ represents the average observational error of the open cluster sample. The model well agrees with the overall abundance patterns, in particular in the last $\mathrm{3\ Gyr}$ of evolution, as shown by the synthetic model results. Only in the case of Eu in the outer regions the model underproduces the OCs data in the age range of interest, reproducing their pattern too early in the Galactic evolution. The lower metallicity data described by the field stars is correctly reproduced for all elements and in all the region of interest, except for Y in the inner regions where data are overestimated. In the case of Eu, in particular, the agreement with the trends observed for field stars in the low metallicity regime ensures that the prescriptions for the MR-SNe nucleosynthesis are correctly fine-tuned. 

\section{Results}
\label{sec: results}

In this section, we use stellar ages as indicators of the passage of time, instead of look-back time. 
Thus, for relationships between ages and abundance ratios, positive slopes indicate relationships that decrease with the passing of time, while negative slopes indicate relationships that increase. 
Here we show results from our chemical evolution model to be compared to the [s/H], [s/Fe] and [s/$\alpha$] vs. age trends observed in the OCs dataset at different Galactocentric distances. The focus is mainly on the chemical clock trends and on how these are influenced by different infall events and nucleosynthesis prescriptions from AGB and massive stars.

\subsection{The effect of primordial gas inflows}
\label{sec: The effect of primordial gas inflows}

\begin{table*}[h]
    \caption{\label{Tab: model infall} Input parameters for the chemical evolution models represented in Figures \ref{fig: sFe vs Age data and infall}, \ref{fig: sH vs Age data and infall} and \ref{fig: sSi vs Age data and infall}. See text for more details.}
    \centering
    \begin{tabular}{cccc}
    \hline
      Model Name   &    Chemical evolution scenario   &    Yields for AGB    &   Yields for MS \\
      \hline
      Model 1      &    two-infall model                &    FRUITY original   &   \cite{Limongi2018} - $\mathrm{v_{rot}=150\ km~s^{-1}}$ \\
      Model 2      &    three-infall model              &    FRUITY original   &   \cite{Limongi2018} - $\mathrm{v_{rot}=150\ km~s^{-1}}$ \\
    \hline
    \end{tabular}
\end{table*}

\begin{figure*}
    \includegraphics[width=1\textwidth]{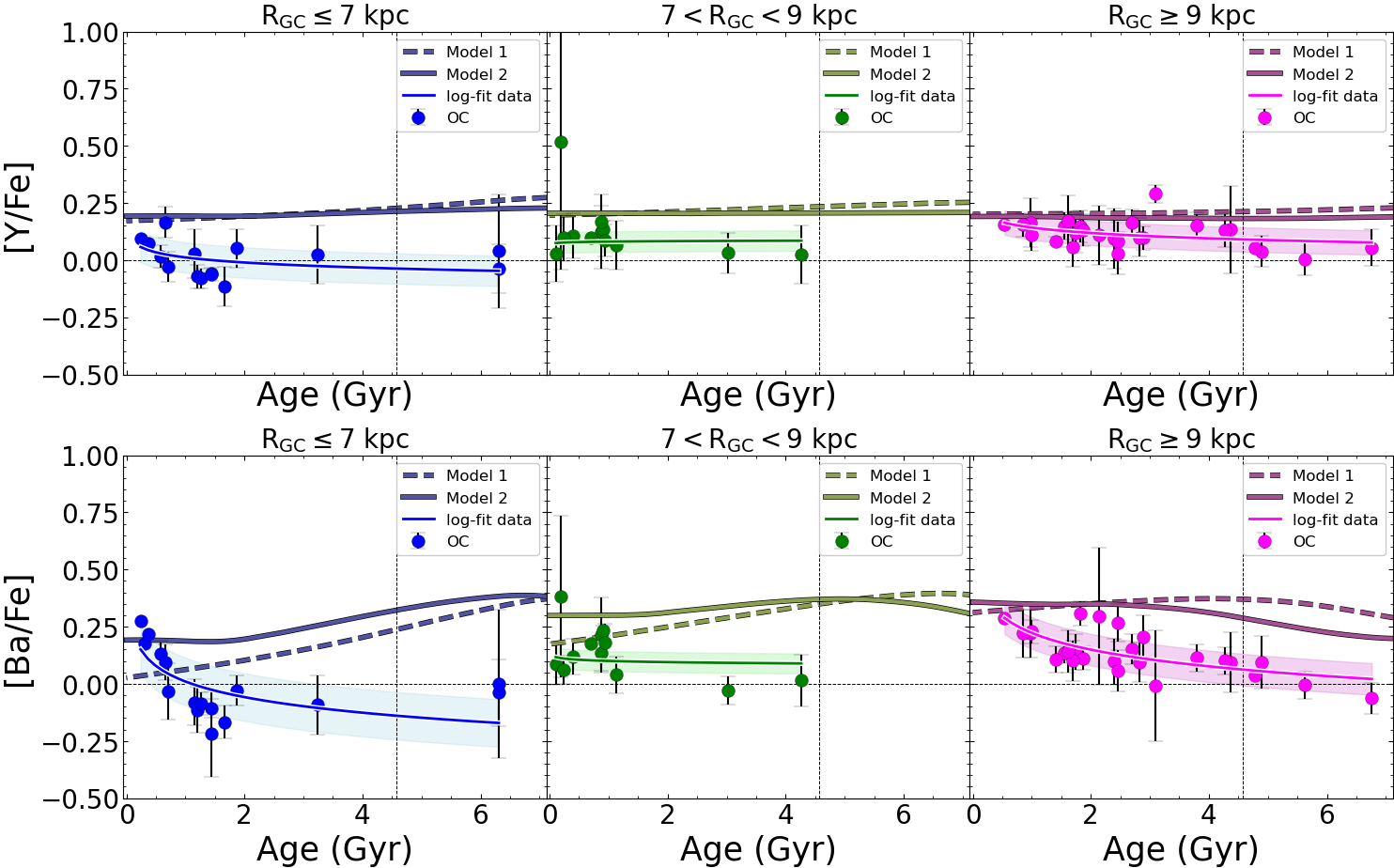}
     \caption{$\mathrm{[s/Fe]}$ vs Age observed trends for Y (upper panel) and Ba (lower panel) divided in the three Galactocentric regions of interested. The logarithmic fits of the OC samples in the three regions are shown as blue (inner region), green (solar region) and magenta (outer region) curves. Predictions of the chemical evolution model in the case of a two- (dashed lines) and of a three-infall (solid lines) scenario are compared to the OC sample. See Table \ref{Tab: model infall} for reference.}%
 \label{fig: sFe vs Age data and infall}%
\end{figure*}

As outlined in \ref{sec: introduction}, different slopes for the $\mathrm{[X/Y]}$ vs. age relations are indications of a different contribution from different nucleosynthesis sources of the two elements $\mathrm{X}$ and $\mathrm{Y}$. It is well-known that, while the observed relations of [$\alpha$/Fe] vs. age have a positive slope, the one of the $\mathrm{[s/Fe]}$ has negative ones. The positive slopes of the $\alpha$ elements represent their production on a shorter timescale with respect to Fe. Type II SNe are indeed producing $\alpha$-elements on short timescales ($\mathrm{\sim 10^{-2}\ Gyr}$) as well as Fe. However, since the bulk of Fe comes from Type Ia SNe over longer timescales, the observed [$\alpha$/Fe] ratio will decrease towards younger ages, producing a positive slope.
On the other hand, the observed negative slope of the $\mathrm{[s/Fe]}$ vs. age relations is attributed to the delayed contribution to the s-process elements production from AGB stars. Nevertheless, although AGB stars are the main producers of s-process elements, rotating massive stars play an important role in interpreting the $\mathrm{[s/Fe]}$ vs. age relations. In Figure \ref{fig: sFe vs Age data and infall}, we show the observed $\mathrm{[s/Fe]}$ vs. age relations for Y, as representative of the first s-process peak elements, and for Ba, as representative of the second peak. The relation is plotted for the three different Galactocentric regions of interest together with a logarithmic fit of the OC sample and the results from the two- and three-infall models described in the previous sections (Model 1 and 2, respectively; see Table \ref{Tab: model infall} for reference). For both Y and Ba, it is possible to see the increase of the $\mathrm{[s/Fe]}$ ratio towards younger ages in the OC set discussed above. However, it is interesting to note that in the case of Y the rise appears only at very young ages, in particular in the outer and inner regions. At older ages the trend is rather flat. This long plateau is most probably due to the production of Y by rotating massive stars which happens on the same timescales as that of Fe. The flat trend, in fact, is less evident in the case of Ba, for which the contribution from massive stars is expected to be weaker. The logarithmic fit in the solar zone is computed excluding the young ($\mathrm{Age\simeq0.2\ Gyr}$) OC NGC 6709, which exhibits a high abundance of s-process elements. NGC 6709 is enriched in Y, Ba, and Nd but shows no significant overabundance in Fe  ($\mathrm{[Fe/H]=-0.025\ dex}$) or $\alpha$-elements such as Mg (e.g., $\mathrm{[Mg/H]=-0.038\ dex}$). However, due to the presence of only two confirmed members and the significant uncertainties in the measured s-process element abundances, NGC 6709 is excluded from the computation of the logarithmic fit, which, as a consequence, displays an overall flat trend. Interestingly, for Ba in particular, the trend appears to increase towards younger ages up to $\mathrm{Age \simeq 1\ Gyr}$.
\\
\\
The two- and three-infall models shown in Figure \ref{fig: sFe vs Age data and infall} both fail to reproduce the observed growth in the [s/Fe] vs. age relations. Instead of producing an increasing trend towards younger ages, the models produce either a decreasing or a flat trend. The decrease is more pronounced in the case of Ba than Y, indicative of faster Y enrichment due to the contribution of massive stars. The flat trend predicted by the chemical evolution model for Y is more consistent with observations in the outer regions compared to the solar and inner zones. Similarly, for Ba, the model's trend in the outer region aligns better with observed data. However, while the observations reveal a clear increasing trend, none of the model predictions capture this feature in any of the regions of interest. The flat trend predicted for Y is a consequence of the similar production mechanism for Y and Fe, both in terms of timescales and overall amounts. Both elements are produced on comparable timescales, and their ratio remains relatively stable over time. Notably, although metallicity increases from the outer to the inner regions, the [Y/Fe] ratio predicted by the models generally exhibits a flat or only slightly decreasing trend towards younger ages in the three different regions. In contrast, the predicted [Ba/Fe] ratio consistently decreases, with the steepest decline observed towards the inner regions, possibly reflecting a reduced contribution to the Ba production at higher metallicities. The trends predicted by the two- and three-infall models are similar, though the three-infall model slightly mitigates the decline of the [Ba/Fe] at very young ages. The third infall event brings in fresh, metal-poor gas that dilutes the existing metal-rich ISM. As a consequence, it also introduces a new episode of star formation that may alter the chemical evolution dynamics. However, although this newly formed stars have the potential to slow the decrease in [Ba/Fe], it is insufficient to reverse the overall trend. The additional gas inflow impacts both s-process elements and Fe similarly, thereby failing to create the observed increasing trend in the [s/Fe] ratio. Thus, while the third infall event introduces new complexities to the Galaxy chemical evolution, it does not provide a solution to the discrepancy between model predictions and observed [s/Fe] trends. An inversion of the predicted trend might therefore be obtained only from a nucleosynthetic point of view, by modifying either the timescales or the amount (or both) of production of s-process material. 

\begin{figure}
    \includegraphics[width=1\columnwidth]{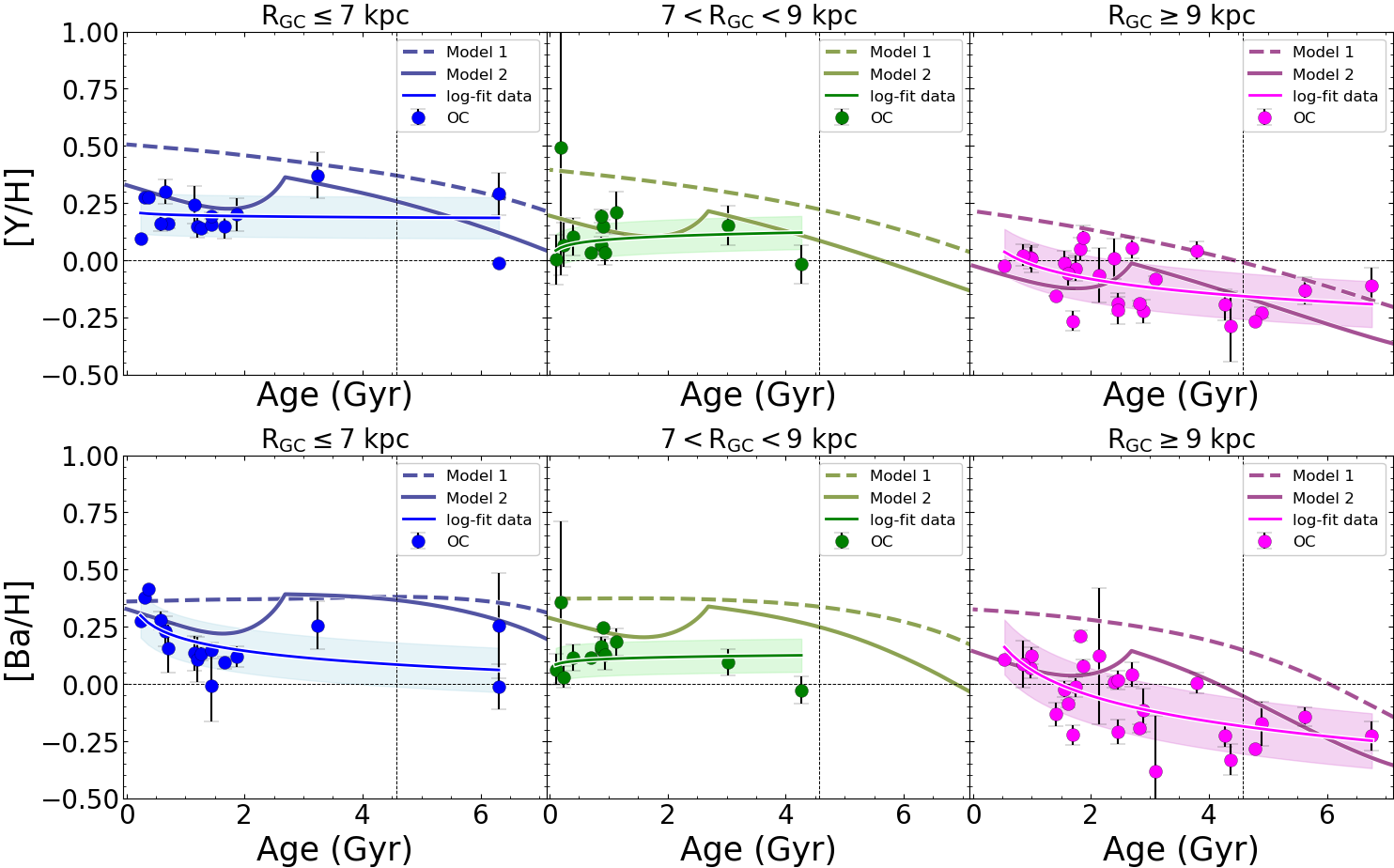}
     \caption{$\mathrm{[s/H]}$ vs age observed trends for Y (upper panel) and Ba (lower panel) divided in the three Galactocentric regions of interested. The logarithmic fits of the OC samples in the three regions are shown as blue (inner region), green (solar region) and magenta (outer region) curves. Predictions of the chemical evolution model in the case of a two- (dashed lines) and of a three-infall (solid lines) scenario are compared to the OC sample. See Table \ref{Tab: model infall} for reference.}%
 \label{fig: sH vs Age data and infall}%
\end{figure}

The three-infall model has nevertheless been demonstrated by \citet{Palla2024} to successfully reproduce the recent dilution at intermediate age ($\mathrm{1 < Age/Gyr < 3}$) in Fe and $\alpha$-elements observed in the same set of OCs used in this work. Specifically, \citet{Palla2024} attribute the observed dilution in Fe to a third, recent gas accretion event of primordial (or slightly enriched) material. This approach allows the authors to accurately replicate the age-metallicity relation observed in the current OC dataset across the entire disc, as well as the present-day metallicity gradient. A similar dilution is observed in the s-process elements Y and Ba, as illustrated in Figure \ref{fig: sH vs Age data and infall}, in which we show both the observed and predicted trends for the [s/H] vs. age relations. When applied to these relations, the extended three-infall model from \citet{Palla2024} (Model 2) achieves a reasonable agreement with the data, maintaining a generally increasing trend. Unlike the two-infall scenario, this model successfully reproduces the dilution observed at $\mathrm{Age\simeq2\ Gyr}$ in the solar and inner regions. However, in some cases, the predicted relations are higher than the observations. In particular, Model 2 nicely fits the [Y/H] in the observed uncertainty ranges, but fails to reproduce the [Ba/H] in the solar and inner regions. 

The overproduction of s-process elements is a well-known issue in Galactic chemical evolution modeling and is often attributed to an excessive production of s-process material from the AGB stars yields set of the FRUITY database. Typically, this issue is addressed by applying a reduction factor to the FRUITY yields across all elements, set independently of the progenitor star (as first done by \citealp{rizzuti2019}). This approach implicitly assumes that the overproduction is constant across mass, metallicity, and elements. However, from Figure \ref{fig: sH vs Age data and infall} it seems that this is not the case. Elements belonging to the first s-process peak, such as Y, are generally less overestimated than those belonging to the second s-process peak, such as Ba. Furthermore, the different discrepancies between our model and the observed trend across different disc regions, show that the necessary reduction factor should not be uniform across the disc, indicating that it should vary with metallicity. Age-related differences also emerge; for instance, in the inner region, the model closely matches the [Ba/H] ratio at young ages, while it overestimates this ratio for ages $\mathrm{\gtrsim 2\ Gyr}$.

\begin{figure}
    \includegraphics[width=1\columnwidth]{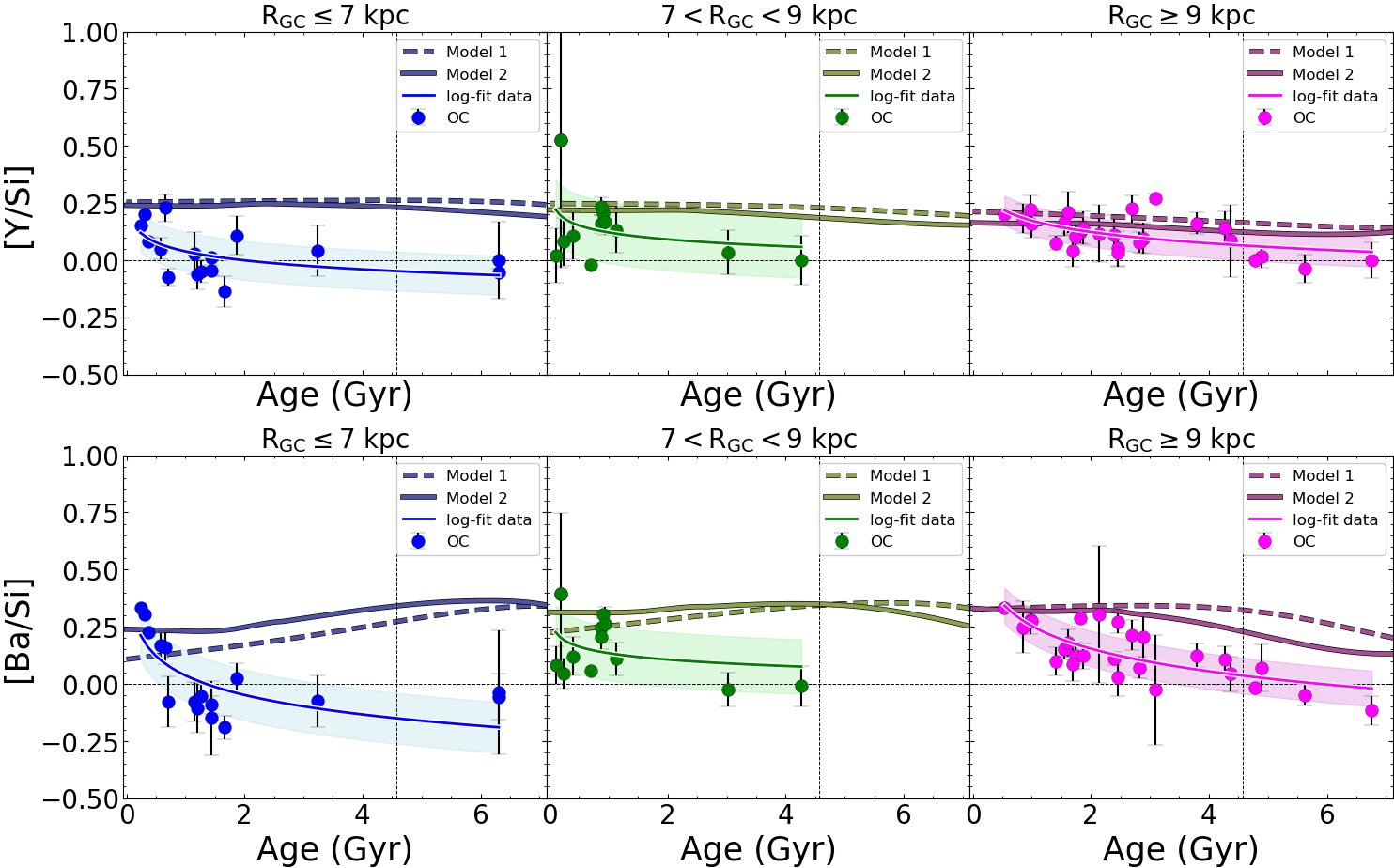}
     \caption{$\mathrm{[s/Si]}$ vs age observed trends for Y (upper panel) and Ba (lower panel) divided in the three Galactocentric regions of interested. The logarithmic fits of the OC samples in the three regions are shown as blue (inner region), green (solar region) and magenta (outer region) curves. Predictions of the chemical evolution model in the case of a two- (dashed lines) and of a three-infall (solid lines) scenario are compared to the OC sample. See Table \ref{Tab: model infall} for reference.}%
 \label{fig: sSi vs Age data and infall}%
\end{figure}

These discrepancies propagate into the [s/Fe] vs. age relations, as previously discussed, and ultimately affect the interpretation of chemical clocks, complicating the understanding of the observed trends. The observed pattern of the [s/Si] vs. Age relations shown in Figure \ref{fig: sSi vs Age data and infall}, shares many similarities with that of [s/Fe]. The logarithmic fit shows a general increase towards younger ages, with a steeper growth for Ba compared to Y. Similar to the [s/Fe], the solar region exhibits an overall flat trend (with an increase only up to $\mathrm{Age\simeq1\ Gyr}$) rather than the increasing trend observed in other regions. We remind that the OC NGC 6709 is excluded from the computation of the fit. The models, both in the case of the two- and three-infall scenarios, struggle to reproduce the observed patterns. Similar to the issues seen with the [s/Fe] ratios, these models exhibit an inverse trend compared to the observed one for [Ba/Si] in the inner and local regions. The trend is better reproduced in the outer region, but the models still show an overproduction of Ba, further highlighting the challenges in accurately modeling the chemical evolution of s-process elements. The models agree only with the data of the [Y/Si] in the outer regions. These discrepancies suggest that the current models may not fully capture the complexities of the s-process nucleosynthesis, particularly in terms of the contributions from different stellar populations.

\subsection{The influence of Asymptotic Giant Branch stars}

\begin{table*}[h]
    \caption{\label{Tab: model AGB} Input parameters for the chemical evolution models represented in Figures \ref{fig: sSi vs Age AGB models}. The models differ by the yields used for AGB stars. In Model 3, AGB yields are reduced by a factor of 10 for metallicities $\mathrm{Z \geq 1.4 \times 10^{-2}}$. In Model 4, a 1.1 solar mass star with the same prescriptions as the 1.3 solar mass star is added to the AGB yields, with its production increased by a factor of 10. In Model 5, the 1.1 solar mass star's production is increased by a factor of 20. Model 2 uses the original FRUITY yields. See text for more details.}
    \centering
    \resizebox{\textwidth}{!}{
    \begin{tabular}{cccc}
    \hline
      Model Name   &    Chemical evolution scenario   &    Yields for AGB    &   Yields for MS \\
      \hline
      Model 2      &    three-infall model              &    FRUITY original   &   \cite{Limongi2018} - $\mathrm{v_{rot}=150\ km~s^{-1}}$ \\
      Model 3      &    three-infall model              &    FRUITY reduced for $\mathrm{Z \geq 1.4 \times 10^{-2}}$ (0.1x)  &   \cite{Limongi2018} - $\mathrm{v_{rot}=150\ km~s^{-1}}$ \\
      Model 4      &    three-infall model              &    FRUITY original + $\mathrm{1.1\ M_\odot}$ star (10x production) &   \cite{Limongi2018} - $\mathrm{v_{rot}=150\ km~s^{-1}}$ \\
      Model 5      &    three-infall model              &    FRUITY original + $\mathrm{1.1\ M_\odot}$ star (20x production) &   \cite{Limongi2018} - $\mathrm{v_{rot}=150\ km~s^{-1}}$ \\
    \hline
    \end{tabular}
    }
\end{table*}

\begin{figure*}[h]
    \includegraphics[width=1.\textwidth]{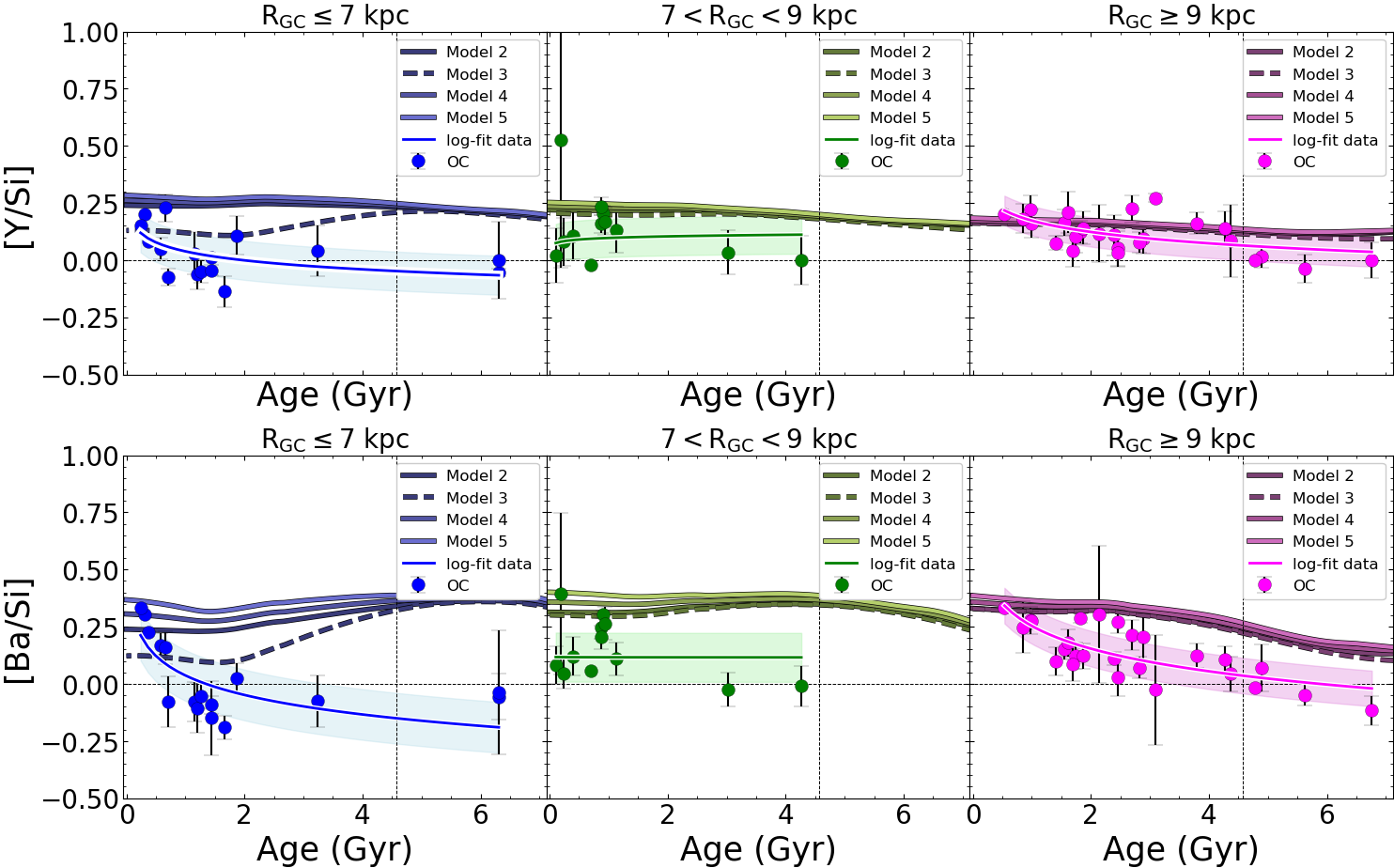}
     \caption{$\mathrm{[s/Si]}$ vs age observed trends for Y (upper panel) and Ba (lower panel) divided in the three Galactocentric regions of interested. The logarithmic fits of the OC samples in the three regions are shown as blue (inner region), green (solar region) and magenta (outer region) curves. Predictions of the chemical evolution model in the case of different assumptions concerning the production of s-process material from AGB stars are compared to the OC sample. See Table \ref{Tab: model AGB} for reference.}%
 \label{fig: sSi vs Age AGB models}%
\end{figure*}

With the current set of nucleosynthesis prescriptions, the issues in reproducing the observed [s/Fe] vs. age and chemical clock trends are twofold: i) the models tend to overestimate the observed patterns in some regions of interest, and ii) they almost always predict an inverted or flat trend compared to the observed increasing ones.

Concerning the first point, predictions from our model appear to be more compatible with the observed patterns in the outer region than in the solar and most inner regions. This observation aligns with findings by \citet{Casali2020A&A...639A.127C}, who noted that the content of neutron-capture elements belonging to the first s-process peak, is lower than expected from chemical evolution models. As a result, they found that the [Y/$\alpha$] ratios in clusters located in the inner region are lower than in clusters of the same age in the solar vicinity. Their direct conclusion was that stellar dating relations between abundance ratios and ages, based on samples of stars in the solar neighborhood, cannot be universally applied across different regions of the disc \citep[see also][]{Viscasillas2022A&A...660A.135V, Casali2023A&A...677A..60C, Ratcliffe2024MNRAS.528.3464R}. According to the authors, one should expect less s-process elements to be produced at high metallicity and tested a set of empirical yields from AGB stars, in which the super-solar metallicity yields where depressed by a factor of 10. This adjustment allowed them to successfully reproduce the chemical abundances of the young OCs located at $\mathrm{R_{GC}\simeq6\ kpc}$. However, when we compare this approach to the more extensive dataset used in this work, it becomes clear that a set of reduced AGB yields at super-solar metallicity does not provide a satisfactory fit across the board. This is illustrated by the dashed line in Figure \ref{fig: sSi vs Age AGB models} (Model 3, see Table \ref{Tab: model AGB}), where, following \citet{Casali2020A&A...639A.127C}, we reduced the yields of AGB stars by a factor of 10 for $\mathrm{Z\geq1.4\times10^{-2}}$ (namely, the solar metallicity as computed by \citealp{Asplund2009}). Although this adjustment improves the model’s match with OCs in the inner regions for ages between approximately 1.0 and 2.0 Gyr, it accentuates the inverted trend compared to the observed data. Moreover, the reduction in AGB yields does not impacts the model's result in the Solar vicinity and in the outer regions, as expected. This suggests that, while reducing the AGB yields at high metallicities may help to some extent in specific cases, it does not provide a comprehensive solution and may introduce new issues in reproducing the observed trends across different Galactic regions.
\\
\\
The second issue, in our opinion, is the most crucial. The inability of the models to capture the correct observed directional trend suggests that fundamental aspects of the chemical evolution, particularly the timing and contribution of s-process elements relative to Fe and $\alpha$-elements, may be inaccurately modeled. This highlights the need for a reassessment of the nucleosynthesis yields. This aspect has emerged also from the recent work of \citet{Ratcliffe2024MNRAS.528.3464R}, where the non-universality of the chemical clock [Ba/Mg] vs. age relation is studied. Once radial migration has been included, the chemical evolution model adopted by \citet{Ratcliffe2024MNRAS.528.3464R} is able to capture the overall trend of the [Ba/Mg] radial gradient as a function of time. However, the same model fails in reproducing the [Ba/Mg] vs. age relation since, similar as our case, it predicts a significantly decrease of the [Ba/Mg] abundance instead of the observed steep increase with time. According to the authors, a possible explanation for this decrease is that the amount of Ba produced in the adopted FRUITY models of high metallicity AGB stars is not high enough. A promising solution proposed by \citet{Ratcliffe2024MNRAS.528.3464R} was to replace the AGB yields for $\mathrm{Z>0.01}$ with those from models at $\mathrm{Z=0.01}$, which, in the case of Ba, are approximately two times larger than those at higher metallicities. This small adjustment to the high metallicity tail of the AGB yields, allowed the authors to mostly resolve the discrepancies observed in the first place. However, here we note that this solution is not universally effective. While it corrects the Ba yields, it fails to address issues with Y. The yields for Y at $\mathrm{Z=0.01}$ are actually lower - by about 1.3 times, depending on the progenitor mass - than those at higher metallicities. This reduction would create a similar discrepancy for Y as that seen in the Model 3 in the [Y/Si] vs. age relation. 

It is important to highlight the complexity of finding a universal solution in terms of AGB yields across different metallicities. The production of s-process elements in AGB stars is influenced by several factors, including the quantity of seed nuclei available and the number of thermal pulses the star undergoes during the AGB phase which, in turn, depend on the metallicity. Moreover, these dependencies do not behave uniformly across all s-process elements. As expected, elements belonging to different s-process peaks respond differently to changes in metallicity which, together with an inside-out scenario of disc formation and a radial variable SF efficiency, makes it challenging to create a single model that accurately predicts the behavior of the chemical clocks across the disc.

Nevertheless, despite variations in efficiency at different Galactocentric distances, the general increasing trend in the [s/$\alpha$] towards younger ages is observed consistently across different regions of the Galactic disc. These similarities in the timing of this trend imply that modifications of the s-process yields from AGB stars may depend more on the initial stellar mass than on the metallicity of the progenitor star. In the context of the FRUITY model database, the lowest progenitor mass included is a star with an initial mass of $\mathrm{1.3\ M_\odot}$. The lifetime of such a star (assuming solar chemical composition) is approximately 5.19 Gyr, meaning it would have enriched the interstellar medium with s-process elements around 8 Gyr ago. Therefore, increasing the yields from a 1.3 $\mathrm{M_\odot}$ star would cause the model to predict higher [s/Si] ratios than observed at each age of interest, implying that stars with even lower masses may be responsible for the observed increases in [s/$\alpha$] ratios (see also \citealp{d'orazi2009}). In Figure \ref{fig: sSi vs Age AGB models}, we show results of models in which we artificially include yields from a $\mathrm{1.1\ M_\odot}$ star. Yields are assumed to be 10-20 times greater than those of a $\mathrm{1.3\ M_\odot}$ (Models 4 and 5). Adding yields from a lower-mass progenitor has different effect on Y and Ba. Specifically, since the Y production is less sensitive to the contribution of AGB stars, the inclusion of the 1.1 $\mathrm{M_\odot}$ progenitor does not significantly impact the model, especially in the outer regions of the Galaxy, where different models produce nearly identical results. In the case of the [Ba/Si], on the other hand, the inclusion of the lower-mass progenitor does have a notable effect, particularly in the solar region, where the models are able to reverse the declining trend in Ba abundance at younger ages. However, while this adjustment leads to an increase in Ba at younger ages, it still falls short of fully invert the modeled trend in accordance to the observational data. This however raises questions about whether lower-mass stars (below 1.3 $\mathrm{M_\odot}$) contribute more significantly to the production of s-process elements than currently accounted for in models. According to \citet{d'orazi2009} and \citet{Maiorca2012}, this might be possible in a scenario where the efficiency of the extra-mixing processes producing the neutron source 13C is anti-correlated with the initial mass of the star. It is also possible that the intermediate-process (i-process) nucleosynthesis takes place during the early AGB phase of low-mass ($\mathrm{1.0\ M_\odot}$) stars. However, as shown in \citet{Choplin2024} (see also \citealp{Choplin2021, Choplin2022}), models are in favor of an i-process operating in AGB stars up only to metallicity $\mathrm{[Fe/H]\simeq-1\ dex}$.

\subsection{The influence of rotating massive stars}
\label{sec: The influence of rotating massive stars}

\begin{table*}[h]
    \caption{\label{Tab: model MS} Input parameters for the chemical evolution models represented in Figures \ref{fig: sSi vs Age MS models}. The models differ by the yields used for rotating massive stars. In Model 6, the initial rotational velocity of massive stars follows the distribution DIS 3 of \cite{Molero2024} ($\mathrm{v_{rot} = 150\ km~s^{-1}}$ for metallicities $\mathrm{Z<3.236\times10^{-3}}$, $\mathrm{v_{rot} = 0\ km~s^{-1}}$ afterwards). In Model 7, it follows the one presented in \cite{Prantzos2018}. Model 2 uses a fixed rotational velocity of $\mathrm{v_{rot} = 150\ km~s^{-1}}$. See text for more details.}
    \centering
    \resizebox{\textwidth}{!}{
    \begin{tabular}{cccc}
    \hline
      Model Name   &    Chemical evolution scenario   &    Yields for AGB    &   Yields for MS \\
      \hline
      Model 2      &    three-infall model              &    FRUITY original   &   \cite{Limongi2018} - $\mathrm{v_{rot}=150\ km~s^{-1}}$\\
      Model 6      &    three-infall model              &    FRUITY original   &   \cite{Limongi2018} - distribution from \cite{Molero2024} \\
      Model 7      &    three-infall model              &    FRUITY original   &   \cite{Limongi2018} - distribution from \citet{Prantzos2018} \\
    \hline
    \end{tabular}
    }
\end{table*}

\begin{figure*}[h]
    \includegraphics[width=1.\textwidth]{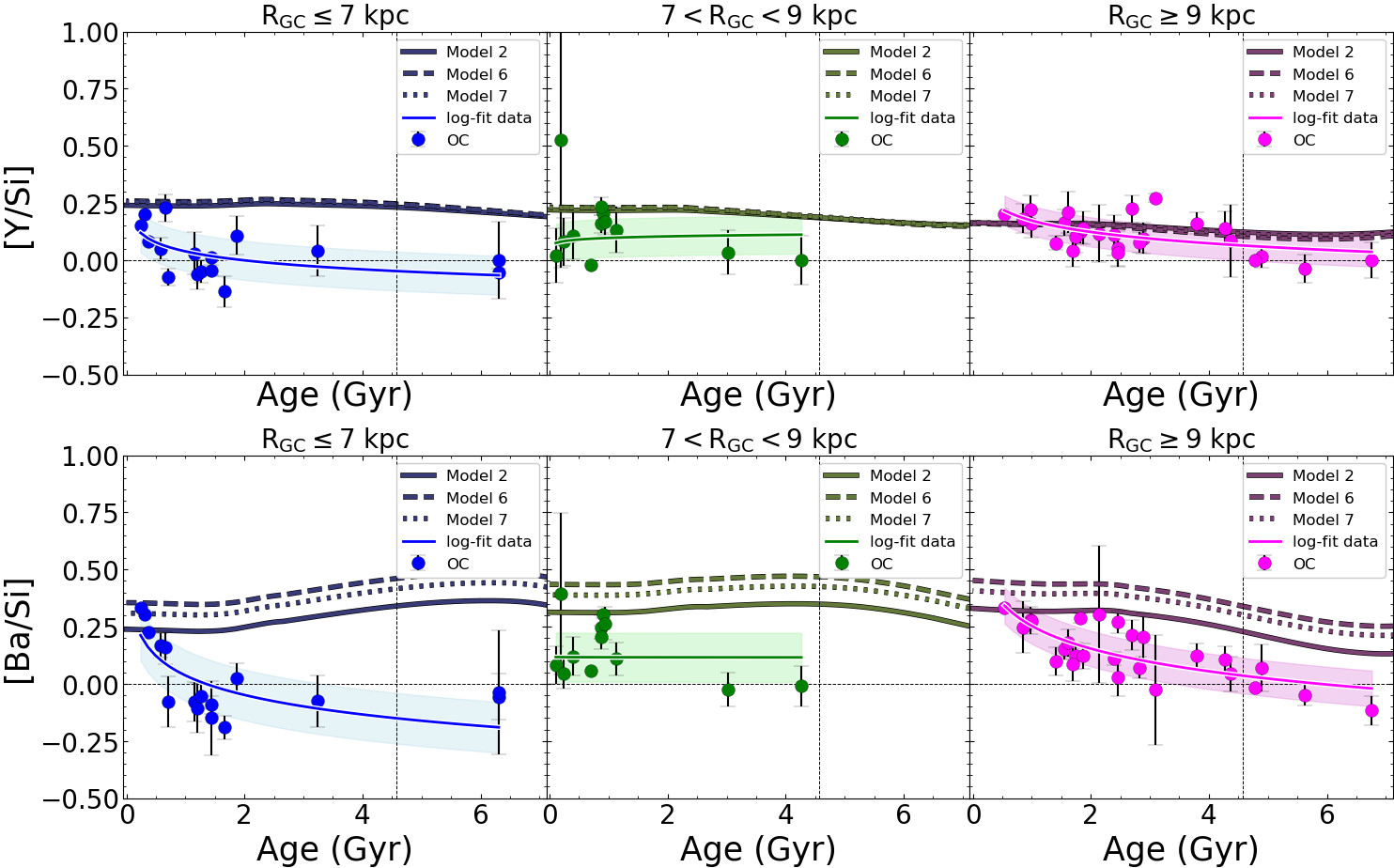}
     \caption{$\mathrm{[s/Si]}$ vs Age observed trends for Y (upper panel) and Ba (lower panel) divided in the three Galactocentric regions of interested. The logarithmic fits of the OC samples in the three regions are shown as blue (inner region), green (solar region) and magenta (outer region) curves. Predictions of the chemical evolution model in the case of a two- (dashed lines) and of a three-infall (solid lines) scenario are compared to the OC sample. Predictions of the chemical evolution model in the case of different assumptions concerning the production of s-process material from massive stars are compared to the OC sample. See Table \ref{Tab: model MS} for reference.}%
 \label{fig: sSi vs Age MS models}%
\end{figure*}

In this Section we shift focus from AGB stars to rotating massive stars as a key nucleosynthetic site for the production of s-process elements in the model. While the previous section demonstrated the challenges in finding a universal model for the chemical clocks based on the s-process yields from AGB stars (due to variations in metallicity and progenitor mass), here we explore the contribution of rotating massive stars to the synthesis of s-process elements. Rotating massive stars play indeed a crucial role in the Galactic chemical evolution, particularly for s-process elements at lower metallicities (see e.g., \citealp{Frischknecht2012, 2016MNRAS.456.1803F, Limongi2018, Prantzos2018, rizzuti2019, Rizzuti2021}
), and their initial rotation speeds influence the production of these elements. 

In our reference model (Model 2) we adopt a fixed initial rotational velocity for massive stars of $\mathrm{150\ km~s^{-1}}$. Even thought such a nucleosynthesis prescription allows us in \citet{Molero2023} to correctly reproduce the majority of the neutron capture-elements abundance patterns observed by \textit{Gaia}-ESO, the adoption of a distribution of rotational velocities is usually favored. A more realistic approach involves adopting a distribution of rotational velocities rather than a single fixed value. \citet{Prantzos2018} firstly implemented the idea of such a distribution in the context of Galactic chemical evolution models, based on the massive star yields from \citet{Limongi2018}. This distribution accounts for different initial rotational velocities and adjusts them as a function of [Fe/H], to match the observed behavior of primary nitrogen (14N) at low metallicities and to avoid an overproduction of s-process elements at higher metallicities. The obtained distribution favors higher rotational velocities at lower [Fe/H] and lower or negligible velocities at solar and supersolar metallicity. This is in agreement with also other distributions in literature which where parametrized later (e.g., \citealp{Romano2019, rizzuti2019, Rizzuti2021, Molero2024}).

Here, we test two different distributions: DIS 3 from \citet{Molero2024} and the one from \cite{Prantzos2018} (Model 6 and 7, respectively; see Table \ref{Tab: model MS} for reference). DIS 3, similar to the distribution adopted by \citet{Romano2019} for studying CNO isotopes, assumes that massive stars rotate at $\mathrm{150\ km~s^{-1}}$ for $\mathrm{Z<3.236\times10^{-3}}$ ($\mathrm{[Fe/H] \lesssim -1.0}$) and rotation becomes negligible beyond the metallicity threshold of $\mathrm{Z=3.236\times10^{-3}}$. The results of such a distribution is reported by the dashed lines in Figure \ref{fig: sSi vs Age MS models}. 

In the case of the [Y/Si], the absence of high rotation in massive stars reduces the production of Y. Si production is however also reduced, and since both Y and Si are reduced by similar amounts, the net effect is negligible in the [Y/Si] ratio across the Galactic disc. In fact, despite this reduction, the overall pattern of the chemical clock remains similar to models where massive stars consistently have higher rotational velocities, showing that even with reduced rotation, the general trend of the chemical clock [Y/Si] can still resemble higher-velocity models due to balanced effects on multiple elements. In the case of [Ba/Si], the impact of adopting a distribution of rotational velocities for massive stars differs from the results seen with [Y/Si]. In all the regions both Ba and Si production are reduced due to the absence of rotation. However, Si is reduced more than Ba, leading to a higher [Ba/Si] ratio compared to the model with constant faster rotation. 

The trends predicted by models with a distribution of rotational velocities for massive stare might be, in general, highly dependent on how the model transitions between populations of rotating and non-rotating massive stars at a fixed metallicity. In Model 6, this transition occurs abruptly at a specific metallicity threshold. A more realistic approach would account for the coexistence of different populations of rotating massive stars across different metallicities, including a small fraction of highly rotating stars (e.g., with initial rotational velocities of 300 km~s$^{-1}$) towards lower Z. This approach is shown by Model 7 (dotted lines in Figure \ref{fig: sSi vs Age MS models}, representing models in which the \citealp{Prantzos2018} distribution is adopted). With this distribution, Model 7 yields similar [s/Si] results to Model 6. For the [Ba/Si], results are slightly under the one of Model 6, as expected, because of the lower percentage of stars with null rotational velocities. Differences are minor, but noticeable. Overall, none of the tested distributions successfully capture the increasing trend observed in chemical clocks toward younger ages. Additionally, in the case of [Ba/Si], the models tend to worsen the overestimation of the observed pattern. Concerning this last point, it appears clear that the general idea according to which reducing the rotational velocity of massive stars would help in reproducing the lower level of [s/$\alpha$] observed in the inner region (due to reduced s-process material production), is challenged. The different contribution to different elements from different progenitor stars are more complex, and their effect are not linear and, as a consequence, it appears difficult to lead back the issue to a specific choice in the distribution or a specific progenitor star without performing a more extensive parametric study.


\section{Discussion and conclusions}
\label{sec: discussion and conclusions}

In this study, we investigated the influence of star formation history/gas infall events, AGB stars and massive stars on the production and distribution of s-process elements across the MW disc, in the context of chemical clocks. We focused on the [Y/Si] and [Ba/Si] vs. age relationships across different Galactic regions, which, as observed in the OC datasets from the \textit{Gaia}-ESO survey used here, exhibit an increasing trend toward younger ages. This trend is more pronounced for second s-process peak elements compared to first peak elements. The discrepancies between our model predictions and observed data -—particularly the overestimation of certain patterns and the prediction of inverted trends—-  highlight the challenges of developing a universal Galactic chemical evolution model for elements belonging to different s-process peaks. 

We showed that neither the two-infall nor the three-infall model can accurately reproduce the increasing trend of the chemical clocks. However, the three-infall model more successfully reproduced the observed depression in the [s/H] vs. age trends at $\mathrm{\simeq2\ Gyr}$.

While AGB stars play a significant role in s-process nucleosynthesis, current models  with state-of-the-art nucleosynthesis prescriptions struggled to accurately reproduce the observed chemical clock trends across the Galactic disc. Our model predictions aligned better with the observed patterns in the outer regions than the solar and inner regions and, in general, with the [Y/Si] than the [Ba/Si]. Attempts to adjust AGB yields for super-solar metallicities, as originally proposed by \citet{Casali2020A&A...639A.127C}, provided some improvement in reproducing the observed data for certain OCs, particularly those in the inner disc aged 1 to 2 Gyr. However, this approach did not yield a satisfactory fit across all regions, instead introducing new discrepancies and failing to provide a universal solution for the entire disc. Suggestions from \citet{Ratcliffe2024MNRAS.528.3464R} are also discussed, underlying the difficulties that the models face in reproducing simultaneously elements belonging to different s-process peaks. A potential solution to reproduce the increasing trend in chemical clocks is the inclusion of lower-mass stars (below $\mathrm{1.3\ M_\odot}$) as contributors to s-process element production, as first suggested by \citealp{d'orazi2009}. However, the s-process yield from such low-mass stars might be substantial yet is likely unrealistic. The potential contribution from the i-process was briefly considered and tentatively ruled out, given its current inability to produce neutron-capture elements above $\mathrm{[Fe/H] \gtrsim -1\ dex}$ (\citealp{Choplin2024}).

\begin{figure*}
    \centering
    \includegraphics[width=\textwidth]{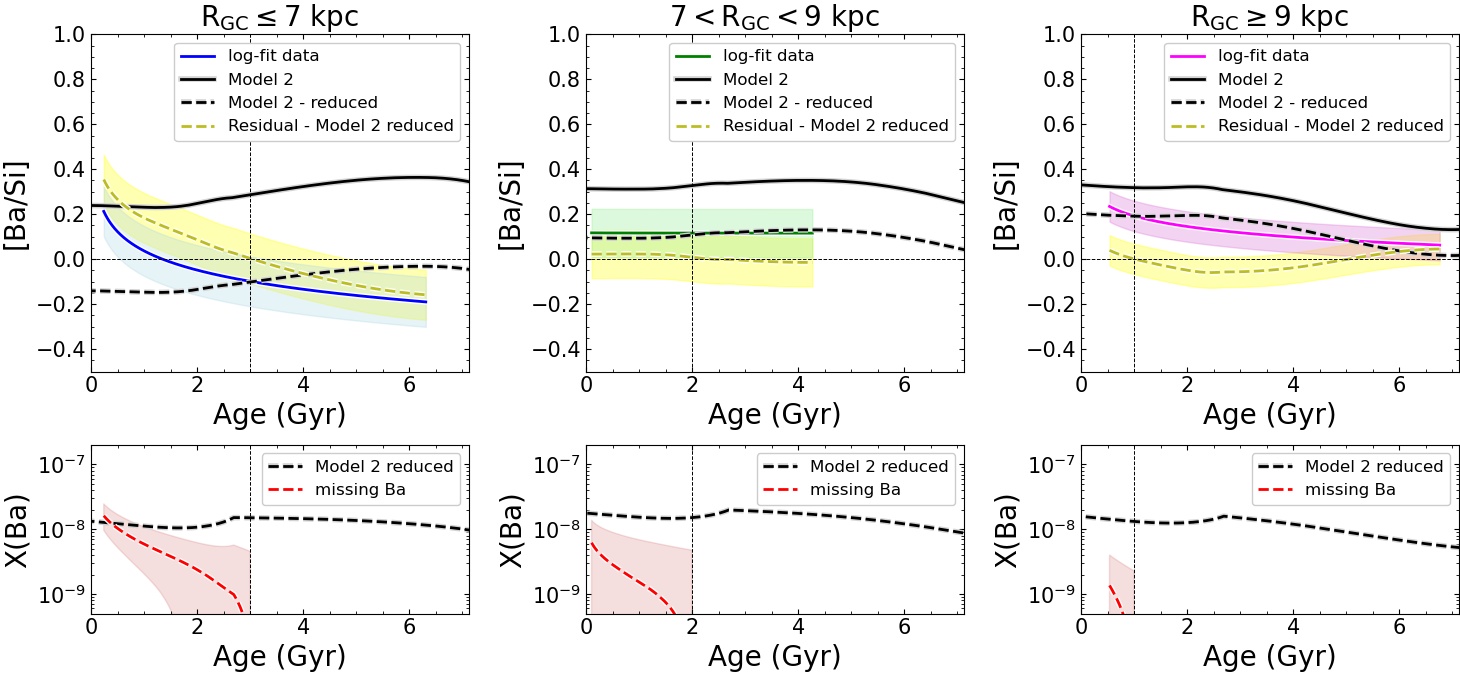}
 \caption{Upper panels: [Ba/Si] vs. Age trend in the inner, solar and outer regions. The observational data trend is represented by the logarithmic fit with the associated 1$\sigma$ error bar curve. Result of Model 2 is shown by the black line. The dashed black and yellow lines represent Model 2 with reduced s-process production and its associated residual curve. Bottom panel: absolute abundance of Ba as predicted by the reduced Model 2 (dashed black line) together with the computed missing abundance of Ba in the age ranges of interest (see text for details).}%
\label{fig: missing Ba}%
\end{figure*}

We recognized the essential role of massive stars in s-process elements production, especially at lower metallicities. The adoption of a fixed initial rotational velocity for massive stars in our starting model (Model 2) proved insufficient to capture the complexities of the chemical clocks. Testing distributions of rotational velocities, as first introduced by \citet{Prantzos2018}, revealed unexpected impacts on the chemical clock patterns, in particular in the case of the [Ba/Si] vs. age. A distribution of initial rotational velocities for massive stars can lead to an overproduction of [Ba/Si], even when null rotational velocities are included, due to the reduced contribution to Si production. The net effect of adopting a rotational velocity distribution strongly depends on the parameter choices for the distribution itself. Overall, adopting a distribution of rotational velocities generally either worsens agreement with observations or leaves it unchanged.
\\
\\
The different models tested in these work show a better agreement with the [Y/Si] rather than with the [Ba/Si]. It is interesting to quantify the overproduction of the overall observed trend shown by our model as well as to compute the amount of Ba, in terms of its surface gas density, that our standard model (Model 2) fails to produce in the most recent few Gyr of chemical evolution, causing the trend to be inverted with respect to the observations. We provide here this calculation for the inner, solar and outer regions. In Figure \ref{fig: missing Ba}, we show the logarithmic fit of the observed chemical clock [Ba/Si] vs. Age together with the predictions from Model 2. As already pointed out in Section \ref{sec: The effect of primordial gas inflows}, the reduction in the nucleosynthesis prescriptions of s-process elements seem not to be constant among the disc. Indeed, reducing the production of s-process material, independently by the source, by a factor of $2.8$, $1.8$ and $0.4$, lead to the results shown by the dashed black lines in the aftermentioned Figure for the inner, solar and outer regions respectively. Implying that, while in the solar region the nucleosynthesis is overestimated approximately by $\mathrm{44\%}$ (consistent with prior estimates by \citealp{rizzuti2019}), in the outer region this percentage is reduced to $\mathrm{28\%}$ while in the inner region it shows the most dramatic value, around $\mathrm{64\%}$. The obtained reduced results are in agreement with the observations for all the age range of interested in the solar and outer regions. On the other hand, in the inner region, the agreement holds for $\mathrm{Age \gtrsim 3.0\ Gyr}$. For younger stellar ages, the model exhibits the insufficient growth trend. To address this, we compute the amount of Ba that our model should produce in the last few Gyr in order to correct its inverted/flat trend with respect to the observations. We compute the residuals between the observed logarithmic fit and our reduced model results, shown in the Figure by the yellow dashed lines. In the solar and outer regions, as expected, the residual curve is consistent with the zero, indicating the nice fit between the reduced model and observations. The residuals becomes positive at very young ages in these regions ($\mathrm{Age \simeq 2\ Gyr}$ in the local and $\mathrm{Age \simeq 1\ Gyr}$ in the outer disc), reflecting the flattening of the curve towards younger ages with the consequent underestimation of the observations. The discrepancies are however not dramatic. In the inner region, on the other hand, the residual curve becomes positive for $\mathrm{Age\lesssim3\ Gyr}$, indicating that the model is increasingly underestimating the observed [Ba/Si] over time as we move toward younger ages. To quantify this discrepancy, we can estimate the missing amount of Ba, both in terms of absolute abundance and of surface mass density (expressed in $\mathrm{M_\odot\ pc^{-2}}$), provided with the assumption that predictions for Si are correct (see Figure \ref{fig: SiH vs Age} for a brief discussion). We write, by definition:
\begin{equation}
 X(Ba)_{\rm O} - X(Ba)_{\rm M} = X(Si)_{\rm M} \times (10^{[Ba/Si]_{\rm O} + {\rm Sun}} - 10^{[Ba/Si]_{\rm M} + {\rm Sun}})
    \label{eq: sb definition}
\end{equation}
where $\mathrm{Sun = \log \left( \frac{X(Ba)}{X(Si)} \right)_\odot}$ (taken from \citet{Asplund2009} solar abundances), $\mathrm{X(Si)_M}$ is the absolute abundance of Si predicted by the model during the relevant time range, $\mathrm{[Ba/Si]_O}$ and $\mathrm{[Ba/Si]_M}$ are the observed and predicted [Ba/Si] ratios, respectively. The obtained missing abundance of Ba is shown in the lower panels of Figure \ref{fig: missing Ba} (dashed red lines) for the reduced model, in the three regions of interest. It is possible to see that in the inner region, nearly half more of the actual predicted amount of Ba, on average, should be produced by the model in order to reproduce the high rise observed towards young ages. In the outer and solar region, on the other hand, the missing amount of Ba is not significantly high and this is reflected by the current calculation. In particular, after multiplying by the surface gas density predicted by the model, we can compute the average local surface mass density of Ba that our model fails to produce in the last 3 Gyr of chemical evolution, which is equal to $\mathrm{\Sigma_{Ba} = 5.71\times10^{-8}\ M_\odot pc^{-2}}$.
\\
\\

In conclusion:
\begin{itemize}
    \item Two- and three-infall chemical evolution models with state-of-the-art nucleosynthesis prescriptions for neutron-capture elements show difficulties in reproducing the high rise observed in the chemical clocks [s/$\alpha$] vs. age toward younger Ages. The deviations between models and data are more evident towards the inner part of the disc and for elements belonging to the second s-process peak (here represented by Ba) than for those of the first peak (here represented by Y). The three-infall model better captures the observed trend, in particular the decrease around $\mathrm{\simeq2\ Gyr}$.

    \item Modifying the production of s-process material from AGB stars, as suggested by previous studies, does not improve the agreement with the data. Modifications that improve the fit for elements of the second s-process peak often fail to match observations for elements of the first peak. An increased s-process contribution from low-mass stars ($\mathrm{\sim 1.1\ M_\odot}$) could potentially help reconcile model predictions with observed abundances, yet such an increase is not supported by current nucleosynthesis models, even with the inclusion of the i-process. Future developments in the nucleosynthesis of AGB that take into account, for example, magnetic buoyancy, are might lead to improvements in the agreement between models and observations because they affect the production of the s-elements of the  first and second peaks differently in terms of both mass and metallicity \citep{Magrini2021A&A...646L...2M, Vescovi2021Univ....8...16V}.

    \item Assuming a distribution of rotational velocities for massive stars, which favors high velocities at low metallicities and no velocity at higher ones, does not provide any improvement for elements of the fist s-process peak due to the balanced effect on multiple elements. In the case of second peak s-process elements, it accentuates the overproduction across the disc due to the lower production of $\alpha$-elements.

    \item Once accounted for the residual between the observed trend shown by the OCs dataset and our model results, it is possible to estimate the missing amount of Ba predicted by our model in the last $\mathrm{3\ Gyr}$ in terms of its surface mass density. After scaling the observed trend to our model result, the average local missing Ba is $\mathrm{\Sigma_{Ba} = 5.71\times10^{-8}\ M_\odot pc^{-2}}$. Namely, the production of Ba in the inner part of the disc should be approximately half more of the current one during the last few Gyrs of chemical evolution.
\end{itemize}
This work gives strong indications that may guide future studies for s-process nucleosynthesis in both AGB and massive stars. Indeed, the theoretical understanding of these elements is important for many aspects of both nucleosynthesis and stellar evolution, as well as the chemical dating of stars, which is becoming increasingly important in view of new large spectroscopic surveys \citep[e.g.][]{Bensby2019Msngr.175...35B,  Jin2024MNRAS.530.2688J} and new instrumentation \citep[e.g.][]{Zhang2018SPIE10702E..7WZ, Magrini2023arXiv231208270M, Mainieri2024arXiv240305398M}.

\begin{acknowledgements}
This work was supported by the Deutsche Forschungsgemeinschaft
(DFG, German Research Foundation) – Project-ID 279384907 – SFB 1245, the State of Hessen within the Research Cluster ELEMENTS (Project ID 500/10.006). M.M. and F.M. thank INAF. for the 1.05.12.06.05 Theory Grant - Galactic archaeology with radioactive and stable nuclei.  This research was partially supported by the Munich Institute for Astro-, Particle and BioPhysics (MIAPbP) which is funded by the Deutsche Forschungsgemeinschaft (DFG, German Research Foundation) under Germany´s Excellence Strategy – EXC-2094 – 390783311.
M.M. thanks Almudena Arcones for the useful discussions which improved the paper.

L.M. and S.R. thank INAF for the support (Large Grant EPOCH), the Mini-Grants Checs and PILOT (1.05.23.04.02), and the financial support under the National Recovery and Resilience Plan (NRRP), Mission 4, Component 2, Investment 1.1, Call for tender No. 104 published on 2.2.2022 by the Italian Ministry of University and Research (MUR), funded by the European Union –NextGenerationEU– Project ‘Cosmic POT’ Grant Assignment Decree No. 2022X4TM3H by the Italian Ministry of Ministry of University and Research (MUR). 
MP acknowledges financial support from the project "LEGO – Reconstructing the building blocks of the Galaxy by chemical tagging" granted by the Italian MUR through contract PRIN2022LLP8TK\_001.
CVV acknowledges funding from the Research Council of Lithuania (LMTLT, grant No. P-MIP-23-24).
\end{acknowledgements}

%
%

\bibliographystyle{aa} 
\bibliography{Bibliography} 

\begin{thebibliography}{116}
\expandafter\ifx\csname natexlab\endcsname\relax\def\natexlab#1{#1}\fi

\bibitem[{{Abbott} {et~al.}(2021){Abbott}, {Abbott}, {Abraham}, {Acernese},
  {Ackley}, {Adams}, {Adams}, {Adhikari}, {Adya}, {Affeldt}, {Agathos},
  {Agatsuma}, {Aggarwal}, {Aguiar}, {Aiello}, {Ain}, {Ajith}, {Allen},
  {Allocca}, {Altin}, {Amato}, {Anand}, {Ananyeva}, {Anderson}, {Anderson},
  {Angelova}, {Ansoldi}, {Antelis}, {Antier}, {Appert}, {Arai}, {Araya},
  {Areeda}, {Ar{\`e}ne}, {Arnaud}, {Aronson}, {Arun}, {Asali}, {Ascenzi},
  {Ashton}, {Aston}, {Astone}, {Aubin}, {Aufmuth}, {AultONeal}, {Austin},
  {Avendano}, {Babak}, {Badaracco}, {Bader}, {Bae}, {Baer}, {Bagnasco},
  {Baird}, {Ball}, {Ballardin}, {Ballmer}, {Bals}, {Balsamo}, {Baltus},
  {Banagiri}, {Bankar}, {Bankar}, {Barayoga}, {Barbieri}, {Barish}, {Barker},
  {Barneo}, {Barnum}, {Barone}, {Barr}, {Barsotti}, {Barsuglia}, {Barta},
  {Bartlett}, {Bartos}, {Bassiri}, {Basti}, {Bawaj}, {Bayley}, {Bazzan},
  {Becher}, {B{\'e}csy}, {Bedakihale}, {Bejger}, {Belahcene}, {Beniwal},
  {Benjamin}, {Bennett}, {Bentley}, {Bergamin}, {Berger}, {Bergmann},
  {Bernuzzi}, {Berry}, {Bersanetti}, {Bertolini}, {Betzwieser}, {Bhandare},
  {Bhandari}, {Bhattacharjee}, {Bidler}, {Bilenko}, {Billingsley}, {Birney},
  {Birnholtz}, {Biscans}, {Bischi}, {Biscoveanu}, {Bisht}, {Bitossi},
  {Bizouard}, {Blackburn}, {Blackman}, {Blair}, {Blair}, {Blair}, {Blanch},
  {Bobba}, {Bode}, {Boer}, {Boetzel}, {Bogaert}, {Boldrini}, {Bondu},
  {Bonilla}, {Bonnand}, {Booker}, {Boom}, {Bork}, {Boschi}, {Bose},
  {Bossilkov}, {Boudart}, {Bouffanais}, {Bozzi}, {Bradaschia}, {Brady},
  {Bramley}, {Branchesi}, {Brau}, {Breschi}, {Briant}, {Briggs}, {Brighenti},
  {Brillet}, {Brinkmann}, {Brockill}, {Brooks}, {Brooks}, {Brown}, {Brunett},
  {Bruno}, {Bruntz}, {Buikema}, {Bulik}, {Bulten}, {Buonanno}, {Buscicchio},
  {Buskulic}, {Byer}, {Cabero}, {Cadonati}, {Caesar}, {Cagnoli}, {Cahillane},
  {Calder{\'o}n Bustillo}, {Callaghan}, {Callister}, {Calloni}, {Camp},
  {Canepa}, {Cannon}, {Cao}, {Cao}, {Carapella}, {Carbognani}, {Carney},
  {Carpinelli}, {Carullo}, {Carver}, {Casanueva Diaz}, {Casentini}, {Caudill},
  {Cavagli{\`a}}, {Cavalier}, {Cavalieri}, {Cella}, {Cerd{\'a}-Dur{\'a}n},
  {Cesarini}, {Chaibi}, {Chakravarti}, {Chan}, {Chan}, {Chandra}, {Chanial},
  {Chao}, {Charlton}, {Chase}, {Chassande-Mottin}, {Chatterjee},
  {Chattopadhyay}, {Chaturvedi}, {Chatziioannou}, {Chen}, {Chen}, {Chen},
  {Chen}, {Cheng}, {Cheong}, {Chia}, {Chiadini}, {Chierici}, {Chincarini},
  {Chiummo}, {Cho}, {Cho}, {Cho}, {Choate}, {Christensen}, {Chu}, {Chua},
  {Chung}, {Chung}, {Ciani}, {Ciecielag}, {Cie{\'s}lar}, {Cifaldi}, {Ciobanu},
  {Ciolfi}, {Cipriano}, {Cirone}, {Clara}, {Clark}, {Clark}, {Clarke},
  {Clearwater}, {Clesse}, {Cleva}, {Coccia}, {Cohadon}, {Cohen}, {Colleoni},
  {Collette}, {Collins}, {Colpi}, {Constancio}, {Conti}, {Cooper}, {Corban},
  {Corbitt}, {Cordero-Carri{\'o}n}, {Corezzi}, {Corley}, {Cornish}, {Corre},
  {Corsi}, {Cortese}, {Costa}, {Cotesta}, {Coughlin}, {Coughlin}, {Coulon},
  {Countryman}, {Couvares}, {Covas}, {Coward}, {Cowart}, {Coyne}, {Coyne},
  {Creighton}, {Creighton}, {Croquette}, {Crowder}, {Cudell}, {Cullen},
  {Cumming}, {Cummings}, {Cunningham}, {Cuoco}, {Curylo}, {Dal Canton},
  {D{\'a}lya}, {Dana}, {DaneshgaranBajastani}, {D'Angelo}, {Danilishin},
  {D'Antonio}, {Danzmann}, {Darsow-Fromm}, {Dasgupta}, {Datrier}, {Dattilo},
  {Dave}, {Davier}, {Davies}, {Davis}, {Daw}, {Dean}, {DeBra}, {Deenadayalan},
  {Degallaix}, {De Laurentis}, {Del{\'e}glise}, {Del Favero}, {De Lillo}, {De
  Lillo}, {Del Pozzo}, {DeMarchi}, {De Matteis}, {D'Emilio}, {Demos}, {Denker},
  {Dent}, {Depasse}, {De Pietri}, {De Rosa}, {De Rossi}, {DeSalvo}, {de
  Varona}, {Dhurandhar}, {D{\'\i}az}, {Diaz-Ortiz}, {Didio}, {Dietrich}, {Di
  Fiore}, {DiFronzo}, {Di Giorgio}, {Di Giovanni}, {Di Giovanni}, {Di
  Girolamo}, {Di Lieto}, {Ding}, {Di Pace}, {Di Palma}, {Di Renzo},
  {Divakarla}, {Dmitriev}, {Doctor}, {D'Onofrio}, {Donovan}, {Dooley},
  {Doravari}, {Dorrington}, {Downes}, {Drago}, {Driggers}, {Du}, {Ducoin},
  {Dupej}, {Durante}, {D'Urso}, {Duverne}, {Dwyer}, {Easter}, {Eddolls},
  {Edelman}, {Edo}, {Edy}, {Effler}, {Eichholz}, {Eikenberry}, {Eisenmann},
  {Eisenstein}, {Ejlli}, {Errico}, {Essick}, {Estell{\'e}s}, {Estevez},
  {Etienne}, {Etzel}, {Evans}, {Evans}, {Ewing}, {Fafone}, {Fair}, {Fairhurst},
  {Fan}, {Farah}, {Farinon}, {Farr}, {Farr}, {Fauchon-Jones}, {Favata}, {Fays},
  {Fazio}, {Feicht}, {Fejer}, {Feng}, {Fenyvesi}, {Ferguson},
  {Fernandez-Galiana}, {Ferrante}, {Ferreira}, {Fidecaro}, {Figura}, {Fiori},
  {Fiorucci}, {Fishbach}, {Fisher}, {Fishner}, {Fittipaldi}, {Fitz-Axen},
  {Fiumara}, {Flaminio}, {Floden}, {Flynn}, {Fong}, {Font}, {Forsyth},
  {Fournier}, {Frasca}, {Frasconi}, {Frei}, {Freise}, {Frey}, {Frey},
  {Fritschel}, {Frolov}, {Fronz{\'e}}, {Fulda}, {Fyffe}, {Gabbard}, {Gadre},
  {Gaebel}, {Gair}, {Gais}, {Galaudage}, {Gamba}, {Ganapathy}, {Ganguly},
  {Gaonkar}, {Garaventa}, {Garc{\'\i}a-Quir{\'o}s}, {Garufi}, {Gateley},
  {Gaudio}, {Gayathri}, {Gemme}, {Gennai}, {George}, {George}, {Gergely},
  {Ghonge}, {Ghosh}, {Ghosh}, {Ghosh}, {Giacomazzo}, {Giacoppo}, {Giaime},
  {Giardina}, {Gibson}, {Gier}, {Gill}, {Giri}, {Glanzer}, {Gleckl}, {Godwin},
  {Goetz}, {Goetz}, {Gohlke}, {Goncharov}, {Gonz{\'a}lez}, {Gopakumar},
  {Gossan}, {Gosselin}, {Gouaty}, {Grace}, {Grado}, {Granata}, {Granata},
  {Grant}, {Gras}, {Grassia}, {Gray}, {Gray}, {Greco}, {Green}, {Green},
  {Gretarsson}, {Griggs}, {Grignani}, {Grimaldi}, {Grimes}, {Grimm}, {Grote},
  {Grunewald}, {Gruning}, {Guerrero}, {Guidi}, {Guimaraes}, {Guix{\'e}},
  {Gulati}, {Guo}, {Gupta}, {Gupta}, {Gupta}, {Gustafson}, {Gustafson},
  {Guzman}, {Haegel}, {Halim}, {Hall}, {Hamilton}, {Hammond}, {Haney}, {Hanke},
  {Hanks}, {Hanna}, {Hannuksela}, {Hannuksela}, {Hansen}, {Hansen}, {Hanson},
  {Harder}, {Hardwick}, {Haris}, {Harms}, {Harry}, {Harry}, {Hartwig},
  {Hasskew}, {Haster}, {Haughian}, {Hayes}, {Healy}, {Heidmann}, {Heintze},
  {Heinze}, {Heinzel}, {Heitmann}, {Hellman}, {Hello}, {Helmling-Cornell},
  {Hemming}, {Hendry}, {Heng}, {Hennes}, {Hennig}, {Hennig}, {Hernandez
  Vivanco}, {Heurs}, {Hild}, {Hill}, {Hines}, {Hochheim}, {Hofgard}, {Hofman},
  {Hohmann}, {Holgado}, {Holland}, {Hollows}, {Holmes}, {Holt}, {Holz},
  {Hopkins}, {Horst}, {Hough}, {Howell}, {Hoy}, {Hoyland}, {Huang},
  {H{\"u}bner}, {Huddart}, {Huerta}, {Hughey}, {Hui}, {Husa}, {Huttner},
  {Hutzler}, {Huxford}, {Huynh-Dinh}, {Idzkowski}, {Iess}, {Imperato},
  {Inchauspe}, {Ingram}, {Intini}, {Isi}, {Iyer}, {JaberianHamedan}, {Jacqmin},
  {Jadhav}, {Jadhav}, {James}, {Jani}, {Janssens}, {Janthalur}, {Jaranowski},
  {Jariwala}, {Jaume}, {Jenkins}, {Jeunon}, {Jiang}, {Johns}, {Jones}, {Jones},
  {Jones}, {Jones}, {Jones}, {Jonker}, {Ju}, {Junker}, {Kalaghatgi},
  {Kalogera}, {Kamai}, {Kandhasamy}, {Kang}, {Kanner}, {Kapadia}, {Kapasi},
  {Karathanasis}, {Karki}, {Kashyap}, {Kasprzack}, {Kastaun}, {Katsanevas},
  {Katsavounidis}, {Katzman}, {Kawabe}, {K{\'e}f{\'e}lian}, {Keitel}, {Key},
  {Khadka}, {Khalili}, {Khan}, {Khan}, {Khazanov}, {Khetan}, {Khursheed},
  {Kijbunchoo}, {Kim}, {Kim}, {Kim}, {Kim}, {Kim}, {Kim}, {Kimball}, {King},
  {Kinley-Hanlon}, {Kirchhoff}, {Kissel}, {Kleybolte}, {Klimenko}, {Knowles},
  {Knyazev}, {Koch}, {Koehlenbeck}, {Koekoek}, {Koley}, {Kolstein}, {Komori},
  {Kondrashov}, {Kontos}, {Koper}, {Korobko}, {Korth}, {Kovalam}, {Kozak},
  {Kr{\"a}mer}, {Kringel}, {Krishnendu}, {Kr{\'o}lak}, {Kuehn}, {Kumar},
  {Kumar}, {Kumar}, {Kumar}, {Kuns}, {Kwang}, {Lackey}, {Laghi}, {Lalande},
  {Lam}, {Lamberts}, {Landry}, {Lane}, {Lang}, {Lange}, {Lantz}, {Lanza}, {La
  Rosa}, {Lartaux-Vollard}, {Lasky}, {Laxen}, {Lazzarini}, {Lazzaro}, {Leaci},
  {Leavey}, {Lecoeuche}, {Lee}, {Lee}, {Lee}, {Lee}, {Lehmann}, {Leon},
  {Leroy}, {Letendre}, {Levin}, {Li}, {Li}, {Li}, {Li}, {Li}, {Linde},
  {Linker}, {Linley}, {Littenberg}, {Liu}, {Liu}, {Llorens-Monteagudo}, {Lo},
  {Lockwood}, {London}, {Longo}, {Lorenzini}, {Loriette}, {Lormand}, {Losurdo},
  {Lough}, {Lousto}, {Lovelace}, {L{\"u}ck}, {Lumaca}, {Lundgren}, {Ma},
  {Macas}, {MacInnis}, {Macleod}, {MacMillan}, {Macquet}, {Maga{\~n}a
  Hernandez}, {Maga{\~n}a-Sandoval}, {Magazz{\`u}}, {Magee}, {Majorana},
  {Maksimovic}, {Maliakal}, {Malik}, {Man}, {Mandic}, {Mangano}, {Mansell},
  {Manske}, {Mantovani}, {Mapelli}, {Marchesoni}, {Marion}, {M{\'a}rka},
  {M{\'a}rka}, {Markakis}, {Markosyan}, {Markowitz}, {Maros}, {Marquina},
  {Marsat}, {Martelli}, {Martin}, {Martin}, {Martinez}, {Martinez}, {Martynov},
  {Masalehdan}, {Mason}, {Massera}, {Masserot}, {Massinger}, {Masso-Reid},
  {Mastrogiovanni}, {Matas}, {Mateu-Lucena}, {Matichard}, {Matiushechkina},
  {Mavalvala}, {Maynard}, {McCann}, {McCarthy}, {McClelland}, {McCormick},
  {McCuller}, {McGuire}, {McIsaac}, {McIver}, {McManus}, {McRae}, {McWilliams},
  {Meacher}, {Meadors}, {Mehmet}, {Mehta}, {Melatos}, {Melchor}, {Mendell},
  {Menendez-Vazquez}, {Mercer}, {Mereni}, {Merfeld}, {Merilh}, {Merritt},
  {Merzougui}, {Meshkov}, {Messenger}, {Messick}, {Metzdorff}, {Meyers},
  {Meylahn}, {Mhaske}, {Miani}, {Miao}, {Michaloliakos}, {Michel}, {Middleton},
  {Milano}, {Miller}, {Miller}, {Millhouse}, {Mills}, {Milotti},
  {Milovich-Goff}, {Minazzoli}, {Minenkov}, {Mir}, {Mishkin}, {Mishra},
  {Mistry}, {Mitra}, {Mitrofanov}, {Mitselmakher}, {Mittleman}, {Mo},
  {Mogushi}, {Mohapatra}, {Mohite}, {Molina}, {Molina-Ruiz}, {Mondin},
  {Montani}, {Moore}, {Moraru}, {Morawski}, {Moreno}, {Morisaki}, {Mours},
  {Mow-Lowry}, {Mozzon}, {Muciaccia}, {Mukherjee}, {Mukherjee}, {Mukherjee},
  {Mukherjee}, {Mukund}, {Mullavey}, {Munch}, {Mu{\~n}iz}, {Murray}, {Nadji},
  {Nagar}, {Nardecchia}, {Naticchioni}, {Nayak}, {Neil}, {Neilson}, {Nelemans},
  {Nelson}, {Nery}, {Neunzert}, {Ng}, {Ng}, {Nguyen}, {Nguyen}, {Nguyen},
  {Nichols}, {Nissanke}, {Nocera}, {Noh}, {North}, {Nothard}, {Nuttall},
  {Oberling}, {O'Brien}, {O'Dell}, {Oganesyan}, {Ogin}, {Oh}, {Oh}, {Ohme},
  {Ohta}, {Okada}, {Olivetto}, {Oppermann}, {Oram}, {O'Reilly}, {Ormiston},
  {Ormsby}, {Ortega}, {O'Shaughnessy}, {Ossokine}, {Osthelder}, {Ottaway},
  {Overmier}, {Owen}, {Pace}, {Pagano}, {Page}, {Pagliaroli}, {Pai}, {Pai},
  {Palamos}, {Palashov}, {Palomba}, {Pan}, {Panda}, {Pang}, {Pankow},
  {Pannarale}, {Pant}, {Paoletti}, {Paoli}, {Paolone}, {Parker}, {Pascucci},
  {Pasqualetti}, {Passaquieti}, {Passuello}, {Patel}, {Patricelli}, {Payne},
  {Pechsiri}, {Pedraza}, {Pegoraro}, {Pele}, {Penn}, {Perego}, {Perez},
  {P{\'e}rigois}, {Perreca}, {Perri{\`e}s}, {Petermann}, {Petterson},
  {Pfeiffer}, {Pham}, {Phukon}, {Piccinni}, {Pichot}, {Piendibene},
  {Piergiovanni}, {Pierini}, {Pierro}, {Pillant}, {Pilo}, {Pinard}, {Pinto},
  {Piotrzkowski}, {Pirello}, {Pitkin}, {Placidi}, {Plastino}, {Pluchar},
  {Poggiani}, {Polini}, {Pong}, {Ponrathnam}, {Popolizio}, {Porter},
  {Poverman}, {Powell}, {Pracchia}, {Prajapati}, {Prasai}, {Prasanna},
  {Pratten}, {Prestegard}, {Principe}, {Prodi}, {Prokhorov}, {Prosposito},
  {Puecher}, {Punturo}, {Puosi}, {Puppo}, {P{\"u}rrer}, {Qi}, {Quetschke},
  {Quinonez}, {Quitzow-James}, {Raab}, {Raaijmakers}, {Radkins}, {Radulesco},
  {Raffai}, {Rafferty}, {Rail}, {Raja}, {Rajan}, {Rajbhandari}, {Rakhmanov},
  {Ramirez}, {Ramirez}, {Ramos-Buades}, {Rana}, {Rao}, {Rapagnani}, {Rapol},
  {Ratto}, {Raymond}, {Razzano}, {Read}, {Regimbau}, {Rei}, {Reid}, {Reitze},
  {Rettegno}, {Ricci}, {Richardson}, {Richardson}, {Richardson}, {Ricker},
  {Riemenschneider}, {Riles}, {Rizzo}, {Robertson}, {Robinet}, {Rocchi},
  {Rocha}, {Rodriguez}, {Rodriguez-Soto}, {Rolland}, {Rollins}, {Roma},
  {Romanelli}, {Romano}, {Romel}, {Romero}, {Romero-Shaw}, {Romie}, {Ronchini},
  {Rose}, {Rose}, {Rose}, {Rosell}, {Rosi{\'n}ska}, {Rosofsky}, {Ross},
  {Rowan}, {Rowlinson}, {Roy}, {Roy}, {Ruggi}, {Ryan}, {Sachdev}, {Sadecki},
  {Sakellariadou}, {Salafia}, {Salconi}, {Saleem}, {Samajdar}, {Sanchez},
  {Sanchez}, {Sanchez}, {Sanchis-Gual}, {Sanders}, {Santiago}, {Santos},
  {Saravanan}, {Sarin}, {Sassolas}, {Sathyaprakash}, {Sauter}, {Savage},
  {Savant}, {Sawant}, {Sayah}, {Schaetzl}, {Schale}, {Scheel}, {Scheuer},
  {Schindler-Tyka}, {Schmidt}, {Schnabel}, {Schofield}, {Sch{\"o}nbeck},
  {Schreiber}, {Schulte}, {Schutz}, {Schwarm}, {Schwartz}, {Scott}, {Scott},
  {Seglar-Arroyo}, {Seidel}, {Sellers}, {Sengupta}, {Sennett}, {Sentenac},
  {Sequino}, {Sergeev}, {Setyawati}, {Shaffer}, {Shahriar}, {Sharifi},
  {Sharma}, {Sharma}, {Shawhan}, {Shen}, {Shikauchi}, {Shink}, {Shoemaker},
  {Shoemaker}, {Shukla}, {ShyamSundar}, {Sieniawska}, {Sigg}, {Singer},
  {Singh}, {Singh}, {Singha}, {Singhal}, {Sintes}, {Sipala}, {Skliris},
  {Slagmolen}, {Slaven-Blair}, {Smetana}, {Smith}, {Smith}, {Somala}, {Son},
  {Soni}, {Sorazu}, {Sordini}, {Sorrentino}, {Sorrentino}, {Soulard},
  {Souradeep}, {Sowell}, {Spencer}, {Spera}, {Srivastava}, {Srivastava},
  {Staats}, {Stachie}, {Steer}, {Steinke}, {Steinlechner}, {Steinlechner},
  {Steinmeyer}, {Stevenson}, {Stolle-McAllister}, {Stops}, {Stover}, {Strain},
  {Stratta}, {Strunk}, {Sturani}, {Stuver}, {S{\"u}dbeck}, {Sudhagar},
  {Sudhir}, {Suh}, {Summerscales}, {Sun}, {Sun}, {Sunil}, {Sur}, {Suresh},
  {Sutton}, {Swinkels}, {Szczepa{\'n}czyk}, {Tacca}, {Tait}, {Talbot},
  {Tanasijczuk}, {Tanner}, {Tao}, {Tapia}, {Tapia San Martin}, {Tasson},
  {Taylor}, {Tenorio}, {Terkowski}, {Thirugnanasambandam}, {Thomas}, {Thomas},
  {Thomas}, {Thompson}, {Thondapu}, {Thorne}, {Thrane}, {Tiwari}, {Tiwari},
  {Tiwari}, {Toland}, {Tolley}, {Tonelli}, {Tornasi}, {Torres-Forn{\'e}},
  {Torrie}, {Tosta e Melo}, {T{\"o}yr{\"a}}, {Tran}, {Trapananti}, {Travasso},
  {Traylor}, {Tringali}, {Tripathee}, {Trovato}, {Trudeau}, {Tsai}, {Tsang},
  {Tse}, {Tso}, {Tsukada}, {Tsuna}, {Tsutsui}, {Turconi}, {Ubhi}, {Udall},
  {Ueno}, {Ugolini}, {Unnikrishnan}, {Urban}, {Usman}, {Utina}, {Vahlbruch},
  {Vajente}, {Vajpeyi}, {Valdes}, {Valentini}, {Valsan}, {van Bakel},
  {Beuzekom}, {van den Brand}, {Van Den Broeck}, {Vander-Hyde}, {van der
  Schaaf}, {van Heijningen}, {Vardaro}, {Vargas}, {Varma}, {Vass},
  {Vas{\'u}th}, {Vecchio}, {Vedovato}, {Veitch}, {Veitch}, {Venkateswara},
  {Venneberg}, {Venugopalan}, {Verkindt}, {Verma}, {Veske}, {Vetrano},
  {Vicer{\'e}}, {Viets}, {Villa-Ortega}, {Vinet}, {Vitale}, {Vo}, {Vocca},
  {Vorvick}, {Vyatchanin}, {Wade}, {Wade}, {Wade}, {Walet}, {Walker},
  {Wallace}, {Wallace}, {Walsh}, {Wang}, {Wang}, {Wang}, {Wang}, {Ward},
  {Warner}, {Was}, {Washington}, {Watchi}, {Weaver}, {Wei}, {Weinert},
  {Weinstein}, {Weiss}, {Wellmann}, {Wen}, {We{\ss}els}, {Westhouse}, {Wette},
  {Whelan}, {White}, {White}, {Whiting}, {Whittle}, {Wilken}, {Williams},
  {Williams}, {Williamson}, {Willis}, {Willke}, {Wilson}, {Wimmer}, {Winkler},
  {Wipf}, {Woan}, {Woehler}, {Wofford}, {Wong}, {Wrangel}, {Wright}, {Wu},
  {Wysocki}, {Xiao}, {Yamamoto}, {Yang}, {Yang}, {Yang}, {Yap}, {Yeeles},
  {Yoon}, {Yu}, {Yu}, {Yuen}, {Zadro{\.z}ny}, {Zanolin}, {Zelenova}, {Zendri},
  {Zevin}, {Zhang}, {Zhang}, {Zhang}, {Zhang}, {Zhao}, {Zhao}, {Zhou}, {Zhou},
  {Zhu}, {Zimmerman}, {Zucker}, {Zweizig}, {LIGO Scientific Collaboration}, \&
  {Virgo Collaboration}}]{Abbott2021}
{Abbott}, R., {Abbott}, T.~D., {Abraham}, S., {et~al.} 2021, \apjl, 913, L7

\bibitem[{{Asplund} {et~al.}(2009){Asplund}, {Grevesse}, {Sauval}, \&
  {Scott}}]{Asplund2009}
{Asplund}, M., {Grevesse}, N., {Sauval}, A.~J., \& {Scott}, P. 2009, \araa, 47,
  481

\bibitem[{{Baratella} {et~al.}(2021){Baratella}, {D'Orazi}, {Sheminova},
  {Spina}, {Carraro}, {Gratton}, {Magrini}, {Randich}, {Lugaro}, {Pignatari},
  {Romano}, {Biazzo}, {Bragaglia}, {Casali}, {Desidera}, {Frasca}, {de Silva},
  {Melo}, {Van der Swaelmen}, {Tautvai{\v{s}}ien{\.{e}}},
  {Jim{\'e}nez-Esteban}, {Gilmore}, {Bensby}, {Smiljanic}, {Bayo},
  {Franciosini}, {Gonneau}, {Hourihane}, {Jofr{\'e}}, {Monaco}, {Morbidelli},
  {Sacco}, {Sbordone}, {Worley}, \& {Zaggia}}]{Baratella2021}
{Baratella}, M., {D'Orazi}, V., {Sheminova}, V., {et~al.} 2021, \aap, 653, A67

\bibitem[{{Bensby} {et~al.}(2019){Bensby}, {Bergemann}, {Rybizki}, {Lemasle},
  {Howes}, {Kovalev}, {Agertz}, {Asplund}, {Barklem}, {Battistini},
  {Casagrande}, {Chiappini}, {Church}, {Feltzing}, {Ford}, {Gerhard},
  {Kushniruk}, {Kordopatis}, {Lind}, {Minchev}, {McMillan}, {Rix}, {Ryde}, \&
  {Traven}}]{Bensby2019Msngr.175...35B}
{Bensby}, T., {Bergemann}, M., {Rybizki}, J., {et~al.} 2019, The Messenger,
  175, 35

\bibitem[{{Berger} {et~al.}(2022){Berger}, {van Saders}, {Huber}, {Gaidos},
  {Schlieder}, \& {Claytor}}]{Berger2022}
{Berger}, T.~A., {van Saders}, J.~L., {Huber}, D., {et~al.} 2022, \apj, 936,
  100

\bibitem[{{Boulet}(2024)}]{Boulet2024A&A...685A..66B}
{Boulet}, T. 2024, \aap, 685, A66

\bibitem[{{Bragaglia} {et~al.}(2022){Bragaglia}, {Alfaro}, {Flaccomio},
  {Blomme}, {Donati}, {Costado}, {Damiani}, {Franciosini}, {Prisinzano},
  {Randich}, {Friel}, {Hatztidimitriou}, {Vallenari}, {Spagna},
  {Balaguer-Nunez}, {Bonito}, {Cantat Gaudin}, {Casamiquela}, {Jeffries},
  {Jordi}, {Magrini}, {Drew}, {Jackson}, {Abbas}, {Caramazza}, {Hayes},
  {Jim{\'e}nez-Esteban}, {Re Fiorentin}, {Wright}, {Bayo}, {Bensby},
  {Bergemann}, {Gilmore}, {Gonneau}, {Heiter}, {Hourihane}, {Pancino}, {Sacco},
  {Smiljanic}, {Zaggia}, \& {Vink}}]{Bragaglia2022A&A...659A.200B}
{Bragaglia}, A., {Alfaro}, E.~J., {Flaccomio}, E., {et~al.} 2022, \aap, 659,
  A200

\bibitem[{{Busso} {et~al.}(2021){Busso}, {Vescovi}, {Palmerini}, {Cristallo},
  \& {Antonuccio-Delogu}}]{Busso2021}
{Busso}, M., {Vescovi}, D., {Palmerini}, S., {Cristallo}, S., \&
  {Antonuccio-Delogu}, V. 2021, \apj, 908, 55

\bibitem[{{Cantat-Gaudin} {et~al.}(2020){Cantat-Gaudin}, {Anders},
  {Castro-Ginard}, {Jordi}, {Romero-G{\'o}mez}, {Soubiran}, {Casamiquela},
  {Tarricq}, {Moitinho}, {Vallenari}, {Bragaglia}, {Krone-Martins}, \&
  {Kounkel}}]{Cantat2020A&A...640A...1C}
{Cantat-Gaudin}, T., {Anders}, F., {Castro-Ginard}, A., {et~al.} 2020, \aap,
  640, A1

\bibitem[{{Cappellaro} {et~al.}(1999){Cappellaro}, {Evans}, \&
  {Turatto}}]{Cappellaro1999}
{Cappellaro}, E., {Evans}, R., \& {Turatto}, M. 1999, \aap, 351, 459

\bibitem[{{Casali} {et~al.}(2023){Casali}, {Grisoni}, {Miglio}, {Chiappini},
  {Matteuzzi}, {Magrini}, {Willett}, {Cescutti}, {Matteucci}, {Stokholm},
  {Tailo}, {Montalb{\'a}n}, {Elsworth}, \&
  {Mosser}}]{Casali2023A&A...677A..60C}
{Casali}, G., {Grisoni}, V., {Miglio}, A., {et~al.} 2023, \aap, 677, A60

\bibitem[{{Casali} {et~al.}(2019){Casali}, {Magrini}, {Tognelli}, {Jackson},
  {Jeffries}, {Lagarde}, {Tautvai{\v{s}}ien{\.{e}}}, {Masseron},
  {Degl'Innocenti}, {Prada Moroni}, {Kordopatis}, {Pancino}, {Randich},
  {Feltzing}, {Sahlholdt}, {Spina}, {Friel}, {Roccatagliata}, {Sanna},
  {Bragaglia}, {Drazdauskas}, {Mikolaitis}, {Minkevi{\v{c}}i{\={u}}t{\.{e}}},
  {Stonkut{\.{e}}}, {Chorniy}, {Bagdonas}, {Jimenez-Esteban}, {Martell}, {Van
  der Swaelmen}, {Gilmore}, {Vallenari}, {Bensby}, {Koposov}, {Korn}, {Worley},
  {Smiljanic}, {Bergemann}, {Carraro}, {Damiani}, {Prisinzano}, {Bonito},
  {Franciosini}, {Gonneau}, {Hourihane}, {Jofre}, {Lewis}, {Morbidelli},
  {Sacco}, {Sousa}, {Zaggia}, {Lanzafame}, {Heiter}, {Frasca}, \&
  {Bayo}}]{Casali2019A&A...629A..62C}
{Casali}, G., {Magrini}, L., {Tognelli}, E., {et~al.} 2019, \aap, 629, A62

\bibitem[{{Casali} {et~al.}(2020){Casali}, {Spina}, {Magrini}, {Karakas},
  {Kobayashi}, {Casey}, {Feltzing}, {Van der Swaelmen}, {Tsantaki},
  {Jofr{\'e}}, {Bragaglia}, {Feuillet}, {Bensby}, {Biazzo}, {Gonneau},
  {Tautvai{\v{s}}ien{\.{e}}}, {Baratella}, {Roccatagliata}, {Pancino}, {Sousa},
  {Adibekyan}, {Martell}, {Bayo}, {Jackson}, {Jeffries}, {Gilmore}, {Randich},
  {Alfaro}, {Koposov}, {Korn}, {Recio-Blanco}, {Smiljanic}, {Franciosini},
  {Hourihane}, {Monaco}, {Morbidelli}, {Sacco}, {Worley}, \&
  {Zaggia}}]{Casali2020A&A...639A.127C}
{Casali}, G., {Spina}, L., {Magrini}, L., {et~al.} 2020, \aap, 639, A127

\bibitem[{{Cescutti} \& {Matteucci}(2022)}]{Cescutti2022}
{Cescutti}, G. \& {Matteucci}, F. 2022, Universe, 8, 173

\bibitem[{{Chaplin} \& {Miglio}(2013)}]{ChaplinMiglio2013}
{Chaplin}, W.~J. \& {Miglio}, A. 2013, \araa, 51, 353

\bibitem[{{Chiappini} {et~al.}(1997){Chiappini}, {Matteucci}, \&
  {Gratton}}]{Chiappini1997}
{Chiappini}, C., {Matteucci}, F., \& {Gratton}, R. 1997, \apj, 477, 765

\bibitem[{{Chiappini} {et~al.}(2001){Chiappini}, {Matteucci}, \&
  {Romano}}]{Chiappini2001}
{Chiappini}, C., {Matteucci}, F., \& {Romano}, D. 2001, \apj, 554, 1044

\bibitem[{{Choplin} {et~al.}(2021){Choplin}, {Siess}, \&
  {Goriely}}]{Choplin2021}
{Choplin}, A., {Siess}, L., \& {Goriely}, S. 2021, \aap, 648, A119

\bibitem[{{Choplin} {et~al.}(2022){Choplin}, {Siess}, \&
  {Goriely}}]{Choplin2022}
{Choplin}, A., {Siess}, L., \& {Goriely}, S. 2022, \aap, 667, A155

\bibitem[{{Choplin} {et~al.}(2024){Choplin}, {Siess}, {Goriely}, \&
  {Martinet}}]{Choplin2024}
{Choplin}, A., {Siess}, L., {Goriely}, S., \& {Martinet}, S. 2024, \aap, 684,
  A206

\bibitem[{{Cristallo} {et~al.}(2011){Cristallo}, {Piersanti}, {Straniero},
  {Gallino}, {Dom{\'\i}nguez}, {Abia}, {Di Rico}, {Quintini}, \&
  {Bisterzo}}]{Cristallo2011ApJS..197...17C}
{Cristallo}, S., {Piersanti}, L., {Straniero}, O., {et~al.} 2011, \apjs, 197,
  17

\bibitem[{{Cristallo} {et~al.}(2009){Cristallo}, {Straniero}, {Gallino},
  {Piersanti}, {Dom{\'\i}nguez}, \& {Lederer}}]{Cristallo2009}
{Cristallo}, S., {Straniero}, O., {Gallino}, R., {et~al.} 2009, \apj, 696, 797

\bibitem[{{Cristallo} {et~al.}(2015){Cristallo}, {Straniero}, {Piersanti}, \&
  {Gobrecht}}]{Cristallo2015}
{Cristallo}, S., {Straniero}, O., {Piersanti}, L., \& {Gobrecht}, D. 2015,
  \apjs, 219, 40

\bibitem[{{Delgado Mena} {et~al.}(2019{\natexlab{a}}){Delgado Mena}, {Moya},
  {Adibekyan}, {Tsantaki}, {Gonz{\'a}lez Hern{\'a}ndez}, {Israelian}, {Davies},
  {Chaplin}, {Sousa}, {Ferreira}, \& {Santos}}]{Delgado2019A&A...624A..78D}
{Delgado Mena}, E., {Moya}, A., {Adibekyan}, V., {et~al.} 2019{\natexlab{a}},
  \aap, 624, A78

\bibitem[{{Delgado Mena} {et~al.}(2019{\natexlab{b}}){Delgado Mena}, {Moya},
  {Adibekyan}, {Tsantaki}, {Gonz{\'a}lez Hern{\'a}ndez}, {Israelian}, {Davies},
  {Chaplin}, {Sousa}, {Ferreira}, \& {Santos}}]{DelgadoMena2019}
{Delgado Mena}, E., {Moya}, A., {Adibekyan}, V., {et~al.} 2019{\natexlab{b}},
  \aap, 624, A78

\bibitem[{{D'Orazi} {et~al.}(2017){D'Orazi}, {De Silva}, \&
  {Melo}}]{d'orazi2017}
{D'Orazi}, V., {De Silva}, G.~M., \& {Melo}, C.~F.~H. 2017, \aap, 598, A86

\bibitem[{{D'Orazi} {et~al.}(2009){D'Orazi}, {Magrini}, {Randich}, {Galli},
  {Busso}, \& {Sestito}}]{d'orazi2009}
{D'Orazi}, V., {Magrini}, L., {Randich}, S., {et~al.} 2009, \apjl, 693, L31

\bibitem[{{Feltzing} {et~al.}(2017){Feltzing}, {Howes}, {McMillan}, \&
  {Stonkut{\.{e}}}}]{Feltzing2017MNRAS.465L.109F}
{Feltzing}, S., {Howes}, L.~M., {McMillan}, P.~J., \& {Stonkut{\.{e}}}, E.
  2017, \mnras, 465, L109

\bibitem[{{Frasca} {et~al.}(2019){Frasca}, {Alonso-Santiago}, {Catanzaro},
  {Bragaglia}, {Carretta}, {Casali}, {D'Orazi}, {Magrini}, {Andreuzzi},
  {Oliva}, {Origlia}, {Sordo}, \& {Vallenari}}]{Frasca2019}
{Frasca}, A., {Alonso-Santiago}, J., {Catanzaro}, G., {et~al.} 2019, \aap, 632,
  A16

\bibitem[{{Frischknecht} {et~al.}(2016){Frischknecht}, {Hirschi}, {Pignatari},
  {Maeder}, {Meynet}, {Chiappini}, {Thielemann}, {Rauscher}, {Georgy}, \&
  {Ekstr{\"o}m}}]{2016MNRAS.456.1803F}
{Frischknecht}, U., {Hirschi}, R., {Pignatari}, M., {et~al.} 2016, \mnras, 456,
  1803

\bibitem[{{Frischknecht} {et~al.}(2012){Frischknecht}, {Hirschi}, \&
  {Thielemann}}]{Frischknecht2012}
{Frischknecht}, U., {Hirschi}, R., \& {Thielemann}, F.~K. 2012, \aap, 538, L2

\bibitem[{{Gaia Collaboration} {et~al.}(2023){Gaia Collaboration},
  {Recio-Blanco}, {Kordopatis}, {de Laverny}, {Palicio}, {Spagna}, {Spina},
  {Katz}, {Re Fiorentin}, {Poggio}, {McMillan}, {Vallenari}, {Lattanzi},
  {Seabroke}, {Casamiquela}, {Bragaglia}, {Antoja}, {Bailer-Jones},
  {Schultheis}, {Andrae}, {Fouesneau}, {Cropper}, {Cantat-Gaudin}, {Bijaoui},
  {Heiter}, {Brown}, {Prusti}, {de Bruijne}, {Arenou}, {Babusiaux}, {Biermann},
  {Creevey}, {Ducourant}, {Evans}, {Eyer}, {Guerra}, {Hutton}, {Jordi},
  {Klioner}, {Lammers}, {Lindegren}, {Luri}, {Mignard}, {Panem}, {Pourbaix},
  {Randich}, {Sartoretti}, {Soubiran}, {Tanga}, {Walton}, {Bastian}, {Drimmel},
  {Jansen}, {van Leeuwen}, {Bakker}, {Cacciari}, {Casta{\~n}eda}, {De Angeli},
  {Fabricius}, {Fr{\'e}mat}, {Galluccio}, {Guerrier}, {Masana}, {Messineo},
  {Mowlavi}, {Nicolas}, {Nienartowicz}, {Pailler}, {Panuzzo}, {Riclet}, {Roux},
  {Sordo}, {Th{\'e}venin}, {Gracia-Abril}, {Portell}, {Teyssier}, {Altmann},
  {Audard}, {Bellas-Velidis}, {Benson}, {Berthier}, {Blomme}, {Burgess},
  {Busonero}, {Busso}, {C{\'a}novas}, {Carry}, {Cellino}, {Cheek},
  {Clementini}, {Damerdji}, {Davidson}, {de Teodoro}, {Nu{\~n}ez Campos},
  {Delchambre}, {Dell'Oro}, {Esquej}, {Fern{\'a}ndez-Hern{\'a}ndez}, {Fraile},
  {Garabato}, {Garc{\'\i}a-Lario}, {Gosset}, {Haigron}, {Halbwachs}, {Hambly},
  {Harrison}, {Hern{\'a}ndez}, {Hestroffer}, {Hodgkin}, {Holl}, {Jan{\ss}en},
  {Jevardat de Fombelle}, {Jordan}, {Krone-Martins}, {Lanzafame},
  {L{\"o}ffler}, {Marchal}, {Marrese}, {Moitinho}, {Muinonen}, {Osborne},
  {Pancino}, {Pauwels}, {Reyl{\'e}}, {Riello}, {Rimoldini}, {Roegiers},
  {Rybizki}, {Sarro}, {Siopis}, {Smith}, {Sozzetti}, {Utrilla}, {van Leeuwen},
  {Abbas}, {{\'A}brah{\'a}m}, {Abreu Aramburu}, {Aerts}, {Aguado}, {Ajaj},
  {Aldea-Montero}, {Altavilla}, {{\'A}lvarez}, {Alves}, {Anders}, {Anderson},
  {Anglada Varela}, {Baines}, {Baker}, {Balaguer-N{\'u}{\~n}ez}, {Balbinot},
  {Balog}, {Barache}, {Barbato}, {Barros}, {Barstow}, {Bartolom{\'e}},
  {Bassilana}, {Bauchet}, {Becciani}, {Bellazzini}, {Berihuete}, {Bernet},
  {Bertone}, {Bianchi}, {Binnenfeld}, {Blanco-Cuaresma}, {Boch}, {Bombrun},
  {Bossini}, {Bouquillon}, {Bramante}, {Breedt}, {Bressan}, {Brouillet},
  {Brugaletta}, {Bucciarelli}, {Burlacu}, {Butkevich}, {Buzzi}, {Caffau},
  {Cancelliere}, {Carballo}, {Carlucci}, {Carnerero}, {Carrasco}, {Castellani},
  {Castro-Ginard}, {Chaoul}, {Charlot}, {Chemin}, {Chiaramida}, {Chiavassa},
  {Chornay}, {Comoretto}, {Contursi}, {Cooper}, {Cornez}, {Cowell}, {Crifo},
  {Crosta}, {Crowley}, {Dafonte}, {Dapergolas}, {David}, {De Luise}, {De
  March}, {De Ridder}, {de Souza}, {de Torres}, {del Peloso}, {del Pozo},
  {Delbo}, {Delgado}, {Delisle}, {Demouchy}, {Dharmawardena}, {Di Matteo},
  {Diakite}, {Diener}, {Distefano}, {Dolding}, {Edvardsson}, {Enke}, {Fabre},
  {Fabrizio}, {Faigler}, {Fedorets}, {Fernique}, {Figueras}, {Fournier},
  {Fouron}, {Fragkoudi}, {Gai}, {Garcia-Gutierrez}, {Garcia-Reinaldos},
  {Garc{\'\i}a-Torres}, {Garofalo}, {Gavel}, {Gavras}, {Gerlach}, {Geyer},
  {Giacobbe}, {Gilmore}, {Girona}, {Giuffrida}, {Gomel}, {Gomez},
  {Gonz{\'a}lez-N{\'u}{\~n}ez}, {Gonz{\'a}lez-Santamar{\'\i}a},
  {Gonz{\'a}lez-Vidal}, {Granvik}, {Guillout}, {Guiraud},
  {Guti{\'e}rrez-S{\'a}nchez}, {Guy}, {Hatzidimitriou}, {Hauser}, {Haywood},
  {Helmer}, {Helmi}, {Sarmiento}, {Hidalgo}, {H{\l}adczuk}, {Hobbs}, {Holland},
  {Huckle}, {Jardine}, {Jasniewicz}, {Jean-Antoine Piccolo},
  {Jim{\'e}nez-Arranz}, {Juaristi Campillo}, {Julbe}, {Karbevska}, {Kervella},
  {Khanna}, {Korn}, {K{\'o}sp{\'a}l}, {Kostrzewa-Rutkowska}, {Kruszy{\'n}ska},
  {Kun}, {Laizeau}, {Lambert}, {Lanza}, {Lasne}, {Le Campion}, {Lebreton},
  {Lebzelter}, {Leccia}, {Leclerc}, {Lecoeur-Taibi}, {Liao}, {Licata},
  {Lindstr{\o}m}, {Lister}, {Livanou}, {Lobel}, {Lorca}, {Loup}, {Madrero
  Pardo}, {Magdaleno Romeo}, {Managau}, {Mann}, {Manteiga}, {Marchant},
  {Marconi}, {Marcos}, {Marcos Santos}, {Mar{\'\i}n Pina}, {Marinoni},
  {Marocco}, {Marshall}, {Martin Polo}, {Mart{\'\i}n-Fleitas}, {Marton},
  {Mary}, {Masip}, {Massari}, {Mastrobuono-Battisti}, {Mazeh}, {Messina},
  {Michalik}, {Millar}, {Mints}, {Molina}, {Molinaro}, {Moln{\'a}r}, {Monari},
  {Mongui{\'o}}, {Montegriffo}, {Montero}, {Mor}, {Mora}, {Morbidelli},
  {Morel}, {Morris}, {Muraveva}, {Murphy}, {Musella}, {Nagy}, {Noval},
  {Oca{\~n}a}, {Ogden}, {Ordenovic}, {Osinde}, {Pagani}, {Pagano}, {Palaversa},
  {Pallas-Quintela}, {Panahi}, {Payne-Wardenaar}, {Pe{\~n}alosa Esteller},
  {Penttil{\"a}}, {Pichon}, {Piersimoni}, {Pineau}, {Plachy}, {Plum},
  {Pr{\v{s}}a}, {Pulone}, {Racero}, {Ragaini}, {Rainer}, {Raiteri}, {Ramos},
  {Ramos-Lerate}, {Regibo}, {Richards}, {Rios Diaz}, {Ripepi}, {Riva}, {Rix},
  {Rixon}, {Robichon}, {Robin}, {Robin}, {Roelens}, {Rogues}, {Rohrbasser},
  {Romero-G{\'o}mez}, {Rowell}, {Royer}, {Ruz Mieres}, {Rybicki}, {Sadowski},
  {S{\'a}ez N{\'u}{\~n}ez}, {Sagrist{\`a} Sell{\'e}s}, {Sahlmann}, {Salguero},
  {Samaras}, {Sanchez Gimenez}, {Sanna}, {Santove{\~n}a}, {Sarasso}, {Sciacca},
  {Segol}, {Segovia}, {S{\'e}gransan}, {Semeux}, {Shahaf}, {Siddiqui},
  {Siebert}, {Siltala}, {Silvelo}, {Slezak}, {Slezak}, {Smart}, {Snaith},
  {Solano}, {Solitro}, {Souami}, {Souchay}, {Spoto}, {Steele},
  {Steidelm{\"u}ller}, {Stephenson}, {S{\"u}veges}, {Surdej}, {Szabados},
  {Szegedi-Elek}, {Taris}, {Taylor}, {Teixeira}, {Tolomei}, {Tonello}, {Torra},
  {Torra}, {Torralba Elipe}, {Trabucchi}, {Tsounis}, {Turon}, {Ulla}, {Unger},
  {Vaillant}, {van Dillen}, {van Reeven}, {Vanel}, {Vecchiato}, {Viala},
  {Vicente}, {Voutsinas}, {Weiler}, {Wevers}, {Wyrzykowski}, {Yoldas}, {Yvard},
  {Zhao}, {Zorec}, {Zucker}, \& {Zwitter}}]{PVP_Ale}
{Gaia Collaboration}, {Recio-Blanco}, A., {Kordopatis}, G., {et~al.} 2023,
  \aap, 674, A38

\bibitem[{{Gallino} {et~al.}(1998){Gallino}, {Arlandini}, {Busso}, {Lugaro},
  {Travaglio}, {Straniero}, {Chieffi}, \& {Limongi}}]{Gallino1998}
{Gallino}, R., {Arlandini}, C., {Busso}, M., {et~al.} 1998, \apj, 497, 388

\bibitem[{{Ghirlanda} {et~al.}(2016){Ghirlanda}, {Salafia}, {Pescalli},
  {Ghisellini}, {Salvaterra}, {Chassande-Mottin}, {Colpi}, {Nappo}, {D'Avanzo},
  {Melandri}, {Bernardini}, {Branchesi}, {Campana}, {Ciolfi}, {Covino},
  {G{\"o}tz}, {Vergani}, {Zennaro}, \& {Tagliaferri}}]{G16}
{Ghirlanda}, G., {Salafia}, O.~S., {Pescalli}, A., {et~al.} 2016, \aap, 594,
  A84

\bibitem[{{Gilmore} {et~al.}(2022){Gilmore}, {Randich}, {Worley}, {Hourihane},
  {Gonneau}, {Sacco}, {Lewis}, {Magrini}, {Fran{\c{c}}ois}, {Jeffries},
  {Koposov}, {Bragaglia}, {Alfaro}, {Allende Prieto}, {Blomme}, {Korn},
  {Lanzafame}, {Pancino}, {Recio-Blanco}, {Smiljanic}, {Van Eck}, {Zwitter},
  {Bensby}, {Flaccomio}, {Irwin}, {Franciosini}, {Morbidelli}, {Damiani},
  {Bonito}, {Friel}, {Vink}, {Prisinzano}, {Abbas}, {Hatzidimitriou}, {Held},
  {Jordi}, {Paunzen}, {Spagna}, {Jackson}, {Ma{\'\i}z Apell{\'a}niz},
  {Asplund}, {Bonifacio}, {Feltzing}, {Binney}, {Drew}, {Ferguson}, {Micela},
  {Negueruela}, {Prusti}, {Rix}, {Vallenari}, {Bergemann}, {Casey}, {de
  Laverny}, {Frasca}, {Hill}, {Lind}, {Sbordone}, {Sousa}, {Adibekyan},
  {Caffau}, {Daflon}, {Feuillet}, {Gebran}, {Gonzalez Hernandez}, {Guiglion},
  {Herrero}, {Lobel}, {Merle}, {Mikolaitis}, {Montes}, {Morel}, {Ruchti},
  {Soubiran}, {Tabernero}, {Tautvai{\v{s}}ien{\.{e}}}, {Traven}, {Valentini},
  {Van der Swaelmen}, {Villanova}, {Viscasillas V{\'a}zquez}, {Bayo}, {Biazzo},
  {Carraro}, {Edvardsson}, {Heiter}, {Jofr{\'e}}, {Marconi}, {Martayan},
  {Masseron}, {Monaco}, {Walton}, {Zaggia}, {Aguirre B{\o}rsen-Koch}, {Alves},
  {Balaguer-Nunez}, {Barklem}, {Barrado}, {Bellazzini}, {Berlanas}, {Binks},
  {Bressan}, {Capuzzo-Dolcetta}, {Casagrande}, {Casamiquela}, {Collins},
  {D'Orazi}, {Dantas}, {Debattista}, {Delgado-Mena}, {Di Marcantonio},
  {Drazdauskas}, {Evans}, {Famaey}, {Franchini}, {Fr{\'e}mat}, {Fu}, {Geisler},
  {Gerhard}, {Gonz{\'a}lez Solares}, {Grebel}, {Guti{\'e}rrez Albarr{\'a}n},
  {Jim{\'e}nez-Esteban}, {J{\"o}nsson}, {Khachaturyants}, {Kordopatis}, {Kos},
  {Lagarde}, {Ludwig}, {Mahy}, {Mapelli}, {Marfil}, {Martell}, {Messina},
  {Miglio}, {Minchev}, {Moitinho}, {Montalban}, {Monteiro}, {Morossi},
  {Mowlavi}, {Mucciarelli}, {Murphy}, {Nardetto}, {Ortolani}, {Paletou},
  {Palou{\v{s}}}, {Pickering}, {Quirrenbach}, {Re Fiorentin}, {Read}, {Romano},
  {Ryde}, {Sanna}, {Santos}, {Seabroke}, {Spina}, {Steinmetz}, {Stonkut{\'e}},
  {Sutorius}, {Th{\'e}venin}, {Tosi}, {Tsantaki}, {Wright}, {Wyse}, {Zoccali},
  {Zorec}, \& {Zucker}}]{Gilmore2022A&A...666A.120G}
{Gilmore}, G., {Randich}, S., {Worley}, C.~C., {et~al.} 2022, \aap, 666, A120

\bibitem[{{Greggio} {et~al.}(2021){Greggio}, {Simonetti}, \&
  {Matteucci}}]{Greggio2021}
{Greggio}, L., {Simonetti}, P., \& {Matteucci}, F. 2021, \mnras, 500, 1755

\bibitem[{{Grisoni} {et~al.}(2018){Grisoni}, {Spitoni}, \&
  {Matteucci}}]{Grisoni2018}
{Grisoni}, V., {Spitoni}, E., \& {Matteucci}, F. 2018, \mnras, 481, 2570

\bibitem[{{Hayden} {et~al.}(2015){Hayden}, {Bovy}, {Holtzman}, {Nidever},
  {Bird}, {Weinberg}, {Andrews}, {Majewski}, {Allende Prieto}, {Anders},
  {Beers}, {Bizyaev}, {Chiappini}, {Cunha}, {Frinchaboy},
  {Garc{\'\i}a-Her{\'n}andez}, {Garc{\'\i}a P{\'e}rez}, {Girardi}, {Harding},
  {Hearty}, {Johnson}, {M{\'e}sz{\'a}ros}, {Minchev}, {O'Connell}, {Pan},
  {Robin}, {Schiavon}, {Schneider}, {Schultheis}, {Shetrone}, {Skrutskie},
  {Steinmetz}, {Smith}, {Wilson}, {Zamora}, \& {Zasowski}}]{Hayden2015}
{Hayden}, M.~R., {Bovy}, J., {Holtzman}, J.~A., {et~al.} 2015, \apj, 808, 132

\bibitem[{{Hayden} {et~al.}(2014){Hayden}, {Holtzman}, {Bovy}, {Majewski},
  {Johnson}, {Allende Prieto}, {Beers}, {Cunha}, {Frinchaboy}, {Garc{\'\i}a
  P{\'e}rez}, {Girardi}, {Hearty}, {Lee}, {Nidever}, {Schiavon}, {Schlesinger},
  {Schneider}, {Schultheis}, {Shetrone}, {Smith}, {Zasowski}, {Bizyaev},
  {Feuillet}, {Hasselquist}, {Kinemuchi}, {Malanushenko}, {Malanushenko},
  {O'Connell}, {Pan}, \& {Stassun}}]{Hayden2014}
{Hayden}, M.~R., {Holtzman}, J.~A., {Bovy}, J., {et~al.} 2014, \aj, 147, 116

\bibitem[{{Isern}(2019)}]{Isern2019}
{Isern}, J. 2019, \apjl, 878, L11

\bibitem[{{Iwamoto} {et~al.}(1999){Iwamoto}, {Brachwitz}, {Nomoto},
  {Kishimoto}, {Umeda}, {Hix}, \& {Thielemann}}]{Iwamoto1999}
{Iwamoto}, K., {Brachwitz}, F., {Nomoto}, K., {et~al.} 1999, \apjs, 125, 439

\bibitem[{{Jacobson} \& {Friel}(2013)}]{Jacobson2013}
{Jacobson}, H.~R. \& {Friel}, E.~D. 2013, \aj, 145, 107

\bibitem[{{Jeffries} {et~al.}(2023){Jeffries}, {Jackson}, {Wright}, {Weaver},
  {Gilmore}, {Randich}, {Bragaglia}, {Korn}, {Smiljanic}, {Biazzo}, {Casey},
  {Frasca}, {Gonneau}, {Guiglion}, {Morbidelli}, {Prisinzano}, {Sacco},
  {Tautvai{\v{s}}ien{\.{e}}}, {Worley}, \&
  {Zaggia}}]{Jeffries2023MNRAS.523..802J}
{Jeffries}, R.~D., {Jackson}, R.~J., {Wright}, N.~J., {et~al.} 2023, \mnras,
  523, 802

\bibitem[{{Jin} {et~al.}(2024){Jin}, {Trager}, {Dalton}, {Aguerri}, {Drew},
  {Falc{\'o}n-Barroso}, {G{\"a}nsicke}, {Hill}, {Iovino}, {Pieri}, {Poggianti},
  {Smith}, {Vallenari}, {Abrams}, {Aguado}, {Antoja}, {Arag{\'o}n-Salamanca},
  {Ascasibar}, {Babusiaux}, {Balcells}, {Barrena}, {Battaglia}, {Belokurov},
  {Bensby}, {Bonifacio}, {Bragaglia}, {Carrasco}, {Carrera}, {Cornwell},
  {Dom{\'\i}nguez-Palmero}, {Duncan}, {Famaey}, {Fari{\~n}a}, {Gonzalez},
  {Guest}, {Hatch}, {Hess}, {Hoskin}, {Irwin}, {Knapen}, {Koposov}, {Kuchner},
  {Laigle}, {Lewis}, {Longhetti}, {Lucatello}, {M{\'e}ndez-Abreu}, {Mercurio},
  {Molaeinezhad}, {Mongui{\'o}}, {Morrison}, {Murphy}, {Peralta de Arriba},
  {P{\'e}rez}, {P{\'e}rez-R{\`a}fols}, {Pic{\'o}}, {Raddi}, {Romero-G{\'o}mez},
  {Royer}, {Siebert}, {Seabroke}, {Som}, {Terrett}, {Thomas}, {Wesson},
  {Worley}, {Alfaro}, {Allende Prieto}, {Alonso-Santiago}, {Amos}, {Ashley},
  {Balaguer-N{\'u}{\~n}ez}, {Balbinot}, {Bellazzini}, {Benn}, {Berlanas},
  {Bernard}, {Best}, {Bettoni}, {Bianco}, {Bishop}, {Blomqvist}, {Boeche},
  {Bolzonella}, {Bonoli}, {Bosma}, {Britavskiy}, {Busarello}, {Caffau},
  {Cantat-Gaudin}, {Castro-Ginard}, {Couto}, {Carbajo-Hijarrubia}, {Carter},
  {Casamiquela}, {Conrado}, {Corcho-Caballero}, {Costantin}, {Deason}, {de
  Burgos}, {De Grandi}, {Di Matteo}, {Dom{\'\i}nguez-G{\'o}mez}, {Dorda},
  {Drake}, {Dutta}, {Erkal}, {Feltzing}, {Ferr{\'e}-Mateu}, {Feuillet},
  {Figueras}, {Fossati}, {Franciosini}, {Frasca}, {Fumagalli}, {Gallazzi},
  {Garc{\'\i}a-Benito}, {Gentile Fusillo}, {Gebran}, {Gilbert}, {Gledhill},
  {Gonz{\'a}lez Delgado}, {Greimel}, {Guarcello}, {Guerra}, {Gullieuszik},
  {Haines}, {Hardcastle}, {Harris}, {Haywood}, {Helmi}, {Hernandez}, {Herrero},
  {Hughes}, {Ir{\v{s}}i{\v{c}}}, {Jablonka}, {Jarvis}, {Jordi}, {Kondapally},
  {Kordopatis}, {Krogager}, {La Barbera}, {Lam}, {Larsen}, {Lemasle}, {Lewis},
  {Lhom{\'e}}, {Lind}, {Lodi}, {Longobardi}, {Lonoce}, {Magrini}, {Ma{\'\i}z
  Apell{\'a}niz}, {Marchal}, {Marco}, {Martin}, {Matsuno}, {Maurogordato},
  {Merluzzi}, {Miralda-Escud{\'e}}, {Molinari}, {Monari}, {Morelli}, {Mottram},
  {Naylor}, {Negueruela}, {O{\~n}orbe}, {Pancino}, {Peirani}, {Peletier},
  {Pozzetti}, {Rainer}, {Ramos}, {Read}, {Rossi}, {R{\"o}ttgering},
  {Rubi{\~n}o-Mart{\'\i}n}, {Sabater}, {San Juan}, {Sanna}, {Schallig},
  {Schiavon}, {Schultheis}, {Serra}, {Shimwell}, {Sim{\'o}n-D{\'\i}az},
  {Smith}, {Sordo}, {Sorini}, {Soubiran}, {Starkenburg}, {Steele}, {Stott},
  {Stuik}, {Tolstoy}, {Tortora}, {Tsantaki}, {Van der Swaelmen}, {van Weeren},
  {Vergani}, {Verheijen}, {Verro}, {Vink}, {Vioque}, {Walcher}, {Walton},
  {Wegg}, {Weijmans}, {Williams}, {Wilson}, {Wright}, {Xylakis-Dornbusch},
  {Youakim}, {Zibetti}, \& {Zurita}}]{Jin2024MNRAS.530.2688J}
{Jin}, S., {Trager}, S.~C., {Dalton}, G.~B., {et~al.} 2024, \mnras, 530, 2688

\bibitem[{{Jofr{\'e}} {et~al.}(2020){Jofr{\'e}}, {Jackson}, \& {Tucci
  Maia}}]{Jofre2020}
{Jofr{\'e}}, P., {Jackson}, H., \& {Tucci Maia}, M. 2020, \aap, 633, L9

\bibitem[{{Jost} {et~al.}(2024){Jost}, {Molero}, {Nav{\'o}}, {Arcones},
  {Obergaulinger}, \& {Matteucci}}]{Jost2024}
{Jost}, F.~P., {Molero}, M., {Nav{\'o}}, G., {et~al.} 2024, arXiv e-prints,
  arXiv:2407.14319

\bibitem[{{Kalogera} {et~al.}(2004){Kalogera}, {Kim}, {Lorimer}, {Burgay},
  {D'Amico}, {Possenti}, {Manchester}, {Lyne}, {Joshi}, {McLaughlin}, {Kramer},
  {Sarkissian}, \& {Camilo}}]{Kalogera2004}
{Kalogera}, V., {Kim}, C., {Lorimer}, D.~R., {et~al.} 2004, \apjl, 601, L179

\bibitem[{{Karakas} \& {Lugaro}(2016)}]{Karakas2016}
{Karakas}, A.~I. \& {Lugaro}, M. 2016, \apj, 825, 26

\bibitem[{{Kawata} \& {Chiappini}(2016)}]{Kawata2016}
{Kawata}, D. \& {Chiappini}, C. 2016, Astronomische Nachrichten, 337, 976

\bibitem[{{Kennicutt}(1998)}]{Kennicutt1998}
{Kennicutt}, Robert~C., J. 1998, \apj, 498, 541

\bibitem[{{Korobkin} {et~al.}(2012){Korobkin}, {Rosswog}, {Arcones}, \&
  {Winteler}}]{Korobkin}
{Korobkin}, O., {Rosswog}, S., {Arcones}, A., \& {Winteler}, C. 2012, \mnras,
  426, 1940

\bibitem[{{Limongi} \& {Chieffi}(2018)}]{Limongi2018}
{Limongi}, M. \& {Chieffi}, A. 2018, \apjs, 237, 13

\bibitem[{{Liu} {et~al.}(2018){Liu}, {Gallino}, {Cristallo}, {Bisterzo},
  {Davis}, {Trappitsch}, \& {Nittler}}]{Liu2018}
{Liu}, N., {Gallino}, R., {Cristallo}, S., {et~al.} 2018, \apj, 865, 112

\bibitem[{{Lugaro} {et~al.}(2003){Lugaro}, {Herwig}, {Lattanzio}, {Gallino}, \&
  {Straniero}}]{Lugaro2003}
{Lugaro}, M., {Herwig}, F., {Lattanzio}, J.~C., {Gallino}, R., \& {Straniero},
  O. 2003, \apj, 586, 1305

\bibitem[{{Magrini} {et~al.}(2023{\natexlab{a}}){Magrini}, {Bensby},
  {Brucalassi}, {Randich}, {Jeffries}, {de Silva}, {Skuladottir}, {Smiljanic},
  {Gonzalez}, {Hill}, {Lagarde}, {Tolstoy}, {Arroyo-Polonio}, {Baratella},
  {Barnes}, {Battaglia}, {Baumgardt}, {Bellazzini}, {Biazzo}, {Bragaglia},
  {Carter}, {Casali}, {Cescutti}, {Danielski}, {Delgado Mena}, {Drazdauskas},
  {Gieles}, {Giribaldi}, {Hawkins}, {Hoeijmakers}, {Jablonka}, {Kamath},
  {Louth}, {Fabiola Marino}, {Martell}, {Merle}, {Montet}, {Murphy}, {Nisini},
  {Nordlander}, {D'Orazi}, {Pino}, {Romano}, {Sacco}, {Sandford}, {Sollima},
  {Spina}, {Tautvaisiene}, {Ting}, {Tozzi}, {Van der Swaelmen}, {Van Eck},
  {Watson}, {Worley}, \& {Zocchi}}]{Magrini2023arXiv231208270M}
{Magrini}, L., {Bensby}, T., {Brucalassi}, A., {et~al.} 2023{\natexlab{a}},
  arXiv e-prints, arXiv:2312.08270

\bibitem[{{Magrini} {et~al.}(2018){Magrini}, {Spina}, {Randich}, {Friel},
  {Kordopatis}, {Worley}, {Pancino}, {Bragaglia}, {Donati},
  {Tautvai{\v{s}}ien{\.{e}}}, {Bagdonas}, {Delgado-Mena}, {Adibekyan}, {Sousa},
  {Jim{\'e}nez-Esteban}, {Sanna}, {Roccatagliata}, {Bonito}, {Sbordone},
  {Duffau}, {Gilmore}, {Feltzing}, {Jeffries}, {Vallenari}, {Alfaro}, {Bensby},
  {Francois}, {Koposov}, {Korn}, {Recio-Blanco}, {Smiljanic}, {Bayo},
  {Carraro}, {Casey}, {Costado}, {Damiani}, {Franciosini}, {Frasca},
  {Hourihane}, {Jofr{\'e}}, {de Laverny}, {Lewis}, {Masseron}, {Monaco},
  {Morbidelli}, {Prisinzano}, {Sacco}, \& {Zaggia}}]{Magrini2018}
{Magrini}, L., {Spina}, L., {Randich}, S., {et~al.} 2018, \aap, 617, A106

\bibitem[{{Magrini} {et~al.}(2021){Magrini}, {Vescovi}, {Casali}, {Cristallo},
  {Viscasillas V{\'a}zquez}, {Cescutti}, {Spina}, {Van Der Swaelmen}, \&
  {Randich}}]{Magrini2021A&A...646L...2M}
{Magrini}, L., {Vescovi}, D., {Casali}, G., {et~al.} 2021, \aap, 646, L2

\bibitem[{{Magrini} {et~al.}(2023{\natexlab{b}}){Magrini}, {Viscasillas
  V{\'a}zquez}, {Spina}, {Randich}, {Romano}, {Franciosini}, {Recio-Blanco},
  {Nordlander}, {D'Orazi}, {Baratella}, {Smiljanic}, {Dantas}, {Pasquini},
  {Spitoni}, {Casali}, {Van der Swaelmen}, {Bensby}, {Stonkute}, {Feltzing},
  {Sacco}, {Bragaglia}, {Pancino}, {Heiter}, {Biazzo}, {Gilmore}, {Bergemann},
  {Tautvai{\v{s}}ien{\.{e}}}, {Worley}, {Hourihane}, {Gonneau}, \&
  {Morbidelli}}]{Magrini2023A&A...669A.119M}
{Magrini}, L., {Viscasillas V{\'a}zquez}, C., {Spina}, L., {et~al.}
  2023{\natexlab{b}}, \aap, 669, A119

\bibitem[{{Mainieri} {et~al.}(2024){Mainieri}, {Anderson}, {Brinchmann},
  {Cimatti}, {Ellis}, {Hill}, {Kneib}, {McLeod}, {Opitom}, {Roth},
  {Sanchez-Saez}, {Smiljanic}, {Tolstoy}, {Bacon}, {Randich}, {Adamo},
  {Annibali}, {Arevalo}, {Audard}, {Barsanti}, {Battaglia}, {Bayo Aran},
  {Belfiore}, {Bellazzini}, {Bellini}, {Beltran}, {Berni}, {Bianchi}, {Biazzo},
  {Bisero}, {Bisogni}, {Bland-Hawthorn}, {Blondin}, {Bodensteiner}, {Boffin},
  {Bonito}, {Bono}, {Bouche}, {Bowman}, {Braga}, {Bragaglia}, {Branchesi},
  {Brucalassi}, {Bryant}, {Bryson}, {Busa}, {Camera}, {Carbone}, {Casali},
  {Casali}, {Casasola}, {Castro}, {Catelan}, {Cavallo}, {Chiappini}, {Cioni},
  {Colless}, {Colzi}, {Contarini}, {Couch}, {D'Ammando}, {d'Assignies D.},
  {D'Orazi}, {da Silva}, {Dainotti}, {Damiani}, {Danielski}, {De Cia}, {de
  Jong}, {Dhawan}, {Dierickx}, {Driver}, {Dupletsa}, {Escoffier}, {Escorza},
  {Fabrizio}, {Fiorentino}, {Fontana}, {Fontani}, {Forero Sanchez}, {Franois},
  {Galindo-Guil}, {Gallazzi}, {Galli}, {Garcia}, {Garcia-Rojas}, {Garilli},
  {Grand}, {Guarcello}, {Hazra}, {Helmi}, {Herrero}, {Iglesias}, {Ilic},
  {Irsic}, {Ivanov}, {Izzo}, {Jablonka}, {Joachimi}, {Kakkad}, {Kamann},
  {Koposov}, {Kordopatis}, {Kovacevic}, {Kraljic}, {Kuncarayakti}, {Kwon}, {La
  Forgia}, {Lahav}, {Laigle}, {Lazzarin}, {Leaman}, {Leclercq}, {Lee}, {Lee},
  {Lehnert}, {Lira}, {Loffredo}, {Lucatello}, {Magrini}, {Maguire}, {Mahler},
  {Zahra Majidi}, {Malavasi}, {Mannucci}, {Marconi}, {Martin}, {Marulli},
  {Massari}, {Matsuno}, {Mattheee}, {McGee}, {Merc}, {Merle}, {Miglio},
  {Migliorini}, {Minchev}, {Minniti}, {Miret-Roig}, {Monreal Ibero}, {Montano},
  {Montet}, {Moresco}, {Moretti}, {Moscardini}, {Moya}, {Mueller},
  {Nanayakkara}, {Nicholl}, {Nordlander}, {Onori}, {Padovani}, {Pala}, {Panda},
  {Pandey-Pommier}, {Pasquini}, {Pawlak}, {Pessi}, {Pisani}, {Popovic},
  {Prisinzano}, {Raddi}, {Rainer}, {Rebassa-Mansergas}, {Richard}, {Rigault},
  {Rocher}, {Romano}, {Rosati}, {Sacco}, {Sanchez-Janssen}, {Sander},
  {Sanders}, {Sargent}, {Sarpa}, {Schimd}, {Schipani}, {Sefusatti}, {Smith},
  {Spina}, {Steinmetz}, {Tacchella}, {Tautvaisiene}, {Theissen}, {Thomas},
  {Ting}, {Travouillon}, {Tresse}, {Trivedi}, {Tsantaki}, {Tsedrik}, {Urrutia},
  {Valenti}, {Van der Swaelmen}, {Van Eck}, {Verdiani}, {Verdier}, {Vergani},
  {Verhamme}, {Vernet}, {Verza}, {Viel}, {Vielzeuf}, {Vietri}, {Vink},
  {Viscasillas Vazquez}, {Wang}, {Weilbacher}, {Wendt}, {Wright}, {Ye},
  {Yeche}, {Yu}, {Zafar}, {Zibetti}, {Ziegler}, \&
  {Zinchenko}}]{Mainieri2024arXiv240305398M}
{Mainieri}, V., {Anderson}, R.~I., {Brinchmann}, J., {et~al.} 2024, arXiv
  e-prints, arXiv:2403.05398

\bibitem[{{Maiorca} {et~al.}(2012){Maiorca}, {Magrini}, {Busso}, {Randich},
  {Palmerini}, \& {Trippella}}]{Maiorca2012}
{Maiorca}, E., {Magrini}, L., {Busso}, M., {et~al.} 2012, \apj, 747, 53

\bibitem[{{Maiorca} {et~al.}(2011){Maiorca}, {Randich}, {Busso}, {Magrini}, \&
  {Palmerini}}]{Maiorca2011}
{Maiorca}, E., {Randich}, S., {Busso}, M., {Magrini}, L., \& {Palmerini}, S.
  2011, \apj, 736, 120

\bibitem[{{Manea} {et~al.}(2023){Manea}, {Hawkins}, {Ness}, {Buder}, {Martell},
  \& {Zucker}}]{Manea2023arXiv231015257M}
{Manea}, C., {Hawkins}, K., {Ness}, M.~K., {et~al.} 2023, arXiv e-prints,
  arXiv:2310.15257

\bibitem[{{Masseron} \& {Gilmore}(2015)}]{Masseron2015MNRAS.453.1855M}
{Masseron}, T. \& {Gilmore}, G. 2015, \mnras, 453, 1855

\bibitem[{{Matteucci}(2012)}]{Matteucci2012}
{Matteucci}, F. 2012, {Chemical Evolution of Galaxies}

\bibitem[{{Matteucci} \& {Francois}(1989)}]{Matteucci1989}
{Matteucci}, F. \& {Francois}, P. 1989, \mnras, 239, 885

\bibitem[{{Matteucci} {et~al.}(2014){Matteucci}, {Romano}, {Arcones},
  {Korobkin}, \& {Rosswog}}]{Matteucci2014}
{Matteucci}, F., {Romano}, D., {Arcones}, A., {Korobkin}, O., \& {Rosswog}, S.
  2014, \mnras, 438, 2177

\bibitem[{{McKee} {et~al.}(2015){McKee}, {Parravano}, \&
  {Hollenbach}}]{McKee2015}
{McKee}, C.~F., {Parravano}, A., \& {Hollenbach}, D.~J. 2015, \apj, 814, 13

\bibitem[{{Miglio} {et~al.}(2021){Miglio}, {Chiappini}, {Mackereth}, {Davies},
  {Brogaard}, {Casagrande}, {Chaplin}, {Girardi}, {Kawata}, {Khan}, {Izzard},
  {Montalb{\'a}n}, {Mosser}, {Vincenzo}, {Bossini}, {Noels}, {Rodrigues},
  {Valentini}, \& {Mandel}}]{Miglio2021}
{Miglio}, A., {Chiappini}, C., {Mackereth}, J.~T., {et~al.} 2021, \aap, 645,
  A85

\bibitem[{{Mikolaitis} {et~al.}(2017){Mikolaitis}, {de Laverny},
  {Recio-Blanco}, {Hill}, {Worley}, \& {de Pascale}}]{Mikolaitis2017}
{Mikolaitis}, {\v{S}}., {de Laverny}, P., {Recio-Blanco}, A., {et~al.} 2017,
  \aap, 600, A22

\bibitem[{{Mints} \& {Hekker}(2018)}]{Mints2018A&A...618A..54M}
{Mints}, A. \& {Hekker}, S. 2018, \aap, 618, A54

\bibitem[{{Mishenina} {et~al.}(2015){Mishenina}, {Pignatari}, {Carraro},
  {Kovtyukh}, {Monaco}, {Korotin}, {Shereta}, {Yegorova}, \&
  {Herwig}}]{Mishenina2015}
{Mishenina}, T., {Pignatari}, M., {Carraro}, G., {et~al.} 2015, \mnras, 446,
  3651

\bibitem[{{Molero} {et~al.}(2023){Molero}, {Magrini}, {Matteucci}, {Romano},
  {Palla}, {Cescutti}, {Viscasillas V{\'a}zquez}, \& {Spitoni}}]{Molero2023}
{Molero}, M., {Magrini}, L., {Matteucci}, F., {et~al.} 2023, \mnras, 523, 2974

\bibitem[{{Molero} {et~al.}(2024){Molero}, {Matteucci}, {Spitoni},
  {Rojas-Arriagada}, \& {Rich}}]{Molero2024}
{Molero}, M., {Matteucci}, F., {Spitoni}, E., {Rojas-Arriagada}, A., \& {Rich},
  R.~M. 2024, arXiv e-prints, arXiv:2405.12585

\bibitem[{{Molero} {et~al.}(2021){Molero}, {Romano}, {Reichert}, {Matteucci},
  {Arcones}, {Cescutti}, {Simonetti}, {Hansen}, \& {Lanfranchi}}]{Molero2021}
{Molero}, M., {Romano}, D., {Reichert}, M., {et~al.} 2021, \mnras, 505, 2913

\bibitem[{{Montalb{\'a}n} {et~al.}(2021){Montalb{\'a}n}, {Mackereth}, {Miglio},
  {Vincenzo}, {Chiappini}, {Buldgen}, {Mosser}, {Noels}, {Scuflaire}, {Vrard},
  {Willett}, {Davies}, {Hall}, {Nielsen}, {Khan}, {Rendle}, {van Rossem},
  {Ferguson}, \& {Chaplin}}]{Montalbal2021}
{Montalb{\'a}n}, J., {Mackereth}, J.~T., {Miglio}, A., {et~al.} 2021, Nature
  Astronomy, 5, 640

\bibitem[{{Mor} {et~al.}(2019){Mor}, {Robin}, {Figueras}, {Roca-F{\`a}brega},
  \& {Luri}}]{Mor2019}
{Mor}, R., {Robin}, A.~C., {Figueras}, F., {Roca-F{\`a}brega}, S., \& {Luri},
  X. 2019, \aap, 624, L1

\bibitem[{{Moya} {et~al.}(2022){Moya}, {Sarro}, {Delgado-Mena}, {Chaplin},
  {Adibekyan}, \& {Blanco-Cuaresma}}]{Moya2022A&A...660A..15M}
{Moya}, A., {Sarro}, L.~M., {Delgado-Mena}, E., {et~al.} 2022, \aap, 660, A15

\bibitem[{{Nishimura} {et~al.}(2017){Nishimura}, {Sawai}, {Takiwaki}, {Yamada},
  \& {Thielemann}}]{Nishimura2017}
{Nishimura}, N., {Sawai}, H., {Takiwaki}, T., {Yamada}, S., \& {Thielemann},
  F.~K. 2017, \apjl, 836, L21

\bibitem[{{Nissen} {et~al.}(2017){Nissen}, {Silva Aguirre},
  {Christensen-Dalsgaard}, {Collet}, {Grundahl}, \& {Slumstrup}}]{Nissen2017}
{Nissen}, P.~E., {Silva Aguirre}, V., {Christensen-Dalsgaard}, J., {et~al.}
  2017, \aap, 608, A112

\bibitem[{{Palla} {et~al.}(2024){Palla}, {Magrini}, {Spitoni}, {Matteucci},
  {Viscasillas V{\'a}zquez}, {Franchini}, {Molero}, \& {Randich}}]{Palla2024}
{Palla}, M., {Magrini}, L., {Spitoni}, E., {et~al.} 2024, arXiv e-prints,
  arXiv:2408.17395

\bibitem[{{Palla} {et~al.}(2020){Palla}, {Matteucci}, {Spitoni}, {Vincenzo}, \&
  {Grisoni}}]{Palla2020}
{Palla}, M., {Matteucci}, F., {Spitoni}, E., {Vincenzo}, F., \& {Grisoni}, V.
  2020, \mnras, 498, 1710

\bibitem[{{Palla} {et~al.}(2022){Palla}, {Santos-Peral}, {Recio-Blanco}, \&
  {Matteucci}}]{palla2022}
{Palla}, M., {Santos-Peral}, P., {Recio-Blanco}, A., \& {Matteucci}, F. 2022,
  \aap, 663, A125

\bibitem[{{Pancino} {et~al.}(2017){Pancino}, {Lardo}, {Altavilla}, {Marinoni},
  {Ragaini}, {Cocozza}, {Bellazzini}, {Sabbi}, {Zoccali}, {Donati}, {Heiter},
  {Koposov}, {Blomme}, {Morel}, {S{\'\i}mon-D{\'\i}az}, {Lobel}, {Soubiran},
  {Montalban}, {Valentini}, {Casey}, {Blanco-Cuaresma}, {Jofr{\'e}}, {Worley},
  {Magrini}, {Hourihane}, {Fran{\c{c}}ois}, {Feltzing}, {Gilmore}, {Randich},
  {Asplund}, {Bonifacio}, {Drew}, {Jeffries}, {Micela}, {Vallenari}, {Alfaro},
  {Allende Prieto}, {Babusiaux}, {Bensby}, {Bragaglia}, {Flaccomio}, {Hambly},
  {Korn}, {Lanzafame}, {Smiljanic}, {Van Eck}, {Walton}, {Bayo}, {Carraro},
  {Costado}, {Damiani}, {Edvardsson}, {Franciosini}, {Frasca}, {Lewis},
  {Monaco}, {Morbidelli}, {Prisinzano}, {Sacco}, {Sbordone}, {Sousa}, {Zaggia},
  \& {Koch}}]{Pancino2017A&A...598A...5P}
{Pancino}, E., {Lardo}, C., {Altavilla}, G., {et~al.} 2017, \aap, 598, A5

\bibitem[{{Pignatari} {et~al.}(2010){Pignatari}, {Gallino}, {Heil}, {Wiescher},
  {K{\"a}ppeler}, {Herwig}, \& {Bisterzo}}]{pignatari2010}
{Pignatari}, M., {Gallino}, R., {Heil}, M., {et~al.} 2010, \apj, 710, 1557

\bibitem[{{Prantzos} {et~al.}(2020){Prantzos}, {Abia}, {Cristallo}, {Limongi},
  \& {Chieffi}}]{Prantzos2020}
{Prantzos}, N., {Abia}, C., {Cristallo}, S., {Limongi}, M., \& {Chieffi}, A.
  2020, \mnras, 491, 1832

\bibitem[{{Prantzos} {et~al.}(2018){Prantzos}, {Abia}, {Limongi}, {Chieffi}, \&
  {Cristallo}}]{Prantzos2018}
{Prantzos}, N., {Abia}, C., {Limongi}, M., {Chieffi}, A., \& {Cristallo}, S.
  2018, \mnras, 476, 3432

\bibitem[{{Randich} {et~al.}(2022){Randich}, {Gilmore}, {Magrini}, {Sacco},
  {Jackson}, {Jeffries}, {Worley}, {Hourihane}, {Gonneau}, {Viscasillas
  Vazquez}, {Franciosini}, {Lewis}, {Alfaro}, {Allende Prieto}, {Bensby},
  {Blomme}, {Bragaglia}, {Flaccomio}, {Fran{\c{c}}ois}, {Irwin}, {Koposov},
  {Korn}, {Lanzafame}, {Pancino}, {Recio-Blanco}, {Smiljanic}, {Van Eck},
  {Zwitter}, {Asplund}, {Bonifacio}, {Feltzing}, {Binney}, {Drew}, {Ferguson},
  {Micela}, {Negueruela}, {Prusti}, {Rix}, {Vallenari}, {Bayo}, {Bergemann},
  {Biazzo}, {Carraro}, {Casey}, {Damiani}, {Frasca}, {Heiter}, {Hill},
  {Jofr{\'e}}, {de Laverny}, {Lind}, {Marconi}, {Martayan}, {Masseron},
  {Monaco}, {Morbidelli}, {Prisinzano}, {Sbordone}, {Sousa}, {Zaggia},
  {Adibekyan}, {Bonito}, {Caffau}, {Daflon}, {Feuillet}, {Gebran}, {Gonzalez
  Hernandez}, {Guiglion}, {Herrero}, {Lobel}, {Maiz Apellaniz}, {Merle},
  {Mikolaitis}, {Montes}, {Morel}, {Soubiran}, {Spina}, {Tabernero},
  {Tautvai{\v{s}}iene}, {Traven}, {Valentini}, {Van der Swaelmen}, {Villanova},
  {Wright}, {Abbas}, {Aguirre B{\o}rsen-Koch}, {Alves}, {Balaguer-Nunez},
  {Barklem}, {Barrado}, {Berlanas}, {Binks}, {Bressan}, {Capuzzo-Dolcetta},
  {Casagrande}, {Casamiquela}, {Collins}, {D'Orazi}, {Dantas}, {Debattista},
  {Delgado-Mena}, {Di Marcantonio}, {Drazdauskas}, {Evans}, {Famaey},
  {Franchini}, {Fr{\'e}mat}, {Friel}, {Fu}, {Geisler}, {Gerhard}, {Gonzalez
  Solares}, {Grebel}, {Gutierrez Albarran}, {Hatzidimitriou}, {Held},
  {Jim{\'e}nez-Esteban}, {J{\"o}nsson}, {Jordi}, {Khachaturyants},
  {Kordopatis}, {Kos}, {Lagarde}, {Mahy}, {Mapelli}, {Marfil}, {Martell},
  {Messina}, {Miglio}, {Minchev}, {Moitinho}, {Montalban}, {Monteiro},
  {Morossi}, {Mowlavi}, {Mucciarelli}, {Murphy}, {Nardetto}, {Ortolani},
  {Paletou}, {Palou{\v{s}}}, {Paunzen}, {Pickering}, {Quirrenbach}, {Re
  Fiorentin}, {Read}, {Romano}, {Ryde}, {Sanna}, {Santos}, {Seabroke},
  {Spagna}, {Steinmetz}, {Stonkut{\'e}}, {Sutorius}, {Th{\'e}venin}, {Tosi},
  {Tsantaki}, {Vink}, {Wright}, {Wyse}, {Zoccali}, {Zorec}, {Zucker}, \&
  {Walton}}]{Randich2022}
{Randich}, S., {Gilmore}, G., {Magrini}, L., {et~al.} 2022, \aap, 666, A121

\bibitem[{{Ratcliffe} {et~al.}(2024){Ratcliffe}, {Minchev}, {Cescutti},
  {Spitoni}, {J{\"o}nsson}, {Anders}, {Queiroz}, \&
  {Steinmetz}}]{Ratcliffe2024MNRAS.528.3464R}
{Ratcliffe}, B., {Minchev}, I., {Cescutti}, G., {et~al.} 2024, \mnras, 528,
  3464

\bibitem[{{Recio-Blanco} {et~al.}(2014){Recio-Blanco}, {de Laverny},
  {Kordopatis}, {Helmi}, {Hill}, {Gilmore}, {Wyse}, {Adibekyan}, {Randich},
  {Asplund}, {Feltzing}, {Jeffries}, {Micela}, {Vallenari}, {Alfaro}, {Allende
  Prieto}, {Bensby}, {Bragaglia}, {Flaccomio}, {Koposov}, {Korn}, {Lanzafame},
  {Pancino}, {Smiljanic}, {Jackson}, {Lewis}, {Magrini}, {Morbidelli},
  {Prisinzano}, {Sacco}, {Worley}, {Hourihane}, {Bergemann}, {Costado},
  {Heiter}, {Joffre}, {Lardo}, {Lind}, \& {Maiorca}}]{recio-blanco2014}
{Recio-Blanco}, A., {de Laverny}, P., {Kordopatis}, G., {et~al.} 2014, \aap,
  567, A5

\bibitem[{{Rizzuti} {et~al.}(2019){Rizzuti}, {Cescutti}, {Matteucci},
  {Chieffi}, {Hirschi}, \& {Limongi}}]{rizzuti2019}
{Rizzuti}, F., {Cescutti}, G., {Matteucci}, F., {et~al.} 2019, \mnras, 489,
  5244

\bibitem[{{Rizzuti} {et~al.}(2021){Rizzuti}, {Cescutti}, {Matteucci},
  {Chieffi}, {Hirschi}, {Limongi}, \& {Saro}}]{Rizzuti2021}
{Rizzuti}, F., {Cescutti}, G., {Matteucci}, F., {et~al.} 2021, \mnras, 502,
  2495

\bibitem[{{Romano} {et~al.}(2010){Romano}, {Karakas}, {Tosi}, \&
  {Matteucci}}]{Romano2010}
{Romano}, D., {Karakas}, A.~I., {Tosi}, M., \& {Matteucci}, F. 2010, \aap, 522,
  A32

\bibitem[{{Romano} {et~al.}(2000){Romano}, {Matteucci}, {Salucci}, \&
  {Chiappini}}]{Romano2000}
{Romano}, D., {Matteucci}, F., {Salucci}, P., \& {Chiappini}, C. 2000, \apj,
  539, 235

\bibitem[{{Romano} {et~al.}(2019){Romano}, {Matteucci}, {Zhang}, {Ivison}, \&
  {Ventura}}]{Romano2019}
{Romano}, D., {Matteucci}, F., {Zhang}, Z.-Y., {Ivison}, R.~J., \& {Ventura},
  P. 2019, \mnras, 490, 2838

\bibitem[{{Ruiz-Lara} {et~al.}(2020){Ruiz-Lara}, {Gallart}, {Bernard}, \&
  {Cassisi}}]{Ruiz-Lara2020}
{Ruiz-Lara}, T., {Gallart}, C., {Bernard}, E.~J., \& {Cassisi}, S. 2020, Nature
  Astronomy, 4, 965

\bibitem[{{Shejeelammal} {et~al.}(2024){Shejeelammal}, {Mel{\'e}ndez},
  {Rathsam}, \& {Martos}}]{Sheejeelammal2024}
{Shejeelammal}, J., {Mel{\'e}ndez}, J., {Rathsam}, A., \& {Martos}, G. 2024,
  \aap, 690, A107

\bibitem[{{Sheminova} {et~al.}(2024){Sheminova}, {Baratella}, \&
  {D'Orazi}}]{Sheminova2024}
{Sheminova}, V., {Baratella}, M., \& {D'Orazi}, V. 2024, arXiv e-prints,
  arXiv:2407.14808

\bibitem[{{Simonetti} {et~al.}(2019){Simonetti}, {Matteucci}, {Greggio}, \&
  {Cescutti}}]{Simonetti2019}
{Simonetti}, P., {Matteucci}, F., {Greggio}, L., \& {Cescutti}, G. 2019,
  \mnras, 486, 2896

\bibitem[{{Soderblom}(2010)}]{Soderblom2010ARA&A..48..581S}
{Soderblom}, D.~R. 2010, \araa, 48, 581

\bibitem[{{Spina} {et~al.}(2018){Spina}, {Mel{\'e}ndez}, {Karakas}, {dos
  Santos}, {Bedell}, {Asplund}, {Ram{\'\i}rez}, {Yong}, {Alves-Brito}, {Bean},
  \& {Dreizler}}]{Spina2018}
{Spina}, L., {Mel{\'e}ndez}, J., {Karakas}, A.~I., {et~al.} 2018, \mnras, 474,
  2580

\bibitem[{{Spina} {et~al.}(2017){Spina}, {Randich}, {Magrini}, {Jeffries},
  {Friel}, {Sacco}, {Pancino}, {Bonito}, {Bravi}, {Franciosini}, {Klutsch},
  {Montes}, {Gilmore}, {Vallenari}, {Bensby}, {Bragaglia}, {Flaccomio},
  {Koposov}, {Korn}, {Lanzafame}, {Smiljanic}, {Bayo}, {Carraro}, {Casey},
  {Costado}, {Damiani}, {Donati}, {Frasca}, {Hourihane}, {Jofr{\'e}}, {Lewis},
  {Lind}, {Monaco}, {Morbidelli}, {Prisinzano}, {Sousa}, {Worley}, \&
  {Zaggia}}]{Spina2017}
{Spina}, L., {Randich}, S., {Magrini}, L., {et~al.} 2017, \aap, 601, A70

\bibitem[{{Spitoni} {et~al.}(2024){Spitoni}, {Matteucci}, {Gratton},
  {Ratcliffe}, {Minchev}, \& {Cescutti}}]{Spitoni2024}
{Spitoni}, E., {Matteucci}, F., {Gratton}, R., {et~al.} 2024, \aap, 690, A208

\bibitem[{{Spitoni} {et~al.}(2023){Spitoni}, {Recio-Blanco}, {de Laverny},
  {Palicio}, {Kordopatis}, {Schultheis}, {Contursi}, {Poggio}, {Romano}, \&
  {Matteucci}}]{Spitoni2023}
{Spitoni}, E., {Recio-Blanco}, A., {de Laverny}, P., {et~al.} 2023, \aap, 670,
  A109

\bibitem[{{Spitoni} {et~al.}(2019){Spitoni}, {Silva Aguirre}, {Matteucci},
  {Calura}, \& {Grisoni}}]{Spitoni2019}
{Spitoni}, E., {Silva Aguirre}, V., {Matteucci}, F., {Calura}, F., \&
  {Grisoni}, V. 2019, \aap, 623, A60

\bibitem[{{Spitoni} {et~al.}(2021){Spitoni}, {Verma}, {Silva Aguirre},
  {Vincenzo}, {Matteucci}, {Vai{\v{c}}ekauskait{\.{e}}}, {Palla}, {Grisoni}, \&
  {Calura}}]{Spitoni2021}
{Spitoni}, E., {Verma}, K., {Silva Aguirre}, V., {et~al.} 2021, \aap, 647, A73

\bibitem[{{Spoo} {et~al.}(2022){Spoo}, {Tayar}, {Frinchaboy}, {Cunha}, {Myers},
  {Donor}, {Majewski}, {Bizyaev}, {Garc{\'\i}a-Hern{\'a}ndez}, {J{\"o}nsson},
  {Lane}, {Pan}, {Longa-Pe{\~n}a}, \& {Roman-Lopes}}]{Spoo2022AJ....163..229S}
{Spoo}, T., {Tayar}, J., {Frinchaboy}, P.~M., {et~al.} 2022, \aj, 163, 229

\bibitem[{{Tautvai{\v{s}}ien{\.{e}}} {et~al.}(2021){Tautvai{\v{s}}ien{\.{e}}},
  {Viscasillas V{\'a}zquez}, {Mikolaitis}, {Stonkut{\.{e}}},
  {Minkevi{\v{c}}i{\={u}}t{\.{e}}}, {Drazdauskas}, \&
  {Bagdonas}}]{Tautvaisiene2021}
{Tautvai{\v{s}}ien{\.{e}}}, G., {Viscasillas V{\'a}zquez}, C., {Mikolaitis},
  {\v{S}}., {et~al.} 2021, \aap, 649, A126

\bibitem[{{Tucci Maia} {et~al.}(2016){Tucci Maia}, {Ram{\'\i}rez},
  {Mel{\'e}ndez}, {Bedell}, {Bean}, \& {Asplund}}]{TucciMaia2016}
{Tucci Maia}, M., {Ram{\'\i}rez}, I., {Mel{\'e}ndez}, J., {et~al.} 2016, \aap,
  590, A32

\bibitem[{{Vescovi}(2021)}]{Vescovi2021Univ....8...16V}
{Vescovi}, D. 2021, Universe, 8, 16

\bibitem[{{Vescovi}(2023)}]{Vescovi2023EPJWC.27906001V}
{Vescovi}, D. 2023, in European Physical Journal Web of Conferences, Vol. 279,
  European Physical Journal Web of Conferences, 06001

\bibitem[{{Vescovi} {et~al.}(2021){Vescovi}, {Cristallo}, {Palmerini}, {Abia},
  \& {Busso}}]{Vescovi2021}
{Vescovi}, D., {Cristallo}, S., {Palmerini}, S., {Abia}, C., \& {Busso}, M.
  2021, \aap, 652, A100

\bibitem[{{Viscasillas V{\'a}zquez} {et~al.}(2022){Viscasillas V{\'a}zquez},
  {Magrini}, {Casali}, {Tautvai{\v{s}}ien{\.{e}}}, {Spina}, {Van der Swaelmen},
  {Randich}, {Bensby}, {Bragaglia}, {Friel}, {Feltzing}, {Sacco}, {Turchi},
  {Jim{\'e}nez-Esteban}, {D'Orazi}, {Delgado-Mena}, {Mikolaitis},
  {Drazdauskas}, {Minkevi{\v{c}}i{\={u}}t{\.{e}}}, {Stonkut{\.{e}}},
  {Bagdonas}, {Montes}, {Guiglion}, {Baratella}, {Tabernero}, {Gilmore},
  {Alfaro}, {Francois}, {Korn}, {Smiljanic}, {Bergemann}, {Franciosini},
  {Gonneau}, {Hourihane}, {Worley}, \&
  {Zaggia}}]{Viscasillas2022A&A...660A.135V}
{Viscasillas V{\'a}zquez}, C., {Magrini}, L., {Casali}, G., {et~al.} 2022,
  \aap, 660, A135

\bibitem[{Watson {et~al.}(2019)Watson, Hansen, Selsing, Koch, Malesani,
  Andersen, Fynbo, Arcones, Bauswein, Covino, Grado, Heintz, Hunt, Kouveliotou,
  Leloudas, Levan, Mazzali, \& Pian}]{Watson2019}
Watson, D., Hansen, C., Selsing, J., {et~al.} 2019, Nature, 574, 497

\bibitem[{{Weeks} {et~al.}(2024){Weeks}, {Van Eylen}, {Huber}, {Kawata},
  {Stokholm}, {Aguirre B{\o}rsen-Koch}, {Pinilla}, {Lysgaard R{\o}rsted},
  {Lykke Winther}, \& {Berger}}]{Weeks2024}
{Weeks}, A., {Van Eylen}, V., {Huber}, D., {et~al.} 2024, arXiv e-prints,
  arXiv:2411.17358

\bibitem[{{Yong} {et~al.}(2012){Yong}, {Carney}, \& {Friel}}]{Yong2012}
{Yong}, D., {Carney}, B.~W., \& {Friel}, E.~D. 2012, \aj, 144, 95

\bibitem[{{Zhang} {et~al.}(2018){Zhang}, {Zhou}, {Tang}, {Saunders}, {Venn},
  {Shi}, {McConnachie}, {Szeto}, {Wang}, {Zhu}, \&
  {Hu}}]{Zhang2018SPIE10702E..7WZ}
{Zhang}, K., {Zhou}, Y., {Tang}, Z., {et~al.} 2018, in Society of Photo-Optical
  Instrumentation Engineers (SPIE) Conference Series, Vol. 10702, Ground-based
  and Airborne Instrumentation for Astronomy VII, ed. C.~J. {Evans},
  L.~{Simard}, \& H.~{Takami}, 107027W

\end{thebibliography}

\begin{appendix}

\section{The Si production}

\begin{figure*}[h]
    \includegraphics[width=1.\textwidth]{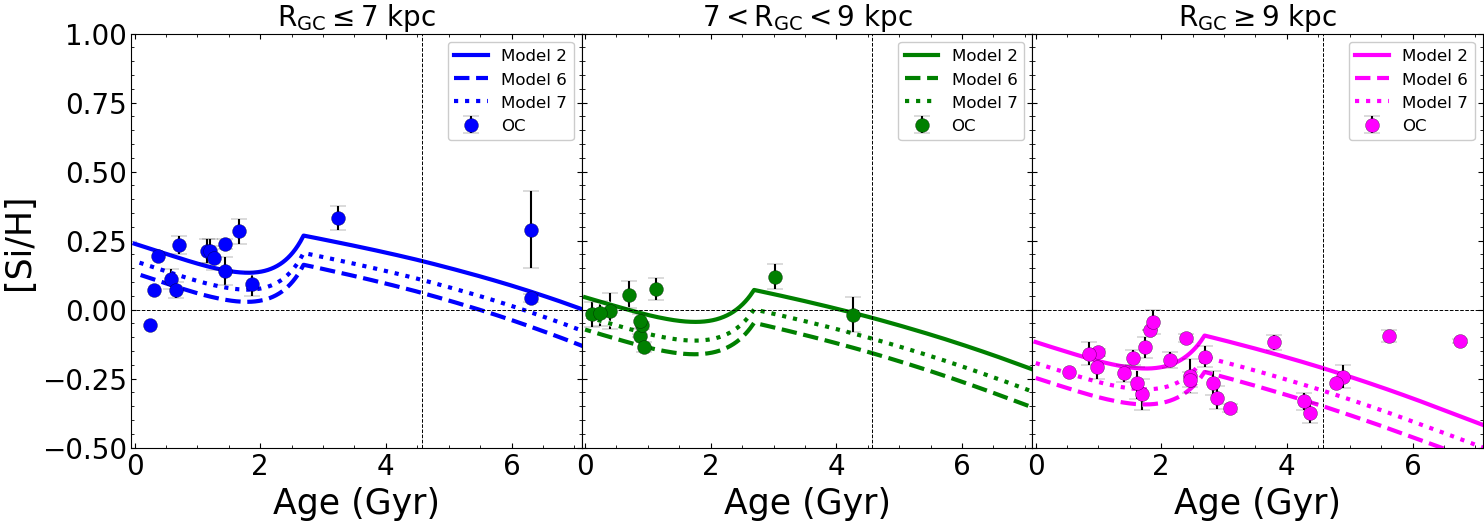}
     \caption{$\mathrm{[Si/H]}$ vs Age observed trends divided in the three Galactocentric regions of interested. Predictions of the chemical evolution model in the case of Model 2, Model 6 and Model 7 are compared to the OC sample. See Table \ref{Tab: model MS} for reference.}%
 \label{fig: SiH vs Age}%
\end{figure*}

The overall conclusions of this work remain robust, even when analyzing different chemical clocks based on various s-process and $\alpha$-elements. However, it is important to note that some discrepancies may still arise. These uncertainties can stem not only from the detailed nucleosynthesis of s-process elements but also from that of $\alpha$-elements. For instance, different $\alpha$-elements exhibit varying production rates from Type Ia supernovae, which can influence the steepness of the chemical clock trends as they evolve toward younger ages. While the general conclusions are strong, deviations may occur, particularly in the final computed missing Ba production. This missing Ba production has been estimated under the assumption that our model accurately reproduces the observed Si in the adopted OC dataset. Indeed, our model provides a good fit for the [Si/Fe] versus [Fe/H] trend, as shown and discussed in Figure \ref{fig: EuFe SiFe vs FeH} in Section \ref{sec: Nucleosynthesis prescriptions}. Figure \ref{fig: SiH vs Age} shows the [Si/H] versus Age trend for Model 2, Model 6, and Model 7, the latter two incorporating different massive star nucleosynthesis prescriptions. Model 2, used to compute the missing Ba production, shows a generally good agreement with the observed trend, particularly in the solar and outer Galactic zones. In particular, in the solar zone, the depression around $\mathrm{\sim 2\ Gyr}$ is well-reproduced. The solar Si abundance predicted by our model is $7.44$, which compares reasonably with the value of $7.51\pm0.03$ from \cite{Asplund2009}. Model 6 and 7, which both assume a distribution of initial velocities of massive stars, predict a lower trend compared with Model 2 as a greater percentage of stars with lower or null rotational velocity. This effect, as discussed in Section \ref{sec: The influence of rotating massive stars} worsen the agreement with chemical clocks based on Ba and left it unchanged in the case of Y. 

\section{The OCs sample}
Abundance ratios of the elements studied in this work, together with Galactocentric distances and ages of the sample of 62 OCs described in section \ref{sec:data} are reported in Table \ref{Tab: OCs parameters}. See \citet{Magrini2023A&A...669A.119M, Palla2024} for further discussion.

\begin{table*}
    \centering
    \captionsetup{width=1.1\textwidth}
    \caption{\label{Tab: OCs parameters} Values of [X/H] with associated uncertainties, R$_{guide}$ and age of the set of OCs from the \textit{Gaia}-ESO survey adopted in this work.}
    \begin{adjustbox}{width=\textwidth}
    \begin{tabular}{ccccccccccccc}
    \hline
       OC  & [Fe/H] & $\sigma$([Fe/H]) & [Si/H] & $\sigma$([Si/H]) & [Y/H] & $\sigma$([Y/H]) & [Ba/H] & $\sigma$([Ba/H]) & [Eu/H] & $\sigma$([Eu/H]) & R$_{guide}$ & Age \\
                     & (dex) & (dex) & (dex) & (dex) & (dex) & (dex) & (dex) & (dex) & (dex) & (dex) & (kpc) & (Gyr) \\
    \hline
Br81       &      0.22 &      0.07 &      0.21 &      0.04 &      0.24 &      0.08 &      0.13 &      0.08 &      0.22 &      0.07 &      5.61 &      1.15 \\
Rup134     &      0.26 &      0.07 &      0.30 &      0.05 &      0.15 &      0.05 &      0.11 &      0.03 &      0.07 &      0.03  &     5.22  &     1.66 \\
Trumpler23 &      0.19 &      0.06 &      0.25 &      0.03 &      0.18 &      0.02 &      0.15 &      0.11 &      0.15 &      0.06  &     5.48  &     0.71 \\
NGC6583    &      0.22 &      0.01 &      0.21 &      0.04 &      0.15 &      0.05 &      0.11 &      0.10 &      0.15 &            &     6.39  &     1.20 \\
NGC6705    &      0.20 &           &      0.07 &           &      0.27 &           &      0.38 &           &      0.17 &            &     5.80  &     0.31 \\
NGC6005    &      0.22 &      0.04 &      0.19 &      0.05 &      0.14 &      0.02 &      0.13 &      0.03 &      0.03 &      0.05  &     5.39  &     1.26 \\
NGC6192    &      0.00 &           &     -0.06 &           &      0.09 &           &      0.28 &           &      0.04 &            &     6.97  &     0.24 \\
NGC6253    &      0.35 &      0.08 &      0.34 &      0.04 &      0.37 &      0.10 &      0.26 &      0.11 &      0.16 &      0.05  &     5.81  &     3.24 \\
Pismis18   &      0.14 &      0.04 &      0.11 &      0.04 &      0.16 &      0.03 &      0.26 &      0.04 &      0.16 &      0.06  &     5.95  &     0.58 \\
Br44       &      0.21 &      0.11 &      0.14 &      0.05 &      0.06 &           &     -0.01 &      0.16 &      0.05 &      0.15  &     5.85  &     1.45 \\
NGC4815    &      0.20 &      0.21 &      0.04 &           &      0.12 &           &      0.27 &           &      0.14 &            &     6.02  &     0.37 \\
NGC6802    &      0.14 &      0.04 &      0.08 &      0.03 &      0.29 &      0.05 &      0.23 &      0.07 &      0.12 &      0.03  &     6.02  &     0.66 \\
Trumpler20 &      0.15 &      0.05 &      0.09 &      0.04 &      0.20 &      0.07 &      0.11 &      0.04 &      0.12 &      0.06  &     6.42  &     1.86 \\
Col261     &      0.03 &      0.11 &      0.04 &      0.01 &     -0.01 &      0.01 &     -0.01 &      0.10 &      0.04 &            &     6.60  &     6.31 \\
NGC4337    &      0.25 &      0.03 &      0.24 &      0.02 &      0.18 &      0.01 &      0.15 &      0.03 &      0.07 &      0.02  &     6.57  &     1.45 \\
NGC6709    &     -0.03 &      0.01 &     -0.03 &           &      0.49 &      0.56 &      0.36 &      0.35 &           &            &     7.07  &     0.19 \\
NGC3960    &     -0.06 &      0.16 &     -0.03 &      0.01 &      0.13 &      0.04 &      0.22 &      0.05 &      0.07 &      0.03  &     7.36  &     0.87 \\
NGC5822    &      0.01 &      0.02 &     -0.06 &      0.01 &      0.15 &      0.03 &      0.25 &      0.01 &      0.07 &      0.05  &     7.37  &     0.91 \\
NGC6791    &      0.25 &      0.23 &      0.29 &      0.14 &      0.36 &      0.09 &      0.25 &      0.23 &      0.37 &      0.04  &     5.56  &     6.31 \\
NGC6633    &     -0.07 &      0.11 &      0.05 &      0.05 &      0.11 &           &      0.19 &           &     -0.14 &            &     7.39  &     0.69 \\
Rup147     &      0.12 &      0.02 &      0.12 &      0.05 &      0.15 &      0.09 &      0.09 &      0.06 &      0.05 &      0.08  &     7.61  &     3.02 \\
NGC3532    &     -0.00 &      0.06 &     -0.00 &      0.06 &      0.10 &      0.08 &      0.11 &      0.06 &      0.05 &      0.14  &     7.74  &     0.40 \\
Blanco1    &     -0.02 &      0.06 &     -0.03 &      0.05 &      0.01 &      0.11 &      0.05 &      0.07 &      0.06 &      0.11  &     8.20  &     0.10 \\
NGC2516    &     -0.03 &      0.04 &     -0.02 &      0.04 &      0.06 &      0.10 &      0.02 &      0.05 &           &            &     7.47  &     0.24 \\
Pismis15   &      0.03 &      0.06 &     -0.04 &      0.03 &      0.22 &      0.03 &      0.19 &      0.04 &      0.13 &      0.04  &     7.26  &     0.87 \\
NGC2477    &      0.14 &      0.05 &      0.07 &      0.04 &      0.21 &      0.09 &      0.18 &      0.06 &      0.10 &      0.04  &     8.91  &     1.12 \\
M67        &     -0.04 &      0.10 &      0.00 &      0.06 &      0.00 &      0.08 &     -0.00 &      0.06 &     -0.01 &      0.12  &     8.36  &     4.27 \\
NGC2660    &     -0.05 &      0.04 &     -0.14 &      0.02 &      0.03 &      0.06 &      0.13 &      0.07 &     -0.02 &      0.06  &     8.43  &     0.93 \\
Melotte71  &     -0.15 &      0.10 &     -0.18 &      0.04 &      0.01 &      0.05 &      0.10 &      0.04 &      0.01 &      0.06  &    10.24  &     0.98 \\
NGC2355    &     -0.10 &      0.04 &     -0.14 &      0.01 &      0.01 &      0.06 &      0.12 &      0.04 &      0.00 &      0.04  &    10.23  &     1.00 \\
Col110     &     -0.10 &      0.05 &     -0.08 &      0.00 &      0.07 &      0.05 &      0.20 &      0.02 &      0.02 &      0.04  &     9.31  &     1.82 \\
NGC2243    &     -0.42 &      0.11 &     -0.35 &      0.04 &     -0.27 &      0.16 &     -0.34 &      0.07 &     -0.14 &      0.18  &    12.22  &     4.37 \\
NGC2420    &     -0.15 &      0.07 &     -0.14 &      0.04 &     -0.05 &      0.06 &     -0.02 &      0.04 &      0.01 &      0.08  &     9.63  &     1.74 \\
Haf10      &     -0.11 &      0.04 &     -0.12 &      0.03 &      0.04 &      0.04 &      0.00 &      0.04 &      0.06 &      0.05  &     9.19  &     3.80 \\
NGC2425    &     -0.09 &      0.10 &     -0.14 &      0.02 &      0.01 &      0.09 &     -0.03 &      0.02 &      0.06 &      0.07  &    10.01  &     2.40 \\
Br32       &     -0.27 &      0.06 &     -0.24 &      0.04 &     -0.25 &      0.02 &     -0.17 &      0.10 &     -0.06 &      0.11  &     9.08  &     4.90 \\
Trumpler5  &     -0.33 &      0.03 &     -0.33 &      0.03 &     -0.19 &      0.07 &     -0.22 &      0.05 &     -0.09 &      0.07  &    11.40  &     4.27 \\
Br39       &     -0.14 &      0.04 &     -0.09 &      0.02 &     -0.12 &      0.06 &     -0.14 &      0.04 &      0.06 &      0.04  &     9.37  &     5.62 \\
Rup4       &     -0.13 &      0.02 &     -0.16 &      0.04 &      0.02 &      0.05 &      0.09 &      0.10 &      0.06 &      0.09  &    10.61  &     0.85 \\
Br36       &     -0.16 &      0.02 &     -0.11 &      0.01 &     -0.11 &      0.08 &     -0.23 &      0.06 &      0.06 &      0.08  &    11.10  &     6.76 \\
NGC2324    &     -0.18 &           &     -0.23 &           &     -0.03 &           &      0.11 &           &     -0.15 &            &    11.54  &     0.54 \\
Cz24       &     -0.11 &      0.04 &     -0.17 &      0.04 &      0.05 &      0.05 &      0.04 &      0.05 &      0.06 &      0.02  &    10.51  &     2.69 \\
NGC2158    &     -0.16 &      0.03 &     -0.17 &      0.03 &     -0.01 &      0.05 &     -0.03 &      0.03 &      0.01 &      0.07  &    11.45  &     1.55 \\
NGC2141    &     -0.04 &      0.04 &     -0.04 &      0.04 &      0.10 &      0.06 &      0.08 &      0.03 &      0.04 &      0.05  &    13.04  &     1.86 \\
Br73       &     -0.24 &      0.02 &     -0.23 &      0.03 &     -0.17 &      0.00 &     -0.13 &      0.05 &     -0.00 &      0.03  &    13.40  &     1.41 \\
Cz30       &     -0.32 &      0.01 &     -0.32 &      0.04 &     -0.22 &      0.05 &     -0.12 &      0.09 &     -0.09 &      0.08  &    11.03  &     2.88 \\
Br25       &     -0.27 &      0.07 &     -0.24 &      0.06 &     -0.22 &      0.05 &     -0.24 &      0.05 &     -0.14 &      0.03  &    10.93  &     2.45 \\
Br22       &     -0.25 &      0.07 &     -0.25 &      0.03 &     -0.23 &      0.06 &      0.05 &      0.04 &      0.07 &      0.01  &    12.81  &     2.45 \\
Br75       &     -0.33 &      0.08 &     -0.31 &      0.06 &     -0.27 &      0.04 &     -0.22 &      0.04 &           &            &    11.25  &     1.70 \\
Br21       &     -0.17 &      0.04 &     -0.18 &      0.03 &     -0.05 &      0.12 &      0.14 &      0.30 &      0.01 &            &    13.72  &     2.14 \\
Br31       &     -0.29 &      0.08 &     -0.29 &      0.04 &     -0.23 &      0.01 &     -0.22 &      0.02 &     -0.05 &      0.06  &    12.73  &     2.82 \\
Tom2       &     -0.23 &      0.09 &     -0.27 &      0.06 &     -0.06 &      0.07 &     -0.14 &      0.02 &      0.01 &      0.12  &    13.34  &     1.62 \\
Br20       &     -0.32 &           &     -0.27 &           &     -0.27 &           &     -0.28 &           &     -0.07 &            &    14.46  &     4.79 \\
Br29       &     -0.37 &      0.04 &     -0.35 &      0.01 &     -0.08 &      0.01 &     -0.37 &      0.24 &      0.18 &            &    17.14  &     3.09 \\
    \hline&
    \end{tabular}
    \end{adjustbox}
\end{table*}

\end{appendix}

\end{document}